\documentclass[11pt,a4paper]{article} 
\usepackage[utf8]{inputenc}
\usepackage[english]{babel}
\usepackage{amsmath,bm}
\usepackage{slashed}
\usepackage{tablefootnote} 
\usepackage[colorlinks = true,
            linkcolor = blue,
            urlcolor  = blue,
            citecolor = blue,
            anchorcolor = blue]{hyperref}
\usepackage{orcidlink}
\usepackage{amsfonts}
\usepackage{amssymb}
\usepackage{multirow}
\usepackage{booktabs}
\usepackage[numbers,sort&compress]{natbib}
\usepackage{cancel}

\usepackage{makeidx}
\usepackage{authblk}
\usepackage{float}
\usepackage{mathtools}
\usepackage{graphicx}
\usepackage{titlesec}
\titleformat{\section}{\Large\bfseries\MakeUppercase}{\thesection}{1em}{}
\usepackage{subfig}
\usepackage[dvipsnames,svgnames]{xcolor}
\usepackage[left=2cm,right=2cm,top=2cm,bottom=2cm]{geometry}

\newcommand{\bea}{\begin{eqnarray}}
\newcommand{\bec}{\begin{columns}}
\newcommand{\eec}{\end{columns}}
\newcommand{\eea}{\end{eqnarray}}
\newcommand{\beq}{\begin{equation}}
\newcommand{\eeq}{\end{equation}}
\newcommand{\ec}{\end{center}}
\newcommand{\bc}{\begin{center}}

\newcommand{\stl}{\sin\theta_L}
\newcommand{\mtp}{m_{_{T_p}}}
\newcommand{\mgd}{m_{\gamma_d}}
\newcommand{\ppxssecbr}{\sigma({T_p \bar{T}_p}) \times {\rm BR}^2(T_p \to t\gamma_d)}
\newcommand{\spxssecbr}{\sigma({T_p j}) \times {\rm BR}(T_p \to t\gamma_d)}

\MakeUppercase

\title{{\Large {\bf 
Machine learning tagged boosted dark photon: \\A signature of fermionic portal matter at the LHC
}}}

\author[a]{{Shivam Verma\orcidlink{0000-0002-7670-8203}~\thanks{\href{mailto:shivam.59910103@gm.rkmvu.ac.in}
{shivam.59910103@gm.rkmvu.ac.in}}}}

\author[a]{{Sanjoy Biswas\orcidlink{0000-0003-1305-8896}~\thanks{\href{mailto:sanjoy.phy@gm.rkmvu.ac.in}{sanjoy.phy@gm.rkmvu.ac.in}}}}

\author[b]{{Tanumoy Mandal\orcidlink{0000-0001-7268-549X}~\thanks{\href{mailto:tanumoy@iisertvm.ac.in}{tanumoy@iisertvm.ac.in}}}}

\author[c, d]{{Subhadip Mitra\orcidlink{0000-0002-7107-0343}~\thanks{\href{mailto:subhadip.mitra@iiit.ac.in}{subhadip.mitra@iiit.ac.in}}}}

\affil[a]{{\it
Department of Physics, Ramakrishna Mission Vivekananda Educational and Research Institute, 
         Belur Math, Howrah 711202, India}}

\affil[b]{{\it Department of Physics, 
Indian Institute of Science Education and Research
Thiruvananthapuram, Vithura, Kerala, 695551, India} }

\affil[c]{{\it
Center for Computational Natural Science and Bioinformatics, International Institute of Information Technology, Hyderabad 500 032, India}}

\affil[d]{{\it
Center for Quantum Science and Technology, International Institute of Information Technology, Hyderabad 500 032, India}}

\date{}

\begin{document}

\maketitle
\thispagestyle{empty}

\begin{abstract}

We use a {hybrid deep neural network} (HDNN) to identify a boosted dark photon jet as a signature of a heavy {vectorlike} fermionic portal matter (PM) connecting the visible and the dark sectors. In this work, the fermionic PM, which mixes only with the Standard Model (SM) third-generation up-type quark, predominantly decays into a top quark and a dark photon pair. The dark photon then {promptly} decays to a pair of standard model fermions via the gauge kinetic mixing. We have analyzed two different final states, namely, (i) exactly one tagged dark photon and exactly one tagged top quark jet, and  (ii) at least two tagged dark photons and at least one tagged top quark jet at the 13 and 14 TeV LHC center of mass energies. Both these final states receive significant contributions from the pair and single production processes of the top partner. The rich event topology of the signal processes, {i.e.}, the presence of a boosted dark photon and top quark jet pair, along with the fact that the invariant mass of the system corresponds to the mass of the top partner, help us to significantly suppress potential SM backgrounds. We have shown that one can set a $2\sigma$ exclusion limit of {$\sim 2.3$ TeV} on the top partner mass with $\sin\theta_L=0.1$ and assuming $100\%$ branching ratio of the top partner in the final state with exactly one tagged dark photon and exactly one tagged top quark jet at the 14 TeV LHC center of mass energy assuming 300 fb$^{-1}$ of integrated luminosity.

\end{abstract}

\clearpage

\pagenumbering{arabic}

\pagestyle{plain}
\section{Introduction}

Extending the Standard Model (SM) of particle physics with an additional dark sector (DS) gives a theoretically well-motivated framework to explain the existence of dark matter (DM) and its associated phenomenology. 
So far, various theoretical models within this framework have been proposed in the literature. Many of these scenarios consider the DS interacting with the SM sector via portal interactions \cite{PhysRevD.99.115024, Rizzo:2022qan, Qin:2021cxl, Wojcik:2022rtk}. A gauge kinetic mixing portal is one such example for the DS-SM interaction. The simplest way to realize it would be to assume that the dark sector contains an {Abelian} dark $U(1)_d$ gauge group under which all the SM particles are neutral. The corresponding gauge boson, i.e., the dark photon, becomes massive via the (dark) Higgs mechanism.

Any ultraviolet completion of such a model will introduce additional {particlelike} states, known as the portal matter, which are charged under both the SM and the $U(1)_d$ gauge groups. These states can be either scalars or fermions. Here, we consider an additional {vectorlike} fermion, specifically a {vectorlike} quark (VLQ), as a portal matter that produces the gauge kinetic mixing via loop contribution. The VLQ can also mix with the SM fermions depending on its charge assignment. 
We assume the VLQ to be a singlet under the $SU(2)_L$ interaction and carry $+2/3$ unit of $U(1)_Y$ charge. 
As it also has a {nonzero} $U(1)_d$ charge, a Yukawa-type interaction involving the VLQ, a SM up-type quark, 
and the Higgs is not allowed. However, a similar Yukawa-type interaction term can be introduced with a charged dark Higgs, which can generate the mixing between SM fermions and the {vectorlike} quark as it acquires a {nonzero} vacuum expectation value (VEV). To avoid the current constraints on the VLQ mixing with the first two generations of quarks, we assume the VLQ mixes only with the third-generation quark with the same electric charge, i.e., the top quark. In a broader setup, a VLQ that transforms {nontrivially}
under $SU(2)_L$ is another possibility, but it is beyond the scope of the present article.
Earlier, Refs.~\cite{Kim:2019oyh, Verma:2022nyd} showed that, in such a setup, the VLQ could predominantly 
decay to a top quark and a dark photon or a dark Higgs, while its traditional decay modes, such as $T_{p} 
\to bW$, $tZ$, and $th$, became suppressed. The origin of the large branching ratio into these {nonstandard} 
modes can be traced to the hierarchies between (i) the SM top quark and VLQ mass, (ii) the VEVs of the two Higgs fields, and (iii) the mixing angle of the SM top quark with the VLQ and their mass ratio~\cite{Kim:2019oyh, Verma:2022nyd}. Other {nonstandard} decay modes of VLQs have attracted significant attention in the current literature~\cite{Gopalakrishna:2013hua,Karabacak:2014nca,Serra:2015xfa,Anandakrishnan:2015yfa,Banerjee:2016wls,Kraml:2016eti,Dobrescu:2016pda,Aguilar-Saavedra:2017giu,Chala:2017xgc,Moretti:2017qby,Bizot:2018tds,Colucci:2018vxz,Das:2018gcr,Han:2018hcu,Dermisek:2019vkc,Benbrik:2019zdp,Cacciapaglia:2019zmj,Wang:2020ips,Dasgupta:2021fzw,Bhardwaj:2022nko, Bhardwaj:2022wfz,Bardhan:2022sif,Alves:2023ufm,Banerjee:2023upj,Banerjee:2024zvg}.

VLQs as portal matters have been getting due attention---there are dedicated studies on their searches in collider experiments \cite{PhysRevD.99.115024, Colucci:2018vxz, Dobrescu:2016pda, Dasgupta:2021fzw, Corcella:2021mdl, Dermisek:2021zjd, duPlessis:2021xuc, Rizzo:2022qan, Wojcik:2022rtk,Verma:2022nyd}. The main focus of the current study is to analyze the LHC signatures of such a {vectorlike} quark in the mass range\footnote{{This closely corresponds to the mass range of the top partner or other heavy resonances decaying to a top quark and another boson, as considered in several experimental studies \cite{CMS:2019eqb, ATLAS:2022ozf}. In our case, we have slightly extended this range to account for the possibility that the top partner can decay into {nonstandard} modes with sizable branching ratios.}} of {$800$ GeV to $\sim 2.8$ TeV}, decaying into a highly boosted dark photon and a top quark. In this scenario, the dark photon decays into SM fermions via gauge kinetic mixing, controlled by the corresponding gauge kinetic mixing parameter, $\varepsilon$. Depending on the dark photon mass, this parameter is strongly constrained by various experiments, including low-energy ones \cite{Fabbrichesi:2020wbt,Bauer:2018onh}. For the mass range $1-10$ GeV (relevant for our LHC analysis), the bound on the kinetic mixing parameter is $\varepsilon \lesssim 10^{-3}$. However, a very small $\varepsilon$ could result in the dark photon decaying outside the detector. We fix this parameter by requiring the dark photon to decay promptly into SM final states ($\gamma_d \to f\bar{f}$, where f = leptons, quarks) inside the detector. Nevertheless, a long-lived dark photon decaying inside the detector presents an interesting possibility, which will be explored in future work.

Searching for dark photon signals at collider experiments is an ongoing quest. In this article, we propose a machine learning (ML) based method to identify a highly boosted dark photon in its hadronic decay mode, which can carry the imprint of a heavy VLQ acting as a portal matter at collider experiments. Machine learning techniques have been used widely in the field of particle physics, in particular, in the context of jet physics \cite{Cogan:2014oua, Almeida:2015jua, deOliveira:2015xxd,Kasieczka:2017nvn,ATLAS:2017dfg,Larkoski:2017jix,Datta:2017rhs,Aguilar-Saavedra:2017rzt,Albertsson:2018maf,Aguilar-Saavedra:2020uhm,Chung:2020ysf,Andrews:2021ejw,Kim:2022miv,Aguilar-Saavedra:2023pde}. 
To identify dark photon jets we employ an algorithm based on a {hybrid deep neural network} (HDNN) framework. This approach combines a {convolutional neural network} (CNN) and a {multilayer perceptron} (MLP) \cite{CMS:2022prd,Hammad:2022lzo, Ban:2023jfo}. The CNN models jet images, and the MLP captures the jet-level features of both signal (initiated by dark photons) and background (initiated by other particles) jets. We focus on a light dark photon of mass around $10$ GeV, which is more likely to be mimicked by background jets, making the jet invariant mass a poor discriminator. {We show how the use of HDNN can significantly improve the tagging efficiency of such low-mass dark photons over that obtained by naive cut based selection algorithm.}

We emphasize that one of our objectives is to provide a method for identifying a boosted dark-photon jet among various other jets present in an event. This task essentially involves the {\it binary classification} between a dark photon jet and other {nondark} photon jets, such as those initiated by quarks, gluons, $\tau$ leptons, or top quarks, across a wide range of transverse momenta. Moreover, even though we consider dark photon production in the decay of a heavy VLQ that mixes only with the up-type third-generation quark in this paper, our analysis of dark photon tagging is more general and applicable to VLQs mixing with other generations as well. The tagging algorithm functions effectively across a wide range of VLQ masses, with only a moderate change in the tagging efficiency. The technique employed in this work can be used whenever the decay of a heavy particle produces a highly boosted ``light exotic particle'' (such as the dark photon).

The present work focuses on two different final states: {\it (i)} exactly one tagged dark photon and exactly one tagged top quark jet (ED1ET1), and {\it (ii)} at least two tagged dark photons and at least one tagged top quark jet (AD2AT1) in the context of the $13$ and $14$ TeV LHC center-of-mass energies. Most studies in the literature focus on single and pair production separately while by choosing different final states. However, disentangling these two processes is generally challenging. As we will demonstrate, the final states we consider---mainly to characterize the single and pair productions of the top partner---exhibit a significant admixture of both processes. 
Moreover, the relative admixture of these two processes in a given final state strongly depends on the mixing angle between the top and top partner. Thus, it becomes crucial to analyze different final states rather than individual single and pair production processes.
We propose a set of kinematic observables, based on the topology of the signal events, to suppress the SM background and optimize the signal significance in these final states. In addition, we also consider the current LHC limits on the model and constrain the relevant parameter space.

The rest of the paper is organized as follows. In Sec.~2, we briefly describe the theoretical framework we have considered. 
Sec.~3 consists of collider analysis of the signal and background processes. Results and discussions are presented 
in Sec.~4. Finally, we conclude in Sec.~5.

\section{Theoretical framework}\label{sec:model}

In this section, we will briefly summarize the theoretical framework detailed in \cite{Kim:2019oyh, Verma:2022nyd}.
We have considered an extension of the SM with a VLQ transforming under the $SU(2)_L$ gauge group as a singlet while being 
charged under both the $U(1)_Y$ and an additional $U(1)_d$ gauge groups with corresponding charges, $+2/3$ and $+1$, respectively. 
The gauge boson corresponding to the $U(1)_d$ gauge group is referred to as ``dark photon" and the SM particles are neutral under this 
gauge group.

The model also contains a complex scalar field $\Phi_d$ that is singlet under the SM gauge group and charged under $U(1)_d$. 
The charge assignment of the relevant fields under 
$SU(3)_C \times SU(2)_L\times U(1)_Y\times U(1)_d$ gauge group are presented in Table~\ref{table:charges}.
The complex scalar field $\Phi_d$ is accountable for the mixing of the VLQ with SM fermions having same 
charges and generating the mass for the dark photon via the Higgs mechanism in the dark sector. 
In this work we have considered VLQ mixing with only the top quark to avoid flavor constraints. 

\begin{table}
  \centering
  \resizebox{0.55\columnwidth}{!}{
  \begin{tabular}{ccccc}
    \toprule
    Fields & $SU(3)_C$ & $SU(2)_L$ & $Y$ & $Y_d$\\
    \midrule
    \midrule
    $t^{'}_{R}$ & {\it 3} & {\it 1} & 2/3 & 0 \\
    
    $b_R$ & {\it 3} & {\it 1} & -1/3 &  0 \\
    
    $Q_L=\begin{pmatrix}
    t^{'}_{L}\\
    b_L
    \end{pmatrix}$ & {\it 3} & {\it 2} & 1/6 & 0 \\
    
    $\Phi$ & {\it 1} & {\it 2} & 1/2 & 0 \\
    
    $T^{'}_{L}$ & {\it 3} & {\it 1} & 2/3 & 1 \\
    
    $T^{'}_{R}$ & {\it 3} & {\it 1} & 2/3 & 1 \\
    
    $\Phi_d$ & {\it 1} & {\it 1} & 0 & 1 \\
    \bottomrule
  \end{tabular}
  }
  \caption{{  Representations of the relevant fields under the full symmetry group of the theory. 
          Here, $t^{\prime}$ is the third generation up-type quark of the SM in interaction basis and $T^{\prime}$ is the 
          {vectorlike} top partner also in the interaction basis. }}
\label{table:charges}
\end{table} 

Below we describe three relevant sectors corresponding to the above-mentioned model, namely, the
(i) gauge ($\mathcal{L}_{\rm Gauge}$),
(ii) scalar ($\mathcal{L}_{\rm Scalar}$), and
(iii) fermion ($\mathcal{L}_{\rm Fermion}$) sectors.

\begin{itemize}
  \item {Gauge sector:} 
  \bea\label{eq:L_gauge}
    {\mathcal{L}}_{\rm Gauge}=-\frac{1}{4}G_{\mu\nu}^a G^{a,\mu\nu}-\frac{1}{4}W_{\mu\nu}^i W^{i,\mu\nu}-
    \frac{1}{4}B_{\mu\nu}^{\prime} B^{\prime\mu\nu}+\frac{\varepsilon^{\prime}}{2\cos\theta_W}B_{d,\mu\nu}^{\prime}
    B^{\prime\mu\nu}-\frac{1}{4}B_{d,\mu\nu}^{\prime} B_d^{\prime,\mu\nu},
  \eea
  
  where $G_{\mu\nu}^a$ are the $SU(3)_C$ field strength tensor with $a=1,\cdots,8$; $W_{\mu\nu}^i$ are $SU(2)_L$ field strength 
  tensor with $i=1,2,3$;  $B_{\mu\nu}^{\prime}$ is that of the $U(1)_Y$  and $B_{d,\mu\nu}^{\prime}$ corresponds the field strength 
  tensor of the additional $U(1)_d$ group. The kinetic mixing term between the $U(1)_Y$ and $U(1)_d$ field strength tensors 
  is {parametrized} by $\varepsilon^{\prime}$.

  \item {Scalar sector:} 
  
  \bea\label{eq:L_scalar}
    \mathcal{L}_{\text{Scalar}} = |D_{\mu}\Phi|^2 + |D_{\mu}\Phi_d|^2 - V(\Phi, \Phi_d)
  \eea
    
  Here, $\Phi$ represents the SM Higgs field, while $\Phi_d$ denotes the corresponding Higgs field in the dark sector. 
  The gauge covariant derivative in general has the form:
    
  \bea\label{eq:cov_der}
  D_{\mu} = \partial_{\mu} - ig_{_S} t^a G_{\mu}^a - ig T^i W_{\mu}^i - ig' Y B_{\mu}' - ig_d' Y_d B_{d,\mu}'
  \eea
    
  In this expression, $g_{_S}$, $g$, $g'$, and $g_d'$ are the coupling constants corresponding to the $SU(3)_C$, 
  $SU(2)_L$, $U(1)_Y$, and $U(1)_d$ gauge groups, respectively. 
  Here, $t^a$ and $T^i$ are the generators of the $SU(3)_C$ and $SU(2)_L$ groups, respectively in appropriate representation.
  Also, $Y$ and $Y_d$ corresponds to the $U(1)_Y$ and $U(1)_d$ charges, respectively.

  The scalar potential takes the form
  
  \bea\label{eq:scalar_potential}
  V(\Phi, \Phi_d) = -\mu^2|\Phi|^2 + \lambda |\Phi|^4 - \mu_{h_d}^2 |\Phi_d|^2 + \lambda_{h_d}|\Phi_d|^4 + 
  \lambda_{hh_d}|\Phi|^2|\Phi_d|^2
  \eea
    
  The fluctuation of the two scalar fields around the classical minima can be {parametrized} as
    
  \bea
  \Phi = \begin{pmatrix} 0 \\ \frac{v_{_{\text{EW}}} + h'}{\sqrt{2}} \end{pmatrix}, \quad \Phi_d = \frac{1}{\sqrt{2}}(v_d + h_d')
  \eea
  
  Where $v_{_{\rm EW}}$ is fixed at 246 GeV, by the masses of the electroweak gauge bosons while $v_d$ can be fixed by the 
  dark gauge coupling ($g_d$) and the mass of the dark photon. However, we will work with $v_d$ and $m_{\gamma_d}$ as free parameters. 

  As a consequence of this, $SU(3)_C \times SU(2)_L\times U(1)_Y\times U(1)_d$ is reduced to 
  $SU(3)_C \times U(1)_{\rm EM}$ and the gauge bosons corresponding to the broken generators becomes massive.

  The scalar mass matrix can be diagonalized using the orthogonal transformation described below, 
    
  \bea\label{eq:mass_basis_rot}
  \begin{pmatrix} h \\ h_d \end{pmatrix} = 
  \begin{pmatrix} \cos\theta_S & -\sin\theta_S \\ 
    \sin\theta_S & \cos\theta_S 
  \end{pmatrix} 
  \begin{pmatrix} h' \\ h_d' \end{pmatrix}
  \eea
    
  In the mass eigenbasis, $h$ corresponds to the observed Higgs with mass $m_h = 125$ GeV, while $h_d$ denotes the
  dark Higgs field with mass $m_{h_d}$. The free parameters of the scalar sector are the scalar mixing angle $\theta_S$, 
  $m_{h_d}$, and $v_d$.
    
  \item {Fermion sector:} 
  The lagrangian for the fermion sector focusing solely on the 3rd generation quarks and VLQ is given by,
  \bea\label{eq:Fermion lagrangian}
  \mathcal{L}_{\rm Fermion}=\bar{Q}_Li\slashed{D}Q_L+\bar{t^{\prime}}_{R}i\slashed{D}t^{\prime}_{R}+
  \bar{b}_Ri\slashed{D}b_R+ \overline{T}^{\prime}\slashed{D}T^{\prime}+\mathcal{L}_{\rm Yuk}
  \eea 
 
  where $\mathcal{L}_{\rm Yuk}$ is given by,
  \bea
    \mathcal{L}_{\rm Yuk}=
    -y_{_b} \bar{Q}_L\Phi b_R-y_{_t}\bar{Q}_L \tilde{\Phi}t^{\prime}_{R}-\lambda_T \Phi_d 
    \overline{T}^{\prime}_{L}t^{\prime}_{R}-m_{_{T}}\overline{T}^{\prime}_{L}T^{\prime}_{R}+H.c..
  \eea
 
  When $ \Phi$ and $\Phi_d$ get VEVs, the mass matrix in $t^{\prime}$ and $T^{\prime}$ basis takes the form,
  \bea
    \mathcal{L}_{u3-{\rm mass}}=-\overline{\chi}_L \mathcal{M}\chi_R + H.c.,
  \eea

  where
  \bea
  \chi_{\tau}=
  \begin{pmatrix}
    t^{\prime}_{\tau} \\
    T^{\prime}_{\tau}
    \end{pmatrix},\quad \mathcal{M}=\begin{pmatrix}
    \frac{y_t\, v_{_{\rm EW}}}{\sqrt{2}} & 0 \\
    \frac{\lambda_T \, v_d}{\sqrt{2}} & m_{_{T}}
  \end{pmatrix},
  \eea
  
  and $\tau=L,R$.

  One requires a {biunitary} transformation of the following kind in order to diagonalize the 
  above mass matrix
  \bea
    \label{eq:fermix}
    \begin{pmatrix}
    t_L \\
    T_{p_{_L}}
    \end{pmatrix}=\begin{pmatrix}
    \cos\theta_L & -\sin\theta_L \\
    \sin\theta_L & \cos\theta_L
    \end{pmatrix}
    \begin{pmatrix}
    t^{\prime}_{L} \\
    T^{\prime}_{L}
    \end{pmatrix},\quad
    \begin{pmatrix}
    t_R \\
    T_{p_{_R}}
    \end{pmatrix}=\begin{pmatrix}
    \cos\theta_R & -\sin\theta_R \\
    \sin\theta_R & \cos\theta_R
    \end{pmatrix}
    \begin{pmatrix}
    t^{\prime}_{R} \\
    T^{\prime}_{R}
    \end{pmatrix}
  \eea
  
The lighter mass eigenstate ($t$) appearing in above equation will be identified as the observed top quark having mass,
$m_t = 173.5$ GeV and $T_p$ denotes the heavier mass eigenstate, to be referred as the top partner. 
Mass of the top partner ($m_{_{T_p}}$) and the mixing angle involving the left chiral top and top partner 
fields ($\theta_{L}$) are the two free parameters of the fermion sector. The parameters appearing in the {Lagrangian} of the 
fermion sector can be written in terms of these free parameters as,

\bea
  \label{eq:lam_T}
  \lambda_{T}&=&\frac{(m_{_{T_{p}}}^2-m_t^2)\sin 2\theta_L}{\sqrt{2}v_d\sqrt{m_t^2 \cos^2\theta_L 
  +m_{_{T_{p}}}^2\sin^2\theta_L}} \\
  m_{T}&=&\frac{m_t m_{_{T_{p}}}}{\sqrt{m_t^2 \cos^2\theta_L + m_{_{T_{p}}}^2\sin^2\theta_L}}
\eea

\end{itemize} 

\noindent
In the context of the present work, the relevant independent free parameters of the model described above are
$m_{_{T_p}}$, $\sin \theta_{L} $, $m_{\gamma_d}$, $v_d$, $m_{h_d}$,
$\sin\theta_S$. The gauge kinetic mixing parameter ($\varepsilon$) is also a free parameter\footnote{$\varepsilon$ 
is related to $\varepsilon^{\prime}$ via $\varepsilon = \varepsilon^{\prime} /\sqrt{1-\varepsilon^{\prime 2}/\cos^{2}\hat{\theta}_{W}}$, 
where $\cos\hat{\theta}_{W} = \cos\theta_{W}+\mathcal{O}(\varepsilon^{2})$} of this model, however, 
we set it to a fixed value consistent with the observations. 

\subsection{Constraints}

Even though the model parameters are considered free some of them are subjected to constraints coming from 
observations and theoretical consistency. Below we present three such important constraints on the parameters that 
represent a portal interaction between the SM and the dark sector.

\begin{itemize}

  \item {\it Gauge sector} ($\varepsilon$): 
    The kinetic mixing parameter is strongly constrained by various 
    low-energy observations in the case of a massive dark photon scenario. 
    These constraints on $\varepsilon$ depend significantly on the mass of the dark photon \cite{Bauer:2018onh,Fabbrichesi:2020wbt}. 
    For dark photons with mass $m_{\gamma_d} \gtrsim $ a few GeVs, the allowed range of $\varepsilon$ is $\lesssim 10^{-3}$.
    However, since we will be interested in the case where the dark photon decays inside the detector, 
    we restrict working with arbitrarily small value of $\varepsilon$ consistent as it plays a crucial role in the 
    context of dark photon tagging. For example, if the $\varepsilon$
    $\sim 10^{-7}$, the decay of dark photon will decay with a displaced vertex that can be utilized as an additional feature 
    to enhance its tagging efficiency.
    However, in the present work, we only consider the prompt decay of the dark photon by fixing the value of 
    $\varepsilon$ at $10^{-4}$.  Any other value in the range {$10^{-3}-10^{-6}$} is acceptable as long as it meets 
    the above requirements. We leave the long-lived dark photon scenario for future work.

  \item {\it Scalar sector} ($\sin\theta_S$): 
    The mixing angle between the SM Higgs and the dark Higgs is constrained by various observations such as, 
    direct searches at the LEP experiment, LHC heavy Higgs searches, the  Higgs signal strength measurement, 
    correction to the $W$ boson mass, perturbative unitarity and LHC Higgs invisible searches.
    \cite{Robens:2015gla, Robens:2019ynf,Robens:2022oue,Adhikari:2022yaa,Lane:2024vur,Robens:2022cun}. 
    However, these constraints depend on the mass of the dark Higgs and the dark Higgs VEV ($v_d$).
    For dark Higgs in the mass range ($10 ~{\rm GeV} \leq m_{h_d} < 100$ GeV) and $v_{d}\sim v_{_{\rm EW}}$ the LEP 
    experiment provides a stringent lower limit of $0.9-0.998$ on the allowed values of $|\sin\theta_{S}|$.
    In the range $100 ~{\rm GeV} \leq m_{h_d} \leq 400$ GeV, the strongest limit on  $|\sin\theta_{S}|< 0.2$  
    comes from the LHC heavy Higgs searches \cite{Adhikari:2022yaa,Robens:2022cun}. 
    For higher masses, the stringent limit comes from the correction to the W boson mass. In fact, $|\sin\theta_{S}|\gtrsim 0.14$ 
    can be ruled out for $m_{h_d} \sim 1$ TeV \cite{Robens:2022cun}.  
    The mixing angle is also constrained by the Higgs invisible searches in the case when $h \to \gamma_d \gamma_d$ is allowed.
    The CMS ~\cite{CMS:2023sdw} and ATLAS ~\cite{ATLAS:2023tkt} limits on this invisible branching ratio set a strong constraint
    on the mixing angle as a function of the dark Higgs VEV:
    \beq
    |\sin\theta_S | \leq 4.6\times 10^{-4}~ \left(\frac{v_d}{\rm GeV}\right) \sqrt{{\rm BR}_{\rm lim}}
    \eeq
    where ${\rm BR}_{\rm lim}$ is 0.15 (CMS) and 0.107 (ATLAS).
    For $v_d = 100$ GeV, this translates to $|\sin\theta_S| \leq 0.018$ (CMS) and $0.015$ (ATLAS). Throughout our analyses we set, $|\sin\theta_S| = 10^{-4}$.

  \item {\it Fermion sector} ($\lambda_T$): 
    The perturbative unitarity requirement coming from the process $h_d t\to h_d t$ process controlled by the 
    parameter $\lambda_T$ gives an upper bound on it, $\lambda_{T} < 4\sqrt{2\pi}$. On the other hand the reality condition 
    on the top and top partner mixing angle ($\sin\theta_L$) also gives an independent constraint on $\lambda_T$. 

    The mixing angle can be written as,
    
    \bea
    \left| \sin \theta_{L}\right| =
    \frac{1}{2}\left[\frac{2 m_{_{T_p}}^{2}-2m_{t}^{2}-\lambda_{T}^{2}v_{d}^{2}}{m_{_{T_p}}^{2}-m_{t}^{2}}
    \left(1-\sqrt{1-\frac{8\lambda_{T}^{2}v_{d}^{2}m_{t}^{2}}{\left(2 m_{_{T_p}}^{2}-2m_{t}^{2}-
    \lambda_{T}^{2}v_{d}^{2}\right)^{2}}}\right)\right]^{1/2}
    \eea

    The reality requirement implies $|\lambda_T| < \sqrt{2}(m_{_{T_{p}}}- m_t)/v_d$ \cite{Kim:2019oyh, Verma:2022nyd}
    The combination of both of these can together be given as,
    \bea
    |\lambda_T| < \sqrt{2}\min{\left(\frac{m_{_{T_{p}}}- m_t}{v_d}, 4\sqrt{\pi}\right)}\label{eq:bound_lamT}
    \eea 
  
\end{itemize}

\subsection[Tp decay width]{$T_p$ decay width}

In this model, the top partner decays to both in the standard ($T_p\to bW,~tZ,~th $) and {nonstandard}  
($T_p\to t\gamma_d,~th_d $) modes. Below we present the partial decay widths of the top partner in various decay modes 
in the limit $|\sin\theta_{d}|,~|\sin\theta_{s}|,~\varepsilon,~ m_{t}/ m_{_{T_{p}}}\ll 1$
\cite{Buchkremer:2013bha,Kim:2018mks, Verma:2022nyd}.

\begin{itemize}
  \item $\bm{T_{p} \to bW}$
    \bea
    \label{eq:width_W}
    \Gamma(T_{p}\to bW)& \approx & \frac{1}{16\pi}\frac{m^{3}_{_{T_{p}}}}{v^{2}_{_{EW}}}\sin^{2}\theta_{L}
    \eea

  \item $\bm{T_{p} \to tV (V = Z, ~\gamma_d)}$   
  
  \bea
\label{eq:width_Z}
\Gamma(T_{p}\to tZ)& \approx & \frac{1}{32\pi}\frac{m^{3}_{_{T_{p}}}}{v^{2}_{_{EW}}}\sin^{2}\theta_{L}\cos^{2}\theta_{L}
\eea

and

\bea
\label{eq:width_gamma_d}
\Gamma(T_{p}\to t\gamma_{d})& \approx & \frac{1}{32\pi}\frac{m^{3}_{_{T_{p}}}}{v^{2}_{d}}\sin^{2}\theta_{L}\cos^{2}\theta_{L}\times \left(1+\frac{m^{2}_{_{T_{p}}}m^{2}_{t}}{D^{2}}\right)
\eea

\item $\bm{T_{p} \to tS (S = h,~h_d)}$
\bea
\label{eq:width_h}
\Gamma(T_{p}\to th)& \approx & \frac{1}{32\pi}\frac{m^{3}_{_{T_{p}}}}{v^{2}_{_{EW}}}\sin^{2}\theta_{L}\cos^{2}\theta_{L}
\eea

and,

\bea
\label{eq:width_hd}
\Gamma(T_{p}\to th_{d})& \approx & \frac{1}{32\pi}\frac{m^{3}_{_{T_{p}}}}{v^{2}_{d}}\sin^{2}\theta_{L}\cos^{2}\theta_{L} \nonumber\\
&&\times\frac{m^{4}_{_{T_{p}}}}{D^{2}} \left(\sin^{4}\theta_{L}+\frac{m^{2}_{t}}{m^{2}_{_{T_{p}}}}\cos^{4}\theta_{L} +4 \frac{m^{2}_{t}}{m^{2}_{_{T_{p}}}}\sin^{2}\theta_{L}\cos^{2}\theta_{L}\right)
\eea

where, $D = m^{2}_{t}\cos^{2}\theta_{L}+m^{2}_{_{T_{p}}}\sin^{2}\theta_{L}$.


\end{itemize}

\begin{figure}[htbp]
  \centering
  \resizebox{\columnwidth}{!}{
  \subfloat[\label{subfig:BR_tp_all_800_200_400_10_13}]{\includegraphics[width=0.5\columnwidth]{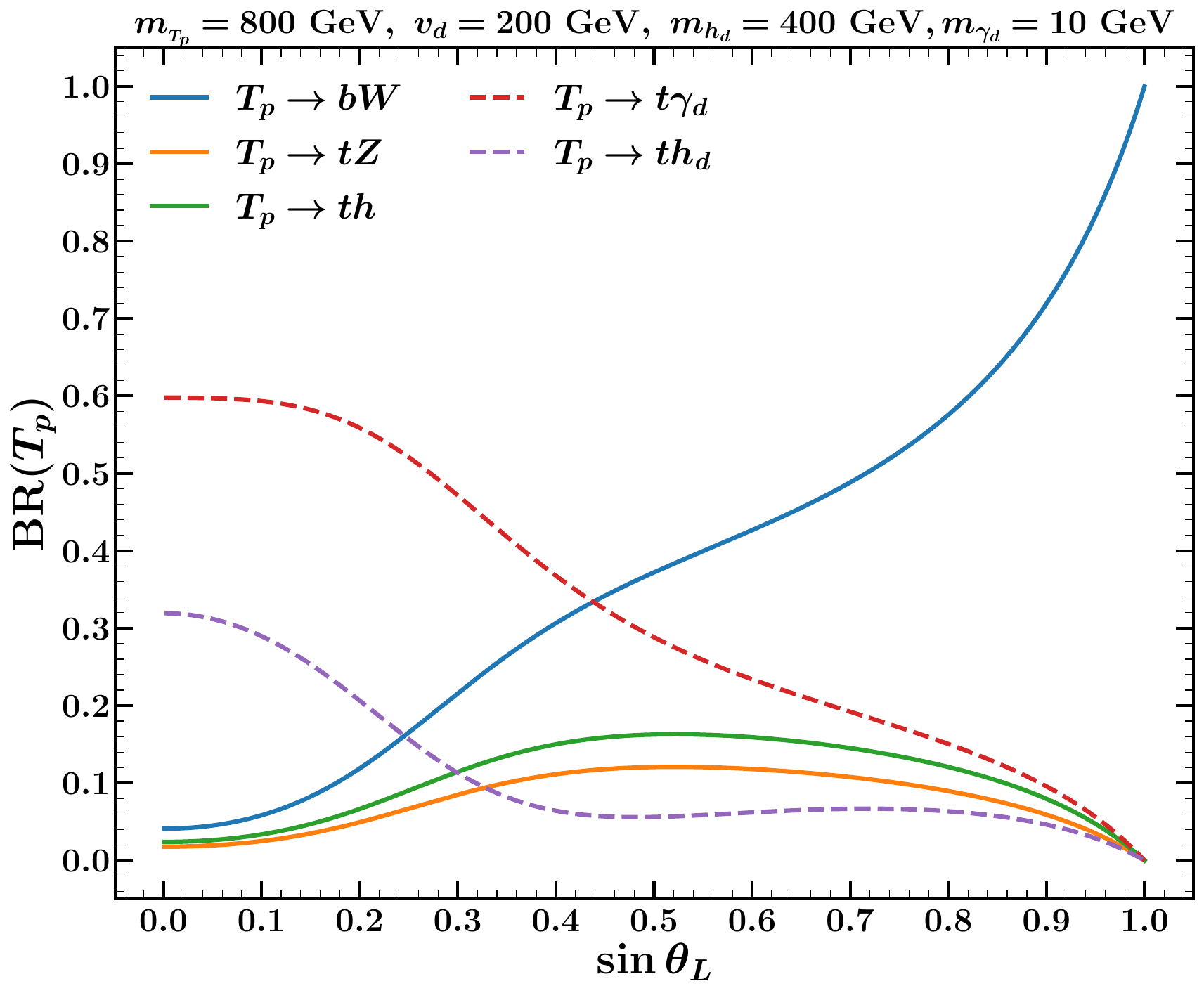}}~
  \subfloat[\label{subfig:BR_tp_all_1800_200_400_10_14}]{\includegraphics[width=0.5\columnwidth]{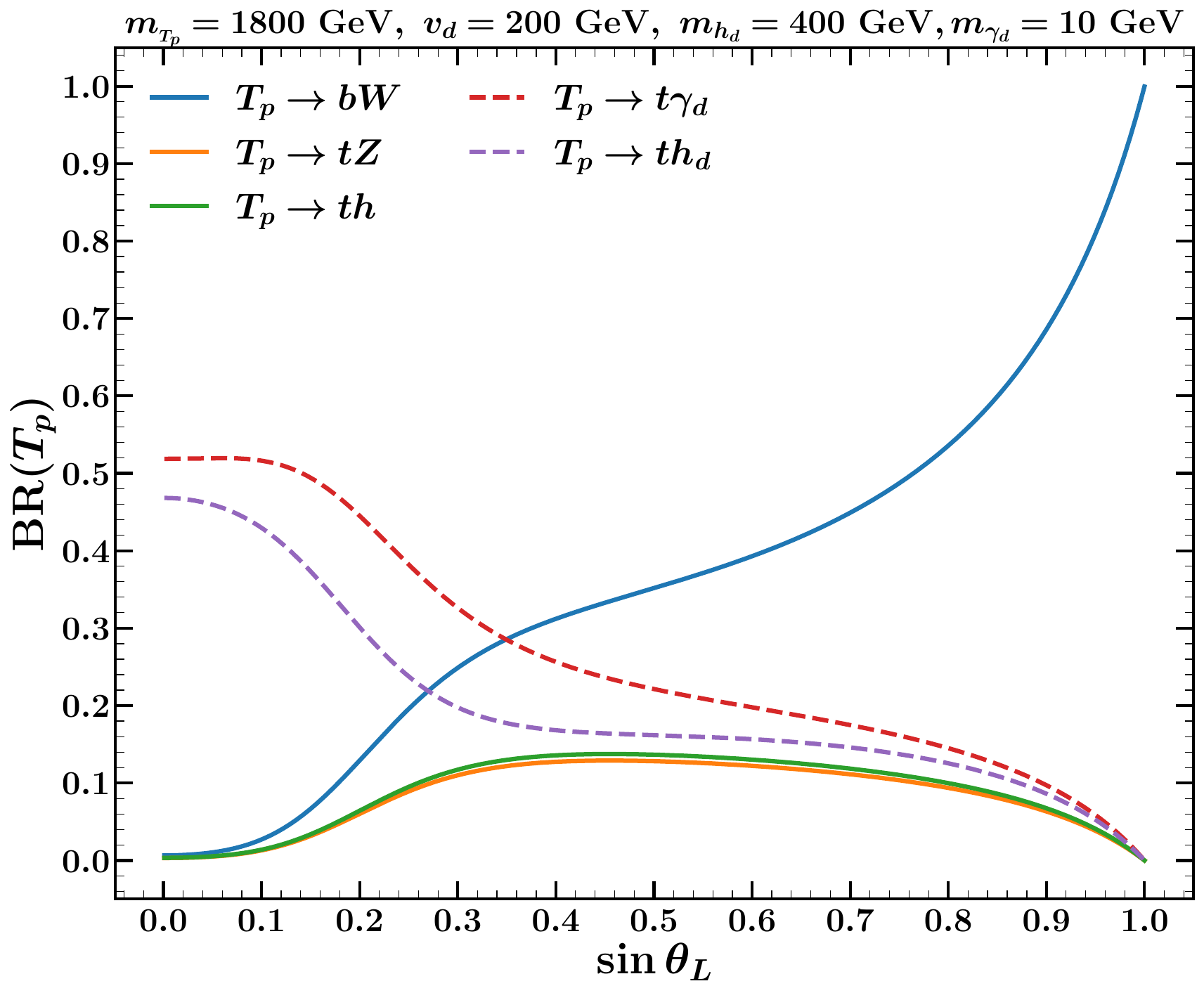}}
  }
  \resizebox{\columnwidth}{!}{
  \subfloat[\label{subfig:BR_tp_all_800_100_400_10_13}]{\includegraphics[width=0.5\columnwidth]{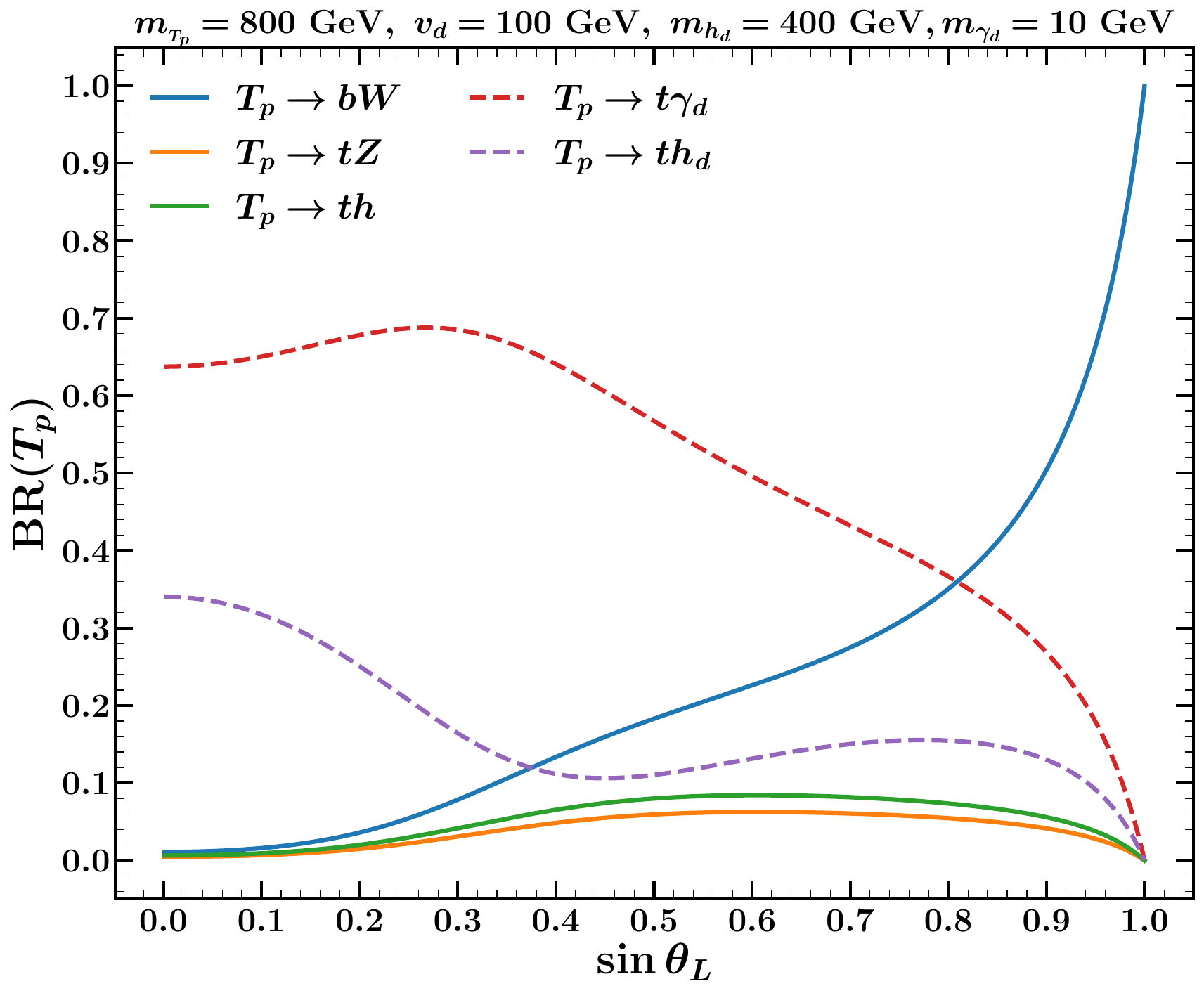}}~
  \subfloat[\label{subfig:BR_tp_all_1800_100_400_10_14}]{\includegraphics[width=0.5\columnwidth]{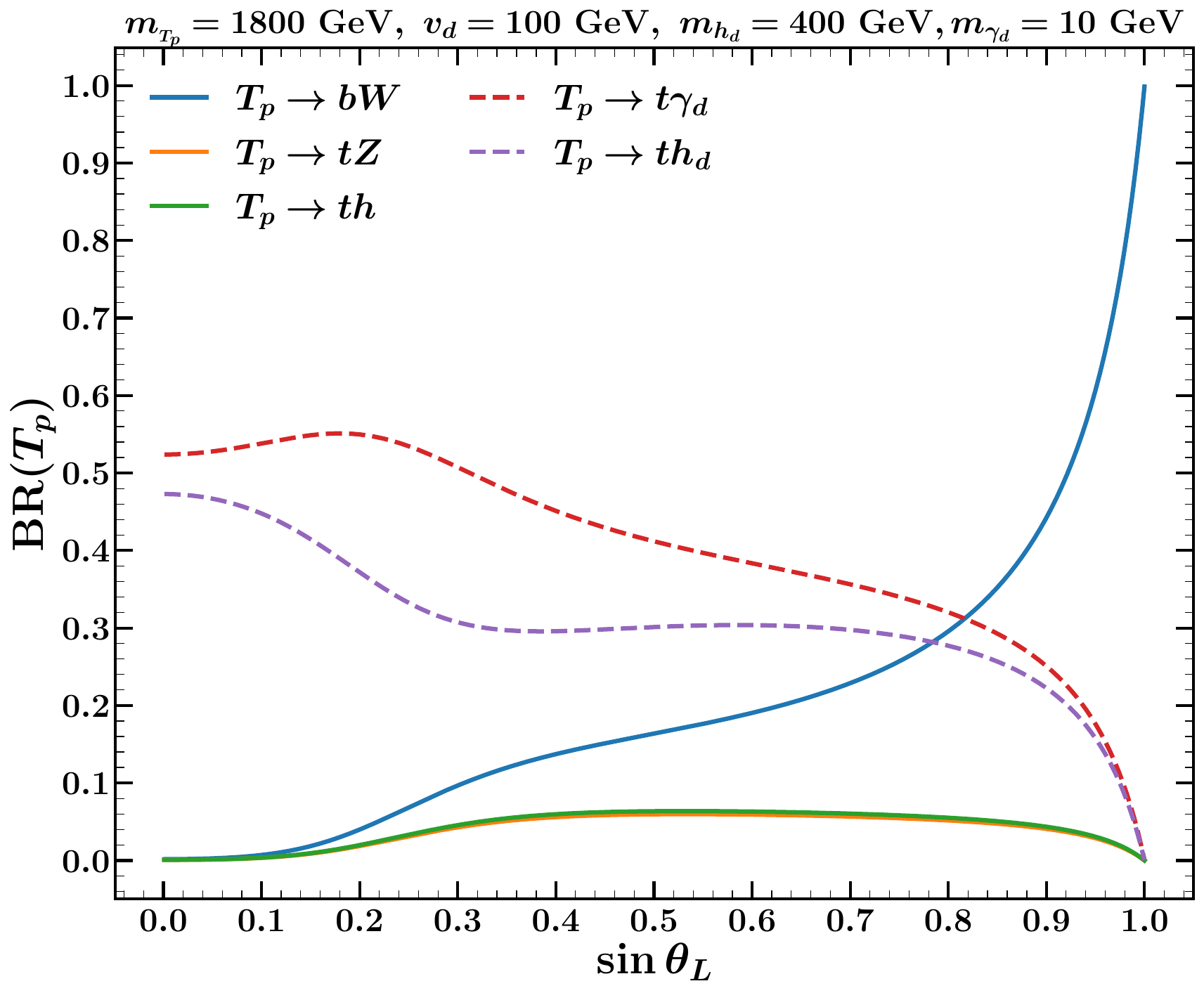}}}
  \caption{Top partner branching ratio in the standard (solid line) and {nonstandard} modes (dashed line) as a function of the mixing angle between top and top partner ($\sin\theta_L$) for different benchmark points,
  (a) $ m_{_{T_p}} = 800$ GeV and (b) $ m_{_{T_p}} = 1.8$ TeV with $ v_d = 200$ GeV and $ m_{h_d} = 400$ GeV. The same for 
  (c) $ m_{_{T_p}} = 800$ GeV and (d) $ m_{_{T_p}} = 1.8$ TeV with $ v_d = 100$ GeV and $ m_{h_d} = 200$ GeV. The dark photon mass is assumed to be  $10$ GeV for all the four plots. }
  \label{fig:br_tp}
\end{figure}

\noindent
{We use the following values for the relevant SM parameters to calculate the top partner decay branching ratios and the {cross section} 
in the later part of our analysis \cite{PDG:2020ssz},

\bea
\label{eq:sm_param_choice}
\nonumber
    m_h = 125.5 {\rm ~ GeV}, ~ 
   	m_t = 173.2 {\rm ~ GeV}, ~
    m_b = 4.18 {\rm ~ GeV} 
\\
    m_W = 80.377 {\rm ~ GeV}, ~
	m_Z = 91.187 {\rm ~ GeV},~
    v_{_{\rm EW}} = 246 {\rm ~ GeV}
    \eea
  
}
In Fig.~\ref{fig:br_tp}, we depict the branching ratio of the top partner in various modes as a function of $\sin\theta_L$ 
for various choices of $v_d,~ m_{\gamma_d},~ m_{h_d},~{\rm and }~ m_{_{T_p}}$

\section{Collider analysis}\label{sec:collider_analysis}
In this section, we detail the collider analysis of signal and background processes. 
We have considered both the single ($pp \to T_p j $) and the pair ($pp \to  T_p \overline{T}_p$) production
of the vector like top partner. We have used \textsc{MadGraph5\_aMC@NLO} (version 3.2.0) \cite{Alwall:2014hca} 
event generator to simulate both the single and pair production processes of top partner at 13 and 
14 TeV LHC energies. In order to achieve this, we have implemented the model described in 
Sec. \ref{sec:model} at the Lagrangian level with the help of \textsc{FeynRules} \cite{Alloul:2013bka}. 
The Universal FeynRules Output \cite{Degrande:2011ua} is then interfaced with \textsc{MadGraph5\_aMC@NLO} 
for event generation. We have used NNPDF (version 2.3) \cite{Ball:2012cx} parton distribution functions
to simulate the parton-parton hard scattering in a $pp$ collision. 
Events generated by MadGraph are then interfaced with \textsc{Pythia} (version 8.307) \cite{Bierlich:2022pfr} 
for parton shower, hadronization, and further analysis. 
The energy and momenta of all the final state objects are smeared with appropriate Gaussian smearing 
function to take into account the finite detector resolution effects. The details of smearing function {parametrization} can be found in \cite{Verma:2022nyd}.

In this work, we are considering the top partner decaying to a top quark and dark photon. 
The dark photon further decays to SM fermions via the gauge kinetic mixing parameter. Since the top partner is expected to be 
heavy ($\gtrsim 0.8$ TeV) in order to satisfy the current LHC bounds, the top and the dark photon produced in the decay of it are 
highly boosted. \emph{One of our primary objectives in this study is to investigate the hadronic decays of the dark photon and the 
top quark, along with their subsequent identification and reconstruction}. 

We use an HDNN framework that uses a combination of convolutional neural network and {multilayer} perceptron to identify and reconstruct a dark photon jet while a 
jet initiated by a top quark is reconstructed using Johns Hopkins top tagger \cite{Kaplan:2008ie}. To ensure the hadronic decays of 
the top quark and the dark photon we have vetoed events with isolated leptons. 
We first proceed with the identification and reconstruction of jets originating from the light exotic particle - specifically, the dark photon in this analysis. In order to achieve this we cluster all the final state hadrons 
to construct jets of radius $0.4$ using anti-$k_T$ jet clustering  algorithm (AK4). We have used \textsc{FastJet} (v3.4.0) \cite{Cacciari:2011ma} to cluster the hadrons. 

Out of all the jets thus formed, we select only those with $p_{_T}>200$ GeV as potential dark photon
jet candidates and prepare jet images and other high level features for HDNN based identification. 
We have utilized two different images of a single high $p_{_T}$ jet of clustering radius $0.4$ in the $\eta-\phi$ plane and a subjet inside it with radius $0.2$ around the jet center,  
using corresponding granularities 0.1 and 0.025, respectively. In addition, a set of crucial jet level features, such as, 
$p_{_T}$, $p_{_T}$ fraction, charge multiplicity and charge multiplicity fraction of the potential dark photon jet candidates help us identify a dark photon-initiated jet very effectively. It is possible to achieve an average tagging efficiency of {$30\%$ to $93\%$} for the dark photon jet in the top partner mass range, {$0.8 - 2.8$ TeV}. The average {mistagging} rate for the {nondark} photon jets, present in both the signal and SM background varies in the range $1.0\%$ to {$2.6\%$}. Details of the HDNN-based identification and reconstruction of the dark photon jets are provided in this appendix \ref{App:dph_tagging}. Events are considered for further analysis only if they contain at least one identified dark photon jet.

The remaining part of the event analysis proceeds in the following manner depending on the number of identified dark photon jets:
\begin{enumerate}
  \item If {\it exactly one} HDNN-tagged dark photon jet is identified, we remove \footnote{We identify and remove the constituents of the dark photon jets before the top reconstruction to make sure same total $E_T$ dependent clustering radius, as used in our previous analysis \cite{Verma:2022nyd}, where the dark photon was treated as an invisible particle.} all of its constituents and {recluster} the remaining hadrons of that event with a total ${E_T}$ dependent clustering radius $R$ using Cambridge-Achen (C-A) clustering algorithm \cite{Bentvelsen:1998ug}.
  The {reclustering} is done only if the total ${E_T}$ of the remaining hadrons is more than 400 GeV. 
  These jets are then analyzed using Johns Hopkins top 
  tagger to identify a boosted top quark structure in it. We keep the event if a top quark initiated jet is obtained. A subset of this event category will give rise to a final state comprising {\it exactly one} dark photon jet and {\it exactly one} top quark jet which predominantly consists of contributions coming from the single production of top partner, with only a small fraction arising from pair production of the top partner.

  \item If {\it more than one} dark photon jets are identified, we remove the constituents of the identified dark photon jets pairwise. If there are {\it two} dark photon jets, we remove the constituents of both - the leading and {subleading} dark photon jets - simultaneously, if it is {\it three} then we have three possible pairs. 
        We consider each possible pair one at a time and remove all the constituents of that pair. The remaining hadrons in all the
        cases are then {reclustered} as prescribed before with a total ${E_{T}}$ dependent clustering radius {\it R} to find a 
        top structure in the jets thus formed using Johns Hopkins top tagger. This has been done keeping in mind the topology of events where the top partner is pair produced and further decays to a top 
        quark and a dark photon giving rise to {\it exactly two} dark photon jets in the final state. We keep an event only if 
        {\it at least one} top quark jet is identified in these events.  
        Therefore, this category primarily contains events from pair production of the top partner, with a small fraction from single production of the top partner. 
  
\end{enumerate}

The Johns Hopkins top tagger mentioned above, iterativelypr{declusters} a C-A clustered jet to identify the substructure inside a fat jet of radius $R$. The J-H top tagging algorithm requires the specification of the following additional parameters: the fraction of the jet $p_{_T}$ carried by a {subjet} and the Manhattan distance (defined as $|\Delta \eta| + |\Delta \phi|$) between two {subjets} to satisfy minimum values, $\delta_p$  and $\delta_r$, respectively to be considered as hard and resolved. We tabulate, in Table~\ref{table:delrvalues}, the values of $R$, $\delta_p$ and $\delta_r$ for different total transverse momentum of an event \cite{Verma:2022nyd}. 

A key motivation for choosing the J-H top tagger is to allow a qualitative comparison with the methodology adopted in Ref.~\cite{Verma:2022nyd}, where a similar approach was used for boosted top tagging with a total $E_T$-dependent jet clustering radius. {While more recent neural networks methods, such as ResNeXt-inspired convolutional networks~\cite{Kasieczka:2019dbj} and ParticleNet~\cite{Qu:2019gqs} architectures that treat jets as point clouds, have demonstrated improved performance in capturing intricate jet substructure, we refrain from introducing an additional ML-based tagger here. Since our work already implements a HDNN-based tagging algorithm for dark photon jets, we adopt the J-H top tagger for boosted top tagging in order to keep the analysis tractable.} This choice also facilitates a direct comparison between the present analysis and that of Ref.~\cite{Verma:2022nyd}, allowing us to validate that the same reconstruction strategy remains effective.

\begin{table}[htbp]
  
  \centering
  \resizebox{0.55\columnwidth}{!}{
  \begin{tabular}{lcccccc}
  \toprule 
  $E_{T}$ & 400 & 600 & 800 & 1000 & 1600 & 2600 \\ 
  \midrule
  \midrule 
  R & 1.4 & 1.2 & 1.0 & 0.8 & 0.6 & 0.4 \\
  
  $\delta p$ & 0.10 & 0.10 & 0.05 & 0.05 & 0.05 & 0.05 \\
    
  $\delta r$ & 0.19 & 0.19 & 0.19 & 0.19 & 0.19 & 0.19\\
  \bottomrule
  \end{tabular} 
  }
  \caption{{  Choices of jet clustering radius(R), minimum jet $p_{_T}$ fraction carried by the {subjet} ($\delta p$), minimum Manhattan distance between two {subjets} ($\delta r$) for various range of total transverse momentum in an event excluding the contribution of dark photon transverse momentum.}}
  \label{table:delrvalues}
  \end{table}

\noindent
In Table~\ref{table:sig_b4_kin_cuts}, we summarize the effect of basic cut, {\it i.e.}~requiring 
{\it at least two} central jets (defined with $p_{_T}>20$ GeV and $|\eta|<2.5$), no isolated leptons and $H_T > 500$ GeV, followed by requirement of tagging {\it at least one} dark photon jet out of jets having $p_{_T}>200$ GeV and {\it at least one} tagged top quark in an event for the signal process.

\noindent

We have analyzed two different final states, namely, 
  \begin{itemize}
    \item {\bf ED1ET1}: {\it exactly one} dark photon jet + {\it exactly one} top quark jet ($\gamma_d+ t$) and,
    \item {\bf AD2AT1}: {\it at least two} dark photon jets + {\it at least one} top quark jet ($\geq 2\gamma_d+\geq t$).
  \end{itemize}

\begin{figure}[htbp]
    \centering
    \resizebox{\columnwidth}{!}{
    \subfloat[\label{subfig:schematic_diag_pp}]{\includegraphics[width=0.5\columnwidth]{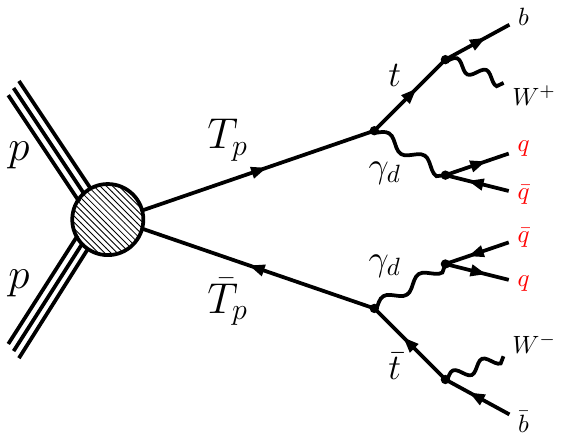}}~
    \subfloat[\label{subfig:schematic_diag_sp}]{\includegraphics[width=0.5\columnwidth]{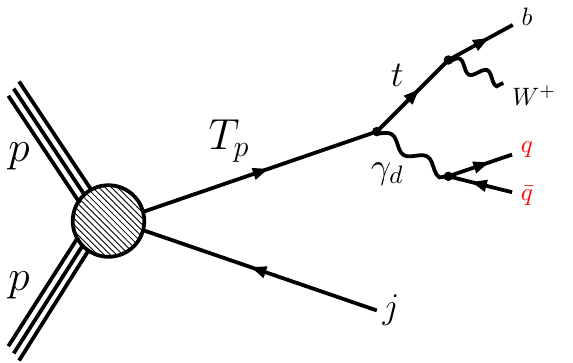}}}
    \caption{{ Schematic diagrams of signal processes: (a) $pp\to T_{p}\bar{T}_p$ and (b) $pp\to T_{p}j$ with subsequent decays of $T_{p}\to t\gamma_d$ and $\gamma_d\to q\bar{q}$. }}
    \label{fig:schematic_diag}
  \end{figure}

  This categorization of final states have been done keeping in mind the event topology of the single and pair production of the top 
  partner. The schematic diagram illustrating the topology of the final states is depicted in Fig.~\ref{fig:schematic_diag}. 
  However, given the finite tagging efficiencies for both the top quark and the dark photon the two final states are populated 
  by events coming from both single and pair production of top partner.  For example, with $\sin\theta_L = 0.1$ the AD2AT1 final state receives contribution dominantly from the pair production of top partner process {(99.8\% to 60.4\% for $m_{_{T_p}}$ in the range $0.8$ TeV to $2.6$ TeV)} with the remaining contribution coming from single production of the top partner process\footnote{{To estimate the relative contribution of these two {subprocesses}, we have assumed 100\% branching ratio for both top partner-decaying to a top and dark photon and dark photon-decaying in hadronic mode}}. In contrast, the ED1ET1 final state consists of a substantial admixture of both single {(21.8\% to 99.8\% for $m_{_{T_p}}$ in the range $0.8$ TeV to $2.6$ TeV)} and pair production of the top partner.

\begin{table}[htbp]
  \centering
  \resizebox{0.8\columnwidth}{!}{
\begin{tabular}{cccccccc}
   \toprule
   \multirow{2}{*}{$\sqrt{s}$}&\multirow{2}{*}{Process} & \multirow{2}{*}{$m_{_{T_p}}$} & Production &  \multirow{2}{*}{Basic cut} &  $p_{_T}\geq 200 $ &  \multirow{2}{*}{$\gamma_d$ tagging} &  \multirow{2}{*}{Top tagging} \\
   & &  & {cross section}   &  &  GeV &  &  \\
   (TeV)& & (TeV) & (fb) &  (fb) & (fb) & (fb) &  (fb) \\
   \midrule
   \midrule
    \multirow{16}{*}{13}&\multirow{5}{*}{$T_{p}\bar{T}_{p}$} 
      &0.8  &     190.00 &      132.19 &    131.44 &         75.10 &           14.28 \\
     && 1.0  &      42.70 &      30.45 &      30.40 &          22.09 &           5.15 \\
     && 1.2 &      11.40 &       8.33 &       8.32 &           6.84 &           1.86 \\
     && 1.4 &       3.42 &       2.58 &       2.58 &           2.25 &           0.66 \\
     && 1.6 &       1.11 &       0.86 &       0.86 &           0.78 &           0.24 \\

     && 1.7 &       0.64 &        0.5 &        0.5 &           0.46 &           0.14 \\

     && 1.8 &       0.38 &       0.30 &       0.30 &           0.28 &           0.09 \\         
     && 2.0 &       0.13 &       0.11 &       0.11 &           0.10 &           0.03 \\
     \cmidrule{2-8}
    &\multirow{6}{*}{$T_{p}j$} 
      &0.8  &      72.03 &      56.88 &      49.95 &          14.74 &           2.86 \\
     && 1.0 &      37.57 &      30.27 &      28.18 &          14.10 &           3.29 \\
     && 1.2 &      20.92 &      17.18 &      16.42 &          10.69 &           2.81 \\
     && 1.4 &      12.16 &      10.13 &       9.82 &           7.33 &           2.02 \\
     && 1.6 &       7.33 &       6.24 &       6.08 &           4.90 &           1.31 \\
     && 1.7 &      5.72 &        4.9 &        4.8 &           3.97 &           1.02 \\
     && 1.8 &       4.50 &       3.90 &       3.82 &           3.22 &           0.84 \\
     && 2.0 &       2.82 &       2.49 &       2.45 &           2.13 &           0.53 \\
   \midrule
    \multirow{24}{*}{14}&\multirow{5}{*}{$T_{p}\bar{T}_{p}$} 
      & 0.8 &      249.61 &     173.12 &     172.35 &          98.47 &          18.71 \\
     && 1.0 &      58.16 &      41.63 &      41.58 &          30.13 &           7.03 \\
     && 1.2 &      16.15 &      11.90 &      11.90 &           9.77 &           2.61 \\
     && 1.4 &       5.06 &       3.82 &       3.82 &           3.32 &           0.97 \\
     && 1.6 &       1.72 &       1.34 &       1.34 &           1.21 &           0.37 \\
     && 1.8 &       0.61 &       0.49 &       0.49 &           0.45 &           0.14 \\
     && 1.9 &       0.38 &       0.30 &       0.30 &           0.28 &           0.09 \\
     && 2.0 &       0.23 &       0.18 &       0.18 &           0.17 &           0.05 \\

     && 2.2 &       0.09 &       0.07 &       0.07 &           0.07 &           0.02 \\
     && 2.4 &       0.034 &     0.029 &     0.029 &         0.027 &         0.008 \\
     && 2.5 &       0.021 &      0.018 &      0.018 &          0.017 &          0.005 \\
     && 2.6 &       0.014 &     0.012 &     0.012 &         0.011 &         0.003 \\
    \cmidrule{2-8}
    &\multirow{5}{*}{$T_{p}j$} 
      &  0.8 &      88.12 &      69.63 &      61.11 &          17.80 &           3.60 \\
     &&  1.0 &      46.89 &      37.72 &      35.05 &          17.55 &           4.14 \\
     && 1.2 &      26.65 &      21.86 &      20.86 &          13.55 &           3.50 \\
     && 1.4 &      15.81 &      13.21 &      12.76 &           9.47 &           2.58 \\
     && 1.6 &       9.70 &       8.26 &       8.06 &           6.48 &           1.72 \\
     && 1.8 &       6.09 &       5.27 &       5.18 &           4.35 &           1.09 \\
     && 1.9 &       4.87 &       4.26 &       4.18 &           3.57 &           0.92 \\
     && 2.0 &       3.91 &       3.45 &       3.40 &           2.95 &           0.74 \\
     && 2.2 &       2.54 &       2.25 &       2.23 &           1.98 &           0.48 \\
     && 2.4 &       1.67 &       1.51 &       1.49 &           1.34 &           0.31 \\
     && 2.5 &        1.36 &       1.23 &       1.22 &           1.11 &           0.26 \\
     && 2.6 &       1.11 &       1.01 &       1.00 &           0.92 &           0.20 \\
   \bottomrule
   \end{tabular}
  }
  \caption{ { The effects of cut flow on {cross sections} for the single and 
  pair production of the top partner at both $\sqrt{s}=13$ and $14$ TeV LHC center of mass energies. 
  $\sin \theta_{L} = 0.1$ is assumed to estimate the {cross section}  for the various benchmark points assuming  
  BR$(T_{p}\to t \gamma_{d}) = 100\%$ and BR$(\gamma_{d}\to  q\bar{q}) = 100\%$. In addition, $v_d = 200$ GeV, $m_{\gamma_d}=10$ GeV and $m_{h_d}= 400$ GeV are used to obtain the above results.} } 
  \label{table:sig_b4_kin_cuts}
  \end{table}
  
{The relevant standard model backgrounds for both the final states are {$V(W/Z)+jets$,}~$t\bar{t}+jets,~t\bar{t}W,~VV+jets ~t\bar{t}Z,~ tW,~tj$ and QCD {multijet} processes, which give rise to one or more top quark jets and jets {mistagged} as dark photon. The QCD {multijet} process~\cite{Verma:2022nyd}, despite its overwhelming {cross section}  can be controlled not only because of the requirement of one or more {mistagged} dark-photon and a boosted top quark but also due to the fact that the {\it invariant mass} of the {\it {mistagged}} top-quark and dark photon jet system to peak at the mass of a heavy particle ( $\sim\mathcal{O}$(TeV) ). Hence, given the rich topology of the signal events, {namely, the presence of one or more boosted dark photon initiated jet(s) along with jets coming from highly boosted top quarks, we focus on SM processes containing either one or more top quark or the presence of an EW vector boson as our potential SM backgrounds. 
}
}

{The other subdominant backgrounds like $t\bar{t}H$, $tZj$ and $tHj$. At the {next-to-leading order} (NLO) level the cross section for these backgrounds in their hadronic final states with $H_T> 500$ GeV are $\sigma_{t\bar{t}H} = 344.5$ fb [NLO+NLL]~\cite{LHCHiggsCrossSectionWorkingGroup:2016ypw}, $\sigma_{tZj} = 217$ fb [NLO$_{\rm QCD+EW}$]~\cite{Pagani:2020mov}, and, $\sigma_{tHj} = 17.3$ fb [NLO$_{\rm QCD+EW}$]~\cite{Pagani:2020mov}, respectively. After dark photon tagging and top tagging requirement, these reduce to {$0.19$} fb ($t\bar{t}H$), {$0.13$} fb ($tZj$), and  {0.01} fb ($tHj$). They will further be reduced after the implementation of the hard cuts (for, e.g., invariant mass of the top--dark photon pair to lie within a window) mentioned later in this section. Hence, their effective contribution is neglected in the estimation of signal significance although we keep the contribution of $t\bar{t}W$ and $t\bar{t}Z$ in our analysis as representatives of these {subdominant} backgrounds.}

In Table~\ref{table:bkg_b4_kin_cuts}, the effect of basic cut followed by the requirement of tagging {\it at least one} dark photon jet out of jets having $p_{_T}>200$ GeV and {\it at least one} tagged top quark in an event for the SM background processes are summarized. The {$V+jets$, $t\bar{t}+jets$ and $VV+jets$} background has been generated with the highest jet multiplicity {\it two} at the matrix element level and implemented MLM \cite{Mangano:2001xp} jet parton matching algorithm to avoid double counting. In this analysis the $V+jets$, $VV+jets$ and $tj$ are considered at LO, $t\bar{t}+jets$ and $~tW$ are considered at aN$^{3}$LO and $t\bar{t}Z$, $t\bar{t}W$ are considered at NLO+NNLL in QCD by multiplying the leading order {cross sections}   obtained from MadGraph with appropriate $k$-factors \cite{Verma:2022nyd}.

  \begin{table}[htbp]
  \centering
  \resizebox{0.8\columnwidth}{!}{
   \begin{tabular}{ccccccc}
   \toprule
   \multirow{2}{*}{$\sqrt{s}$}&\multirow{2}{*}{Process} &  Production &  \multirow{2}{*}{Basic cut} &  $p_{_T}\geq 200 $ &  \multirow{2}{*}{$\gamma_d$ tagging} &  \multirow{2}{*}{Top tagging} \\
   &   & {cross section}  \tablefootnote{The production {cross section}  for SM $V+jets$, $t\bar{t}+jets$, $tW$, $VV+jets$ and $tj$ background processes is quoted with $H_T \geq 500$ GeV for both 13 and 14 TeV LHC center of mass energies. \label{fn:bkg}} &  &  GeV &  &  \\
   (TeV)& & (fb) &  (fb) & (fb) & (fb) &  (fb) \\
   \midrule
   \midrule
   \multirow{5}{*}{13}&
   $V+jets$ & 142980.0 &   140928.14 &    127905.52 &   4342.12 &          140.38 \\
   &$t\bar{t}+jets$ &   41728.84 &   41600.33 &   28001.31 &         504.81 &          92.47 \\
   &$tj$ & 2084.11 &    2072.00 &    1813.11 &          67.43 &          13.02 \\
   &$tW$ & 1537.61 &    1533.13 &    1141.30 &          27.16 &           4.50 \\
   &$VV+jets$ & 1087.00 &   786.08 &    781.80 &          49.40 &          2.21 \\
   &$t\bar{t}Z$ & 863.00 &     490.29 &     138.75 &           3.11 &           0.46 \\
   &$t\bar{t}W$ & 566.00 &     273.52 &      64.43 &           1.67 &           0.30 \\
   \cmidrule{2-7}
   \multirow{5}{*}{14}&
   $V+jets$ & 170950.00 &  168381.60 &  152876.26 &        5132.34 &         162.95  \\
   &$t\bar{t}+jets$ &  51297.49 &   51141.77 &   34459.14 &         620.22 &         105.61 \\
   &$tj$ & 2477.80 &    2461.50 &    2130.53 &          79.01 &          15.38 \\
   &$tW$ & 2188.40 &    2182.11 &    1632.11 &          38.87 &       6.27 \\
   &$VV+jets$ & 1315.00 &   952.07 &    946.64 &          60.00 &          2.69 \\
   &$t\bar{t}Z$ &   1045.00 &     600.27 &     173.52 &           3.99 &           0.56 \\
   &$t\bar{t}W$ &   653.00 &     320.24 &      76.65 &           1.96 &           0.35 \\
   \bottomrule
   \end{tabular}
    }
  \caption{
      {The effects of cut flow on {cross sections}   for the SM background processes relevant for our analysis at both $\sqrt{s}=13$ and $14$ TeV LHC center of mass energies. }
  }
  \label{table:bkg_b4_kin_cuts}
  \end{table}

One can see from Table~\ref{table:bkg_b4_kin_cuts} that the requirement of boosted dark photon jet in the SM background event gives a suppression factor of more than 1 in 100 (can also be seen in {Appendix} \ref{App:dph_tagging}). In addition, requirement of the presence of at least one boosted top quark jet in the final state reduces the SM background by a factor of $4$ to $7$ depending on the particular SM background process.


\subsection{Kinematic variables}
For both the final state considered in this analysis the following set of kinematic variables is found to be useful for separating the signal from the background events.

\subsubsection{Transverse momentum of tagged dark photon}

\begin{figure}[htbp]
  \centering
  \resizebox{\columnwidth}{!}{
  \subfloat[\label{subfig:dph_pt_a2a1_13}]{\includegraphics[width=0.5\columnwidth]{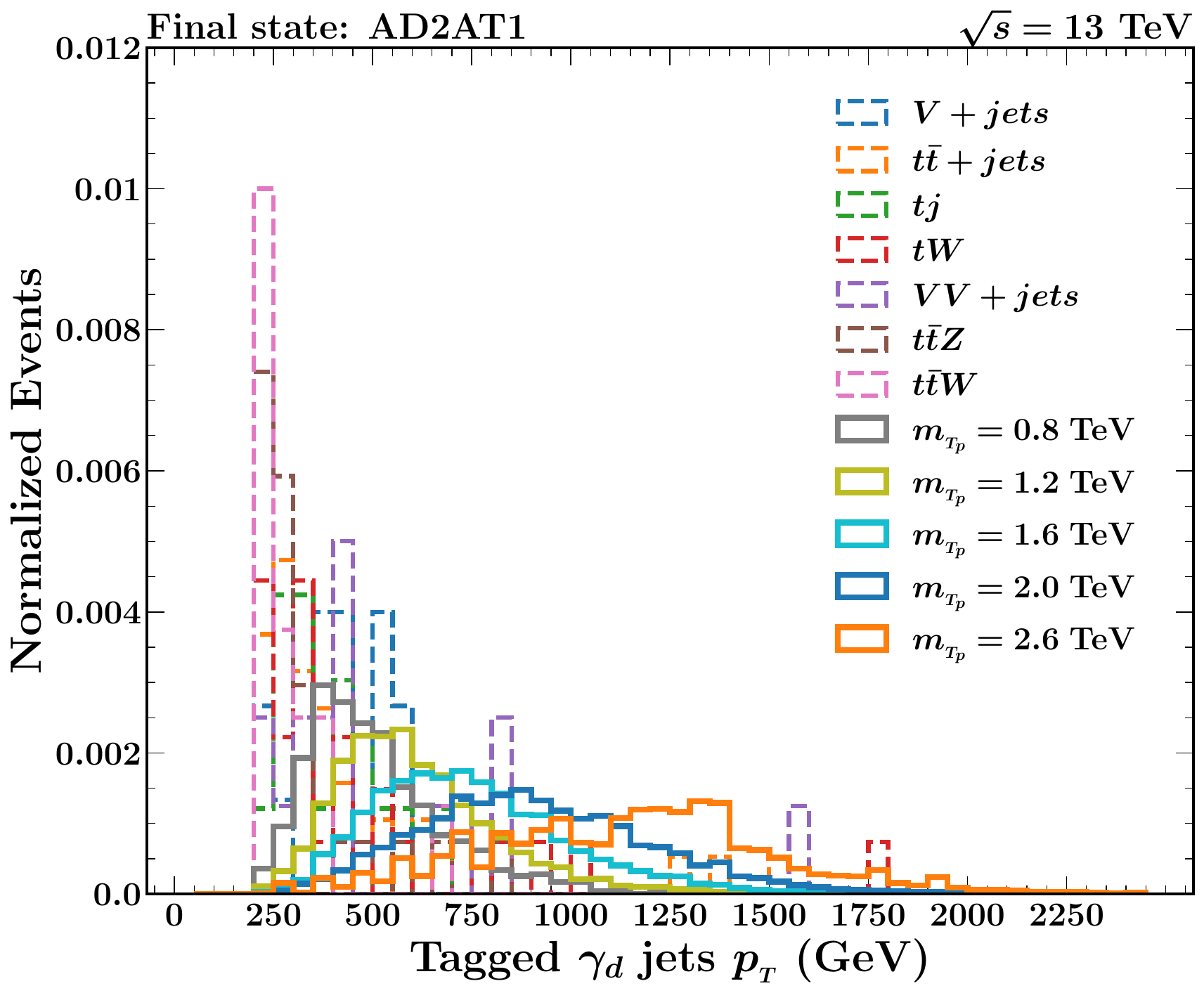}}~
  \subfloat[\label{subfig:dph_pt_e1e1_13}]{\includegraphics[width=0.5\columnwidth]{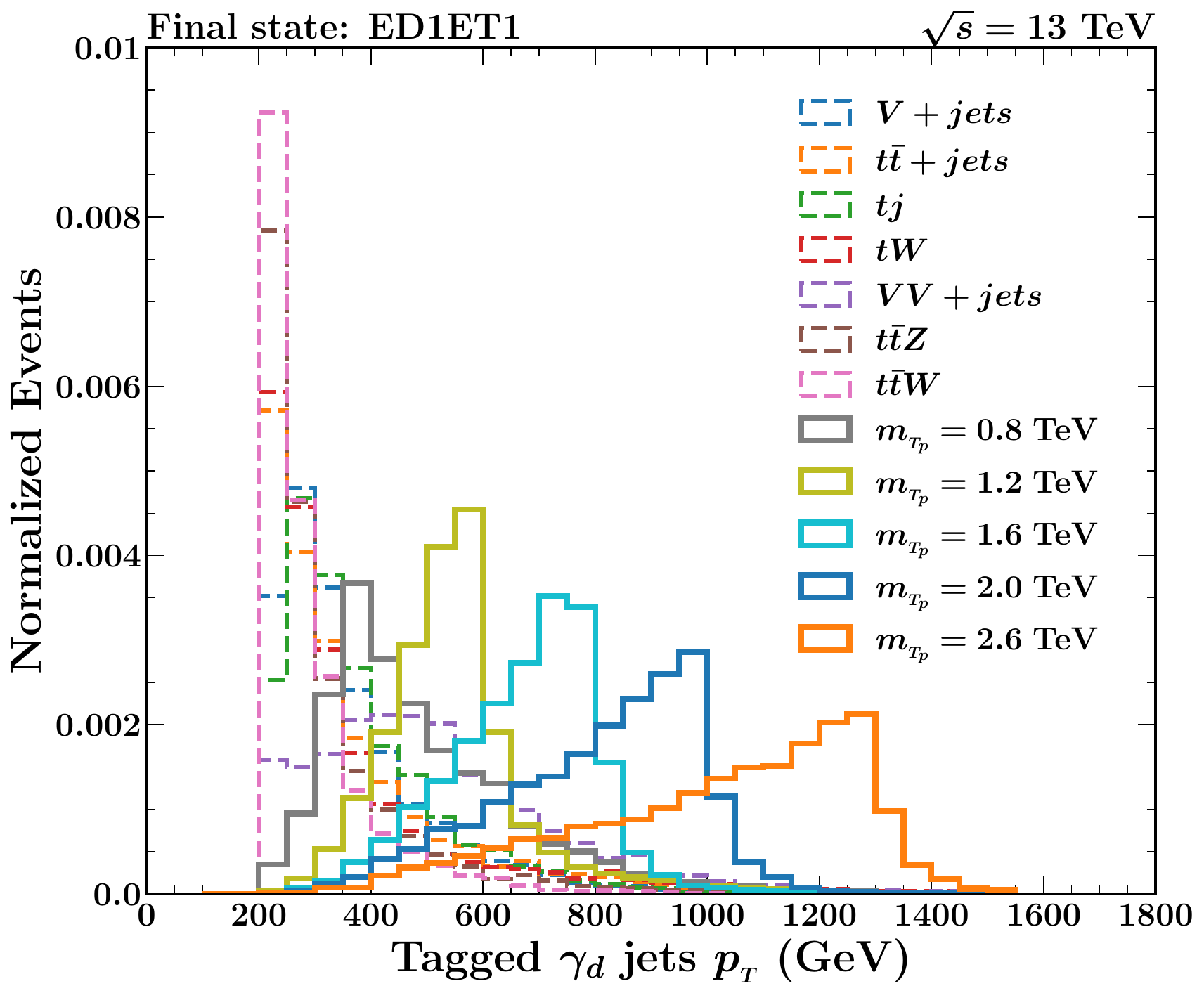}}
  }
  \caption{{ Transverse momentum distributions of the HDNN identified dark photon jet for signal and backgrounds in the (a) AD2AT1 and (b) ED1ET1 final states at 13 TeV LHC center of mass energy.}}
  \label{fig:tag_dph_pt_13}
\end{figure}

\begin{figure}[htbp]
  \centering
  \resizebox{\columnwidth}{!}{
  \subfloat[\label{subfig:dph_pt_a2a1_14}]{\includegraphics[width=0.5\columnwidth]{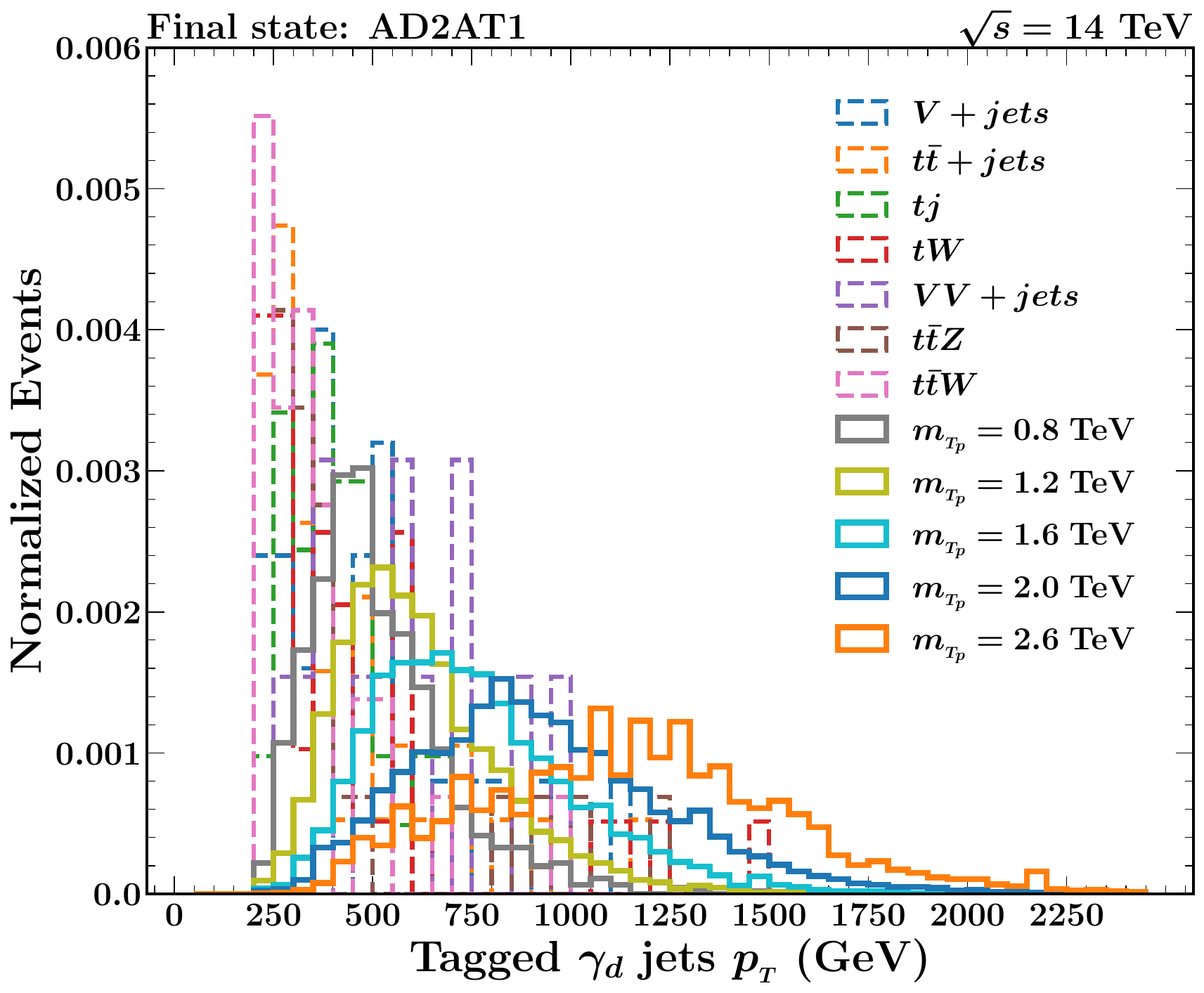}}~
  \subfloat[\label{subfig:dph_pt_e1e1_14}]{\includegraphics[width=0.5\columnwidth]{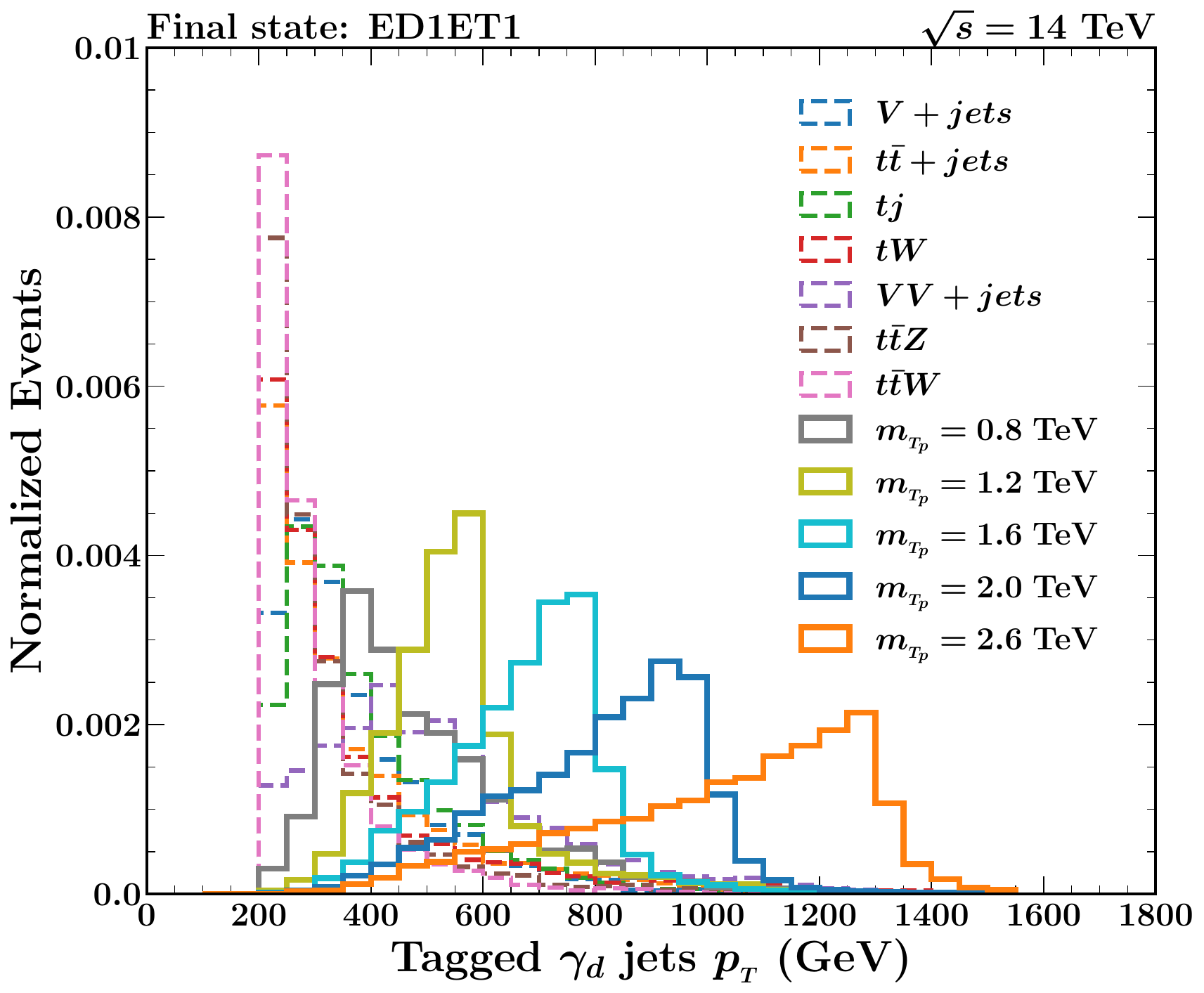}}
  }
  \caption{{ Transverse momentum distributions of the HDNN identified dark photon jet for signal and backgrounds in the (a) AD2AT1 and (b) ED1ET1 final states at 14 TeV LHC center of mass energy.}}
  \label{fig:tag_dph_pt_14}
\end{figure}

In Fig.~\ref{fig:tag_dph_pt_13} and Fig.~\ref{fig:tag_dph_pt_14} we plot the transverse momentum, defined as $ p_{_T} = \sqrt{p_x^2 + p_y^2}$, of the HDNN identified dark photon jet for two different final states considered in this analysis at $\sqrt{s} = 13$ and $14$ TeV, respectively.

\subsubsection{Invariant mass of the top-dark photon pair}

\begin{figure}[htbp]
  \centering
  \resizebox{\columnwidth}{!}{
  \subfloat[\label{subfig:inv_a2a1_13}]{\includegraphics[width=0.5\columnwidth]{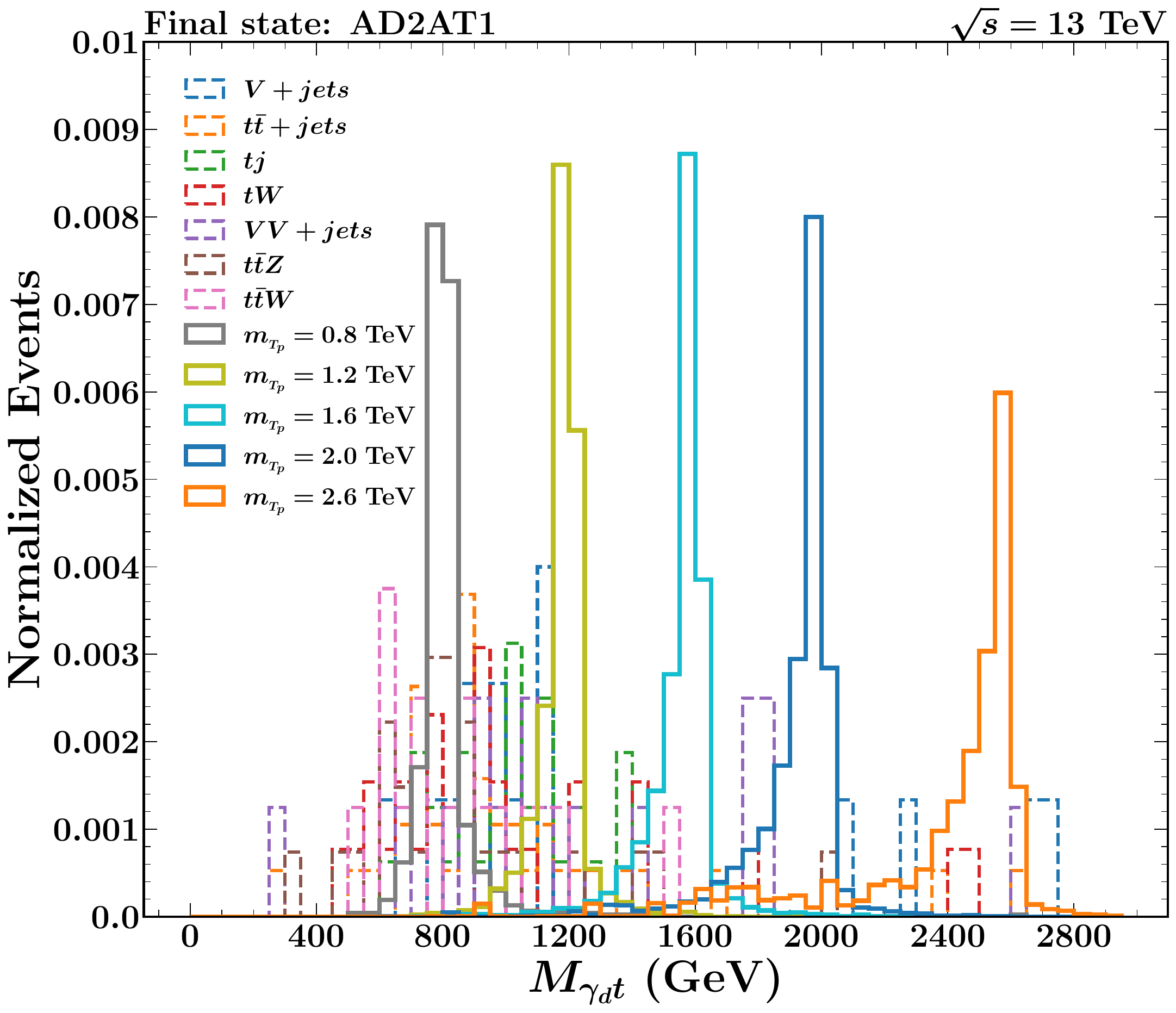}}~
  \subfloat[\label{subfig:inv_e1e1_13}]{\includegraphics[width=0.5\columnwidth]{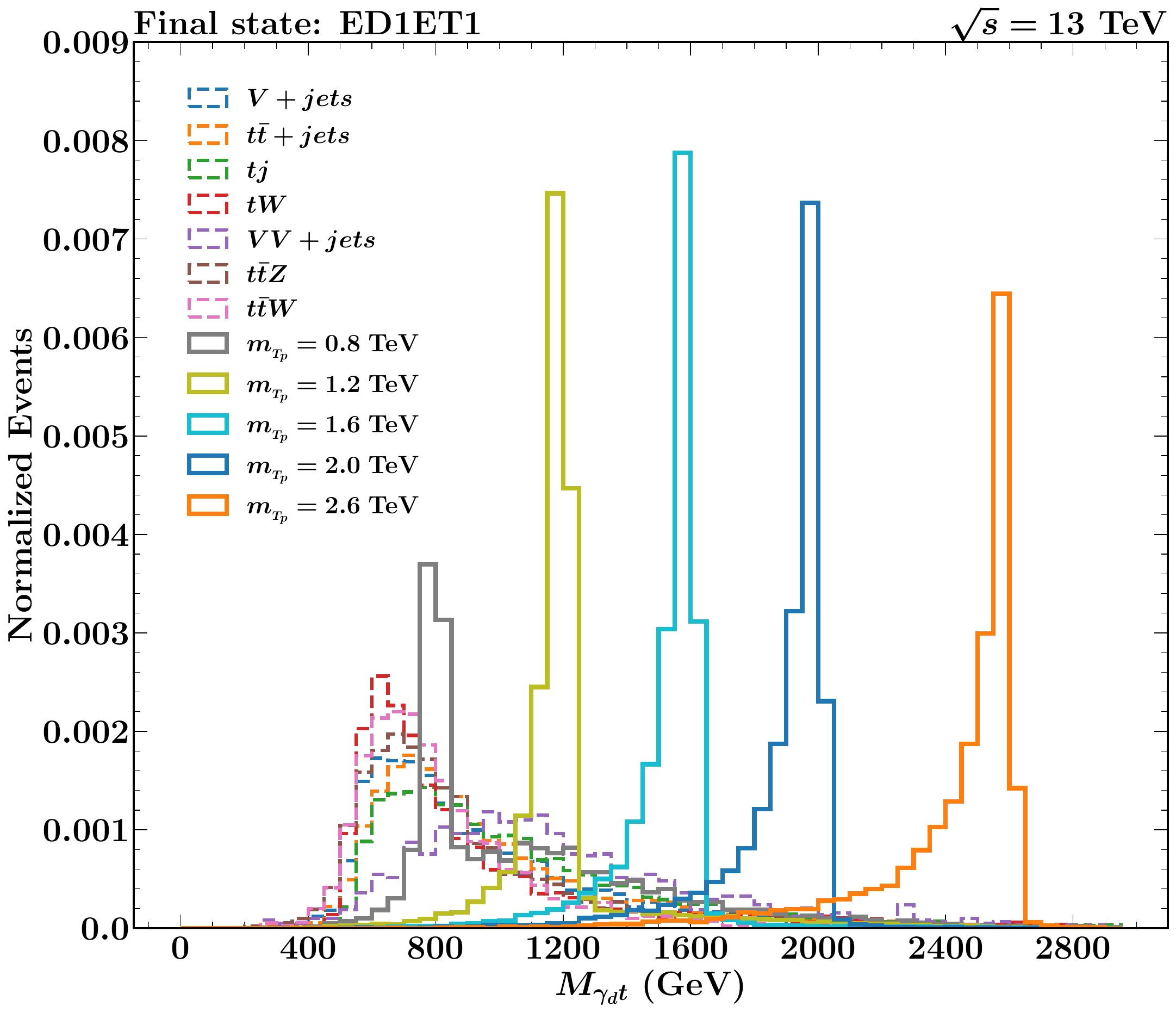}}}
  \caption{{ Invariant mass distribution of the identified dark photon and top quark system for signal and background in the (a) AD2AT1 and (b) ED1ET1 final states  at 13 TeV LHC center of mass energy.}}  
  \label{fig:inv_mass_top_dph_sys_13}
\end{figure}

\begin{figure}[htbp]
  \centering
  \resizebox{\columnwidth}{!}{
  \subfloat[\label{subfig:inv_a2a1_14}]{\includegraphics[width=0.5\columnwidth]{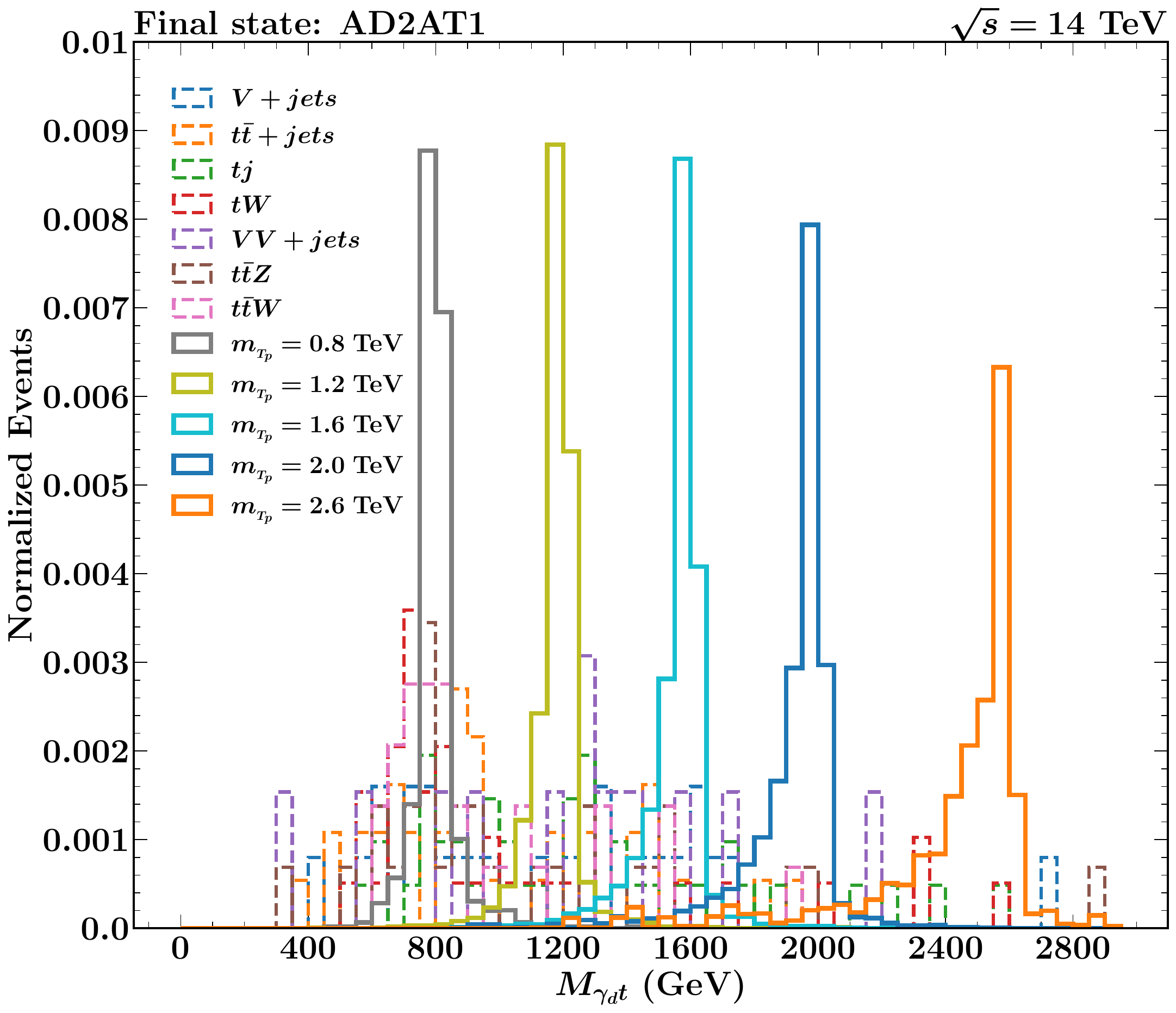}}~
  \subfloat[\label{subfig:inv_e1e1_14}]{\includegraphics[width=0.5\columnwidth]{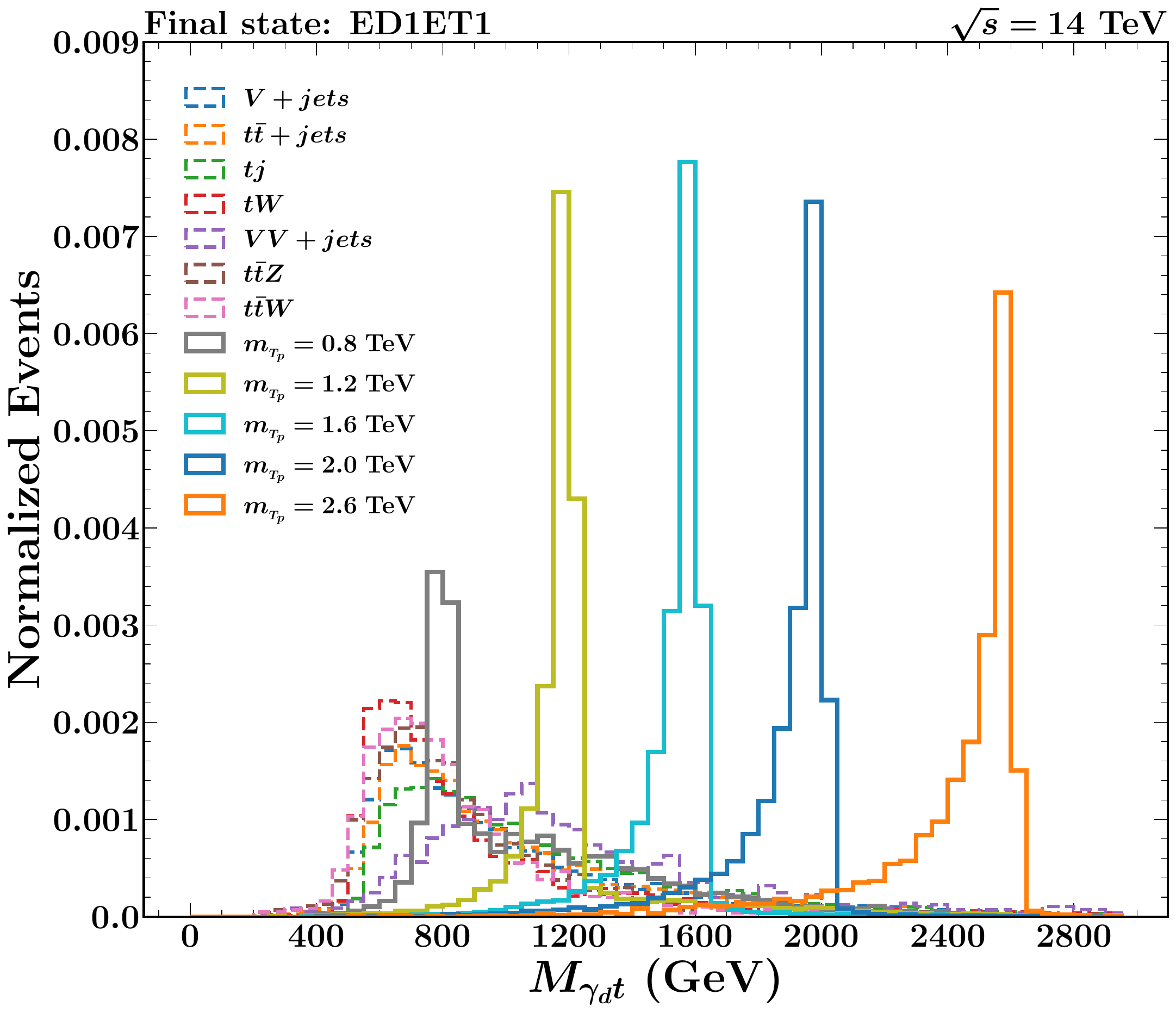}}}
  \caption{{  Invariant mass distribution of the identified dark photon and top quark system for signal and background in the (a) AD2AT1 and (b) ED1ET1 final states  at 14 TeV LHC center of mass energy.}}  
  \label{fig:inv_mass_top_dph_sys_14}
\end{figure}

We show the invariant mass distribution of the top-dark photon jet pair, defined as 
\[ M_{\gamma_d t} = \sqrt{\left\vert \left(p^j_{\gamma_d} +p^j_t\right)^2 \right\vert },\] in Fig.~\ref{fig:inv_mass_top_dph_sys_13} and Fig.~\ref{fig:inv_mass_top_dph_sys_14} for two different final states considered in this analysis at $\sqrt{s} = 13$ and $14$ TeV, respectively.

Based on the distributions of these kinematic variables we decide to apply a set of cuts to optimize the statistical significance of signal over the background in two different final states: 
\begin{itemize}
  \item Transverse momentum of the dark photon jet, $p_{_T}^j > p_{_{T0}}$.
  \item The invariant mass of the dark photon-top quark system to satisfy $\alpha \leq M_{\gamma_d t}\leq \beta$.
\end{itemize}

The values of the cut parameters collectively called $C_0$, are listed in Table~\ref{table:cuts} for various choices of benchmark points.

\begin{table}[htbp]
  \centering
  \resizebox{0.65\columnwidth}{!}{
  \begin{tabular}{ c c |c| c c c| c c c c c c c c}
  \toprule
  \multirow{2}{*}{Final State} & $\sqrt{s} $ &\multirow{2}{*}{$C_{0}$ (GeV)} &\multicolumn{11}{c}{$m_{_{T_p}}$ (TeV)}\\
  & (TeV)& & \multicolumn{3}{c|}{$0.8-1.2$}& \multicolumn{2}{c}{$1.4 -2.6$}\\
  \midrule
  \midrule
  \multirow{2}{*}{AD2AT1} & \multirow{2}{*}{13 $\&$ 14}
  &$\alpha$ & \multicolumn{3}{c|}{$\mtp$-100}  & \multicolumn{8}{c}{$\mtp$-150}\\
  && $\beta$ &\multicolumn{3}{c|}{$\mtp$+100} & \multicolumn{8}{c}{$\mtp$+100} \\
  \midrule 
  \multirow{3}{*}{ED1ET1} 
  & \multirow{3}{*}{13 $\&$ 14}&  $p_{_{T0}}$ & \multicolumn{11}{c}{$0.33 \times m_{_{T_p}}$}\\
  && $\alpha$ & \multicolumn{11}{c}{$\mtp$-100} \\
  & &  $\beta$ & \multicolumn{11}{c}{$\mtp$+50} \\        
  \bottomrule
  \end{tabular}}
  \caption{{ Values of the cut parameters ($C_0$) : top-dark photon invariant mass window ($\alpha$, $\beta$) and minimum dark photon transverse momentum (${ p_{_{T0}}}$) in GeV for various choices of benchmark points in two different final states ({\it i.e.}~AD2AT1 and ED1ET1) at $\sqrt{s} = $ $13$ TeV and $14$ TeV in the context of LHC.}}
  \label{table:cuts}
\end{table}

\section{Results and discussions}\label{sec:results}

In this section we present the results of our analysis. Table~\ref{table:res_ad2at1} and Table~\ref{table:res_ed1et1} contains the event rate for AD2AT1 and ED1ET1 final states, respectively, at both 13 TeV and 14 TeV LHC center of mass energies. 
In Table~\ref{table:res_ad2at1}, we show the effect of cut flow in AD2AT1 final state for both signal and background. As mentioned earlier, the final state $\geq 2\gamma_d+\geq t$ which consists of {\it at least two dark photons and at least one top tagged jets}, has contributions from both the pair and single production of the top partner. The single production of top partner contributes when at least one or more {nondark} photon jets are {mistagged} as dark photon jets. {This final state also receives contributions from standard model background processes {\it e.g.}, $V+jets,~t\bar{t}+jets,~tj,~tW,~VV+jets,~t\bar{t}Z$ and $t\bar{t}W$ as already mentioned, with the dominant contribution coming from the $V+jets$ and $t\bar{t}+jets$ background}. The requirement of {\it at least two} dark photon jets in this final state suppresses the SM background significantly due to the low {mistagging} rate. 
One can see from the Table~\ref{table:res_ad2at1} that the invariant mass of the {\it tagged top quark and dark photon} system can effectively suppress the standard model background in this final state.

In Table~\ref{table:res_ed1et1}, we show the effect of cut flow in ED1ET1 final state for both signal and background. The final state $\gamma_d+t$ which consists of {\it exactly one dark photon and exactly one top tagged jets}, also receive contribution from both the pair and single production of the top partner processes. However, the relative contribution of these two processes strongly depends on the mass ($m_{_{T_p}}$) and the mixing angle ($\sin\theta_L$). 
For example, at $\sin\theta_L = 0.1$ the single production {cross section}  exceeds the pair production {cross section}  beyond $1$ TeV \cite{Kim:2019oyh,Verma:2022nyd}. On the other hand, for very small $\sin\theta_L$ ($\lesssim$ 0.02) both the final state {cross sections}   will be dominated by the pair production process. The SM backgrounds for this final state are identical to those mentioned in the AD2AT1 case.
The combination of the two kinematic variables, namely, {\it (i)} the transverse momentum of the tagged dark photon jet and, {\it (ii)} the invariant mass of the {\it tagged top quark and dark photon} system help us to optimize the signal significance in this case.

{The SM background for $\gamma_d+t$ final state which is dominated by the $V+jets$ and $t\bar{t}+jets$ processes is larger by a factor of $200$ or more than that in the $\geq 2\gamma_d+\geq t$ final state}. However, the estimated signal significance in this final state for a given benchmark point is higher than that in the  $\geq 2\gamma_d+\geq t$ final state. This is due to the fact that for $\sin\theta_L = 0.1$ and $m_{_{T_p}}\geq 1$ TeV, the single production {cross section}  is higher than the pair production cross-section.

The last two columns of Table~\ref{table:res_ad2at1} and \ref{table:res_ed1et1} gives an estimate of signal significance for two different assumptions of the top partner branching ratio. The column {labeled} by {${\rm BR}(T_p)_{100}$} assumes 100\% branching ratio while {${\rm BR}(T_p)_{\rm a}$} represents the actual branching ratio for the top partner, respectively. {The branching ratio for the dark photon decaying hadronically is assumed to be 60\% throughout our analysis.} 

{To set exclusion limits on the parameter space we use the following definition as obtained from the profile likelihood ratio method \cite{Cowan:2010js, Kumar:2015tna,Bhattiprolu:2020mwi},

\bea
\sigma_{\rm excl.} = \sqrt{2\left[s-b\ln\left(1 + \frac{s}{b}\right)\right]}
\eea

\noindent
Here $\sigma_{excl.}$ quantifies the median exclusion statistical significance estimation. 
$s$ is expected number of signal events and $b$ is the expected number of background events.}
We have assumed 139 fb$^{-1}$ (300 fb$^{-1}$ and 3 ab$^{-1}$) of integrated luminosity at 13 (14) TeV LHC center of mass energy to obtain the signal significance.


We also consider implications of existing LHC analyses in the context of an extension of standard model with a singlet {vectorlike} top quark partner by CMS collaboration. The CMS data corresponding to an integrated luminosity of 137 fb$^{-1}$ at 13 TeV center of mass energy has set limits on the ${\sigma(p p \to T_p b q)\times BR (T_p\to tb\bar{b})}$ as a function of the top partner mass in fully hadronic final state consisting a total of seven jets, where five jets come from the decay of the top partner and three of them are b quark jets \cite{CMS:2024qdd}. 
We also use the CMS limit on top partner pair production {cross section}  times branching ratio in the context of the same model in three different final states: {single lepton} channel, same-sign {dilepton} channel, and ``{multilepton}'' channel with at least three leptons, corresponding to an integrated luminosity of 137 fb$^{-1}$ at 13 TeV center of mass energy \cite{CMS:2022fck}. 
We recast these bounds in the context of the model considered in this work where the {vectorlike} top quark acts as a portal matter and the branching ratio of the top partner in the SM mode is not necessarily fixed by the mass of the top partner ($m_{_{T_p}}$) and mixing angle ($\sin\theta_L$) between the top quark and top partner. As already mentioned in sec. \ref{sec:model} and in \cite{Verma:2022nyd}, the top partner branching ratios both in the standard ($bW,~tZ,~th$) and in the {nonstandard} ($t\gamma_d,~ th_d$) modes also depend on the vacuum expectation value of the dark Higgs ($v_d$), mass of the dark photon ($m_{\gamma_d}$) and mass of the dark Higgs($m_{h_d}$). 

\begin{figure}[htbp]
  \centering
  \resizebox{\columnwidth}{!}{
  \subfloat[\label{subfig:exclusion_13_1}]{\includegraphics[width=0.5\columnwidth]{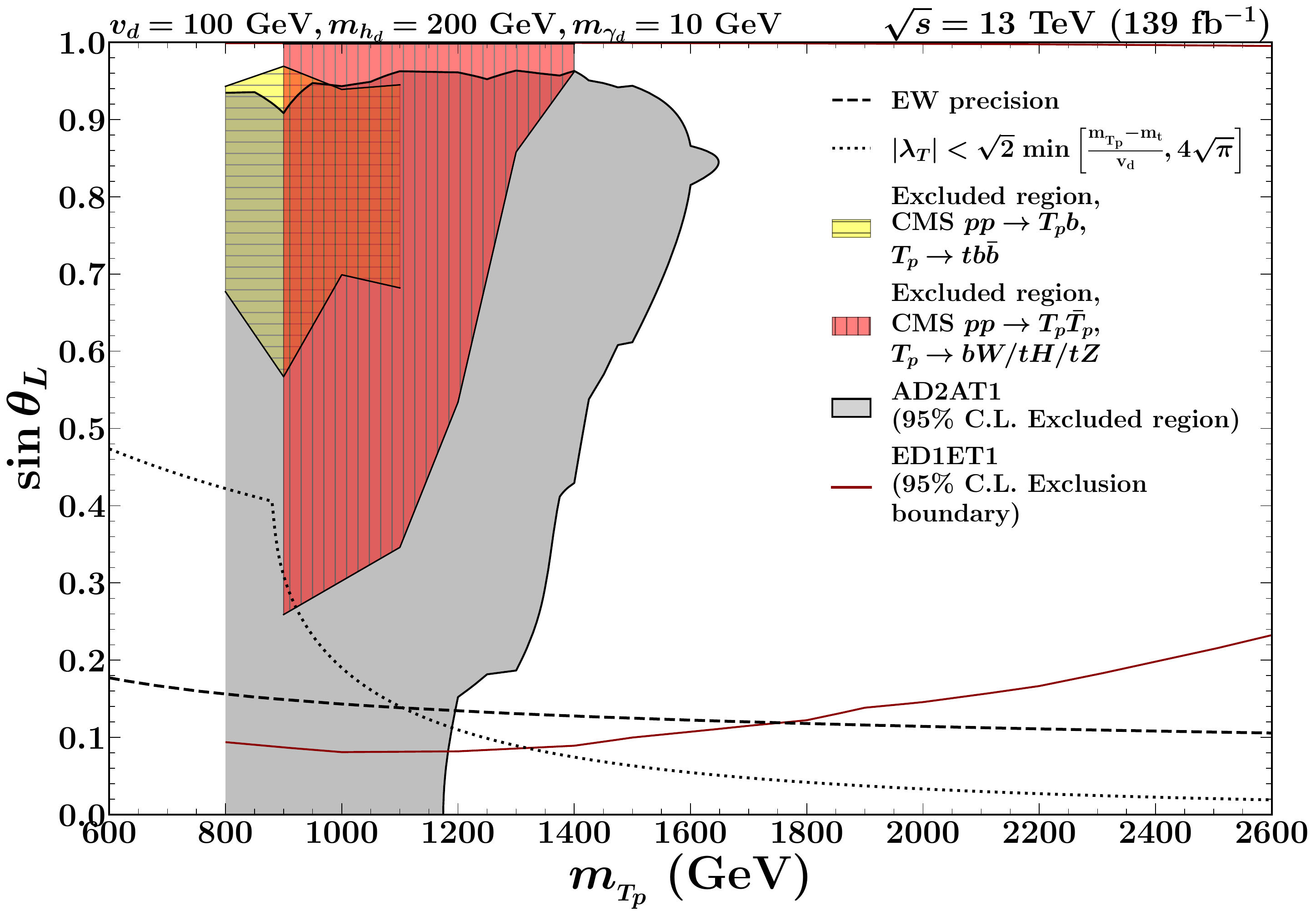}}~
  \subfloat[\label{subfig:exclusion_13_2}]{\includegraphics[width=0.5\columnwidth]{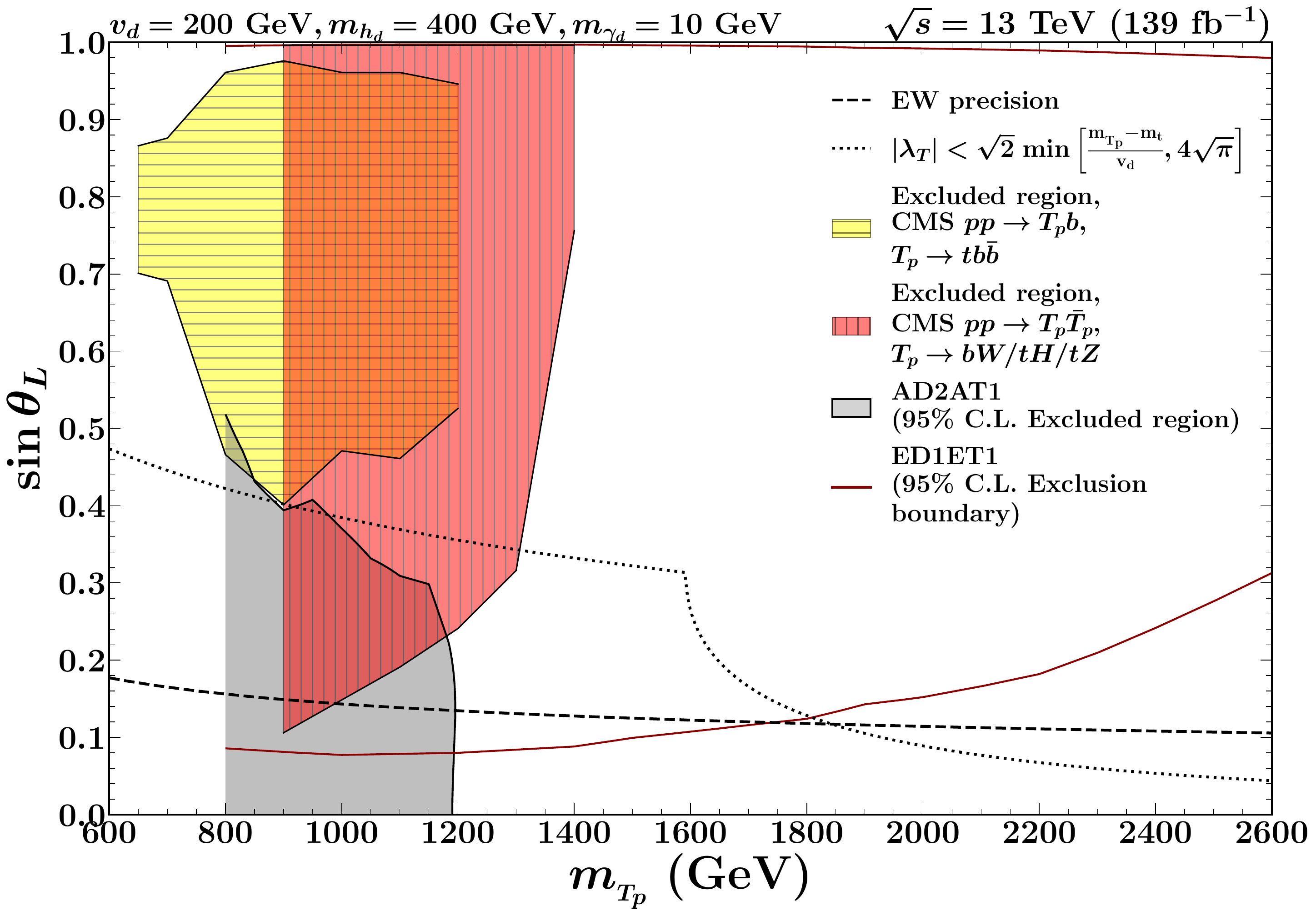}}
  }
  \caption{{
    Various exclusion limits in the $\sin\theta_L - m_{_{T_p}}$ plane. The exclusion limits set by the CMS data \cite{CMS:2024qdd,CMS:2022fck} in the context of our model, as mentioned in the text, are represented by the horizontally hatched (yellow shaded) region and vertically hatched (red shaded) region, respectively. The constraints coming from the Eq.~\eqref{eq:bound_lamT} and the electroweak precision measurements are depicted, respectively by dotted (black) and dashed (black) lines. The region bounded between two solid (red) lines and the gray colored region are {disfavored} with 95\% {CL} using our analyses in the ED1ET1 and AD2AT1 final states, respectively, assuming actual BR($T_p \to t \gamma_d$) at $\sqrt{s}=13$ TeV with 139 fb$^{-1}$ of integrated luminosity. The two illustrative {subfigures} correspond to (a) $v_d = 100$ GeV, $m_{h_d}=200$ GeV and $m_{\gamma_d}= 10$ GeV and (b) $v_d = 200$ GeV, $m_{h_d}=400$ GeV and $m_{\gamma_d}= 10$ GeV.
  }}
  \label{fig:exclusion_plot_13}
\end{figure}

\begin{figure}[htbp]
  \centering
  \resizebox{\columnwidth}{!}{
  \subfloat[\label{subfig:exclusion_14_300fb_1}]{\includegraphics[width=0.5\columnwidth]{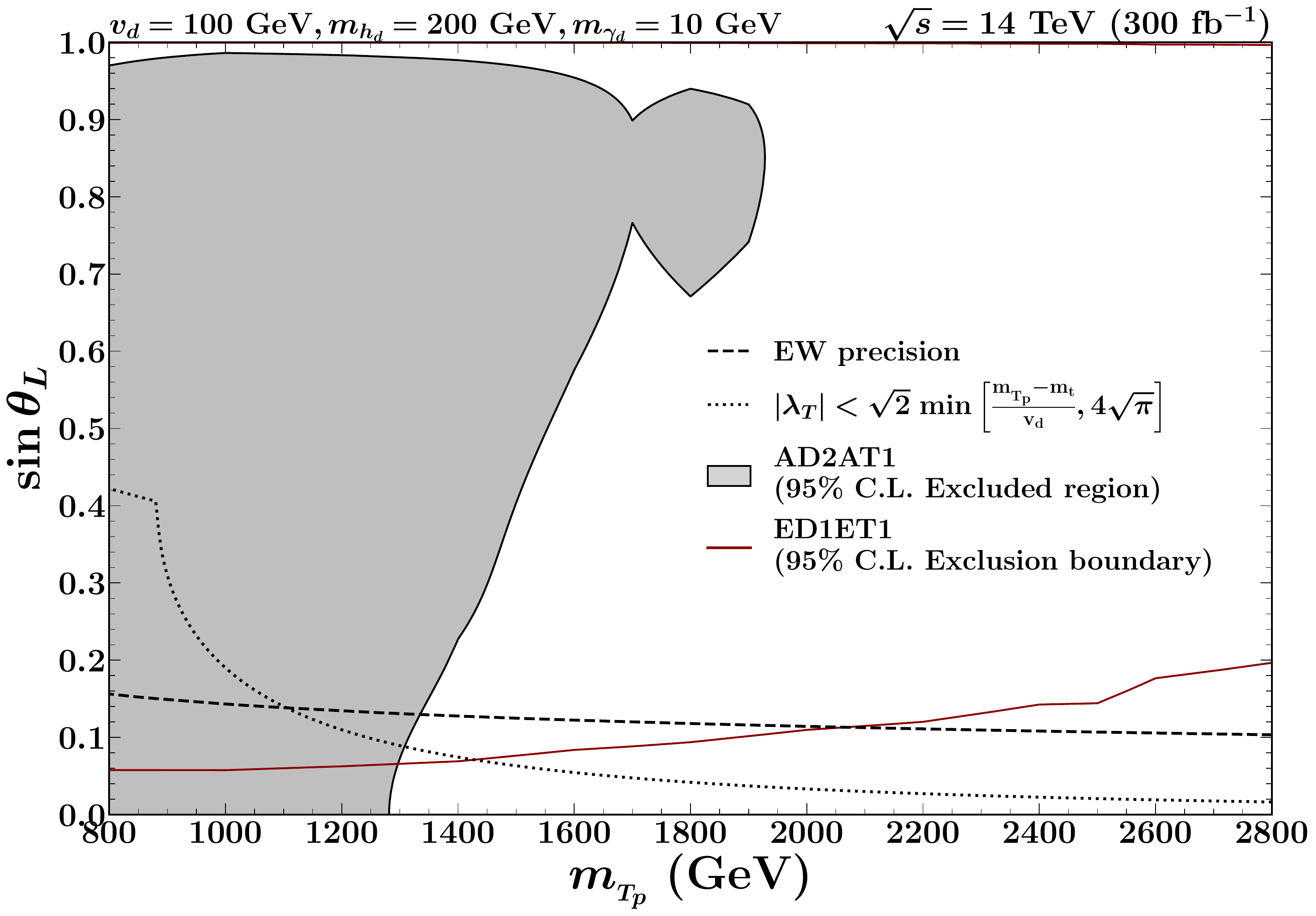}}~
  \subfloat[\label{subfig:exclusion_14_300fb_2}]{\includegraphics[width=0.5\columnwidth]{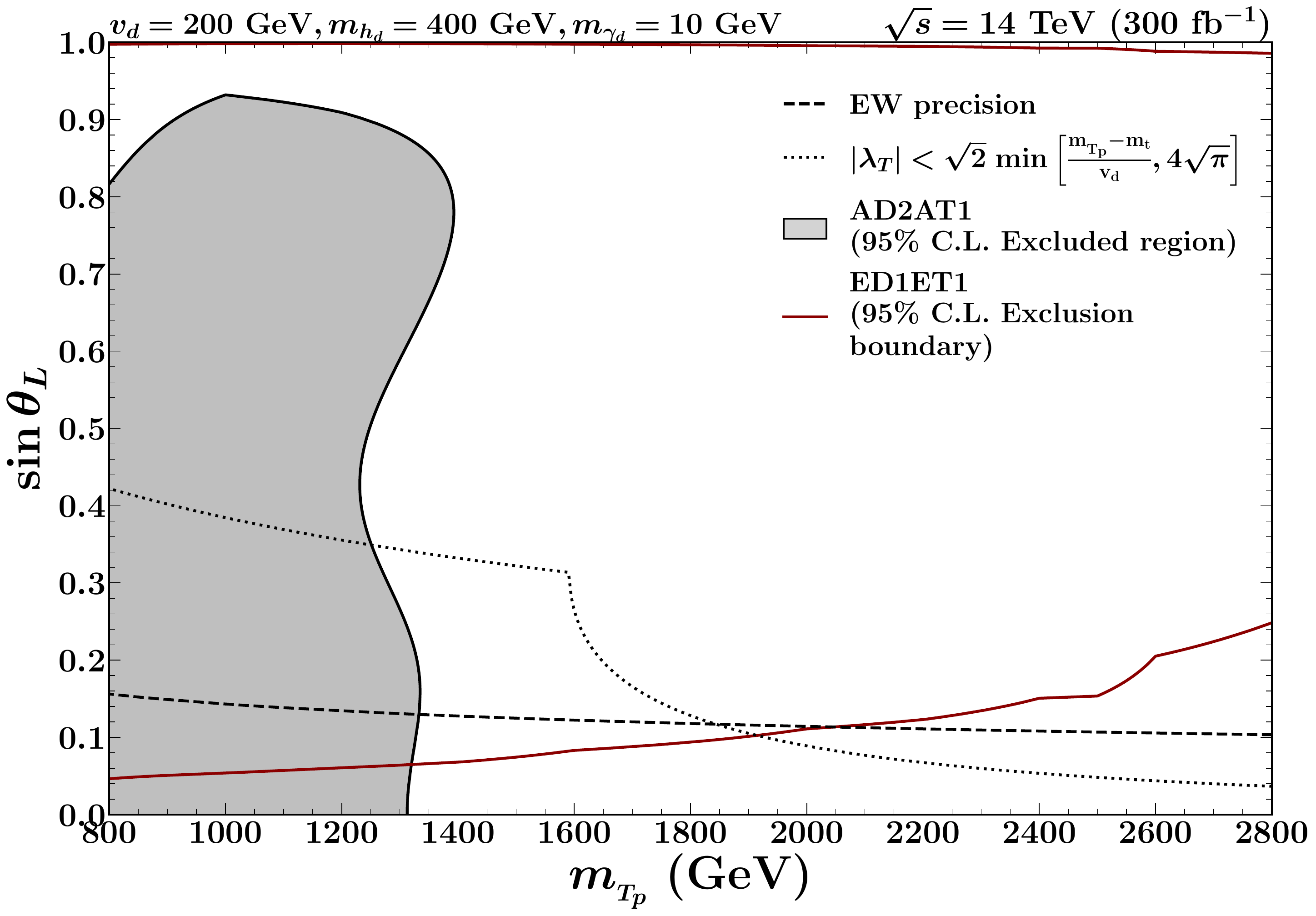}}}
  \caption{{
    Various exclusion limits in the $\sin\theta_L - m_{_{T_p}}$ plane. The constraints coming from the Eq.~\eqref{eq:bound_lamT} and the electroweak precision measurements are depicted, respectively by dotted (black) and dashed (black) lines. The region bounded between two solid (red) lines and the gray colored region are {disfavored} with 95\% {CL} using our analyses in the ED1ET1 and AD2AT1 final states, respectively, assuming actual BR($T_p \to t \gamma_d$) at $\sqrt{s}=14$ TeV with 300 fb$^{-1}$ of integrated luminosity. The two illustrative {subfigures} correspond to (a) ${ v_d}$ =100 GeV, ${ m_{h_d}}$ = 200 GeV, ${ m_{\gamma_d}}$ = 10 GeV and (b) ${\rm v_d}$ =200 GeV, $m_{h_d}$ = 400 GeV, ${ m_{\gamma_d}}$ = 10 GeV. 
  }}
  \label{fig:exclusion_plot_14_300}
\end{figure}

\begin{figure}[htbp]
  \centering
  \resizebox{\columnwidth}{!}{
  \subfloat[\label{subfig:exclusion_14_3ab_1}]{\includegraphics[width=0.5\columnwidth]{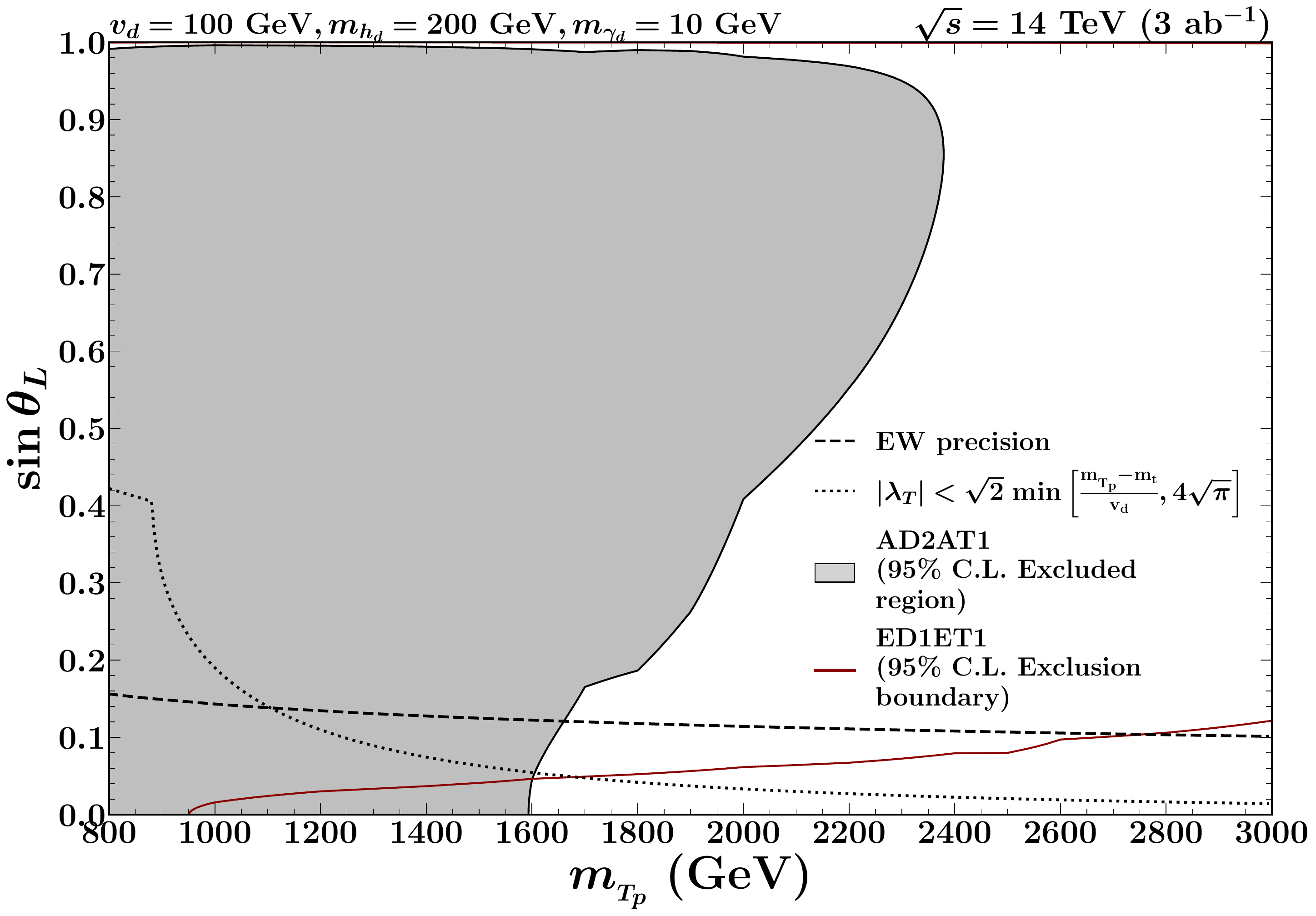}}~
  \subfloat[\label{subfig:exclusion_14_3ab_2}]{\includegraphics[width=0.5\columnwidth]{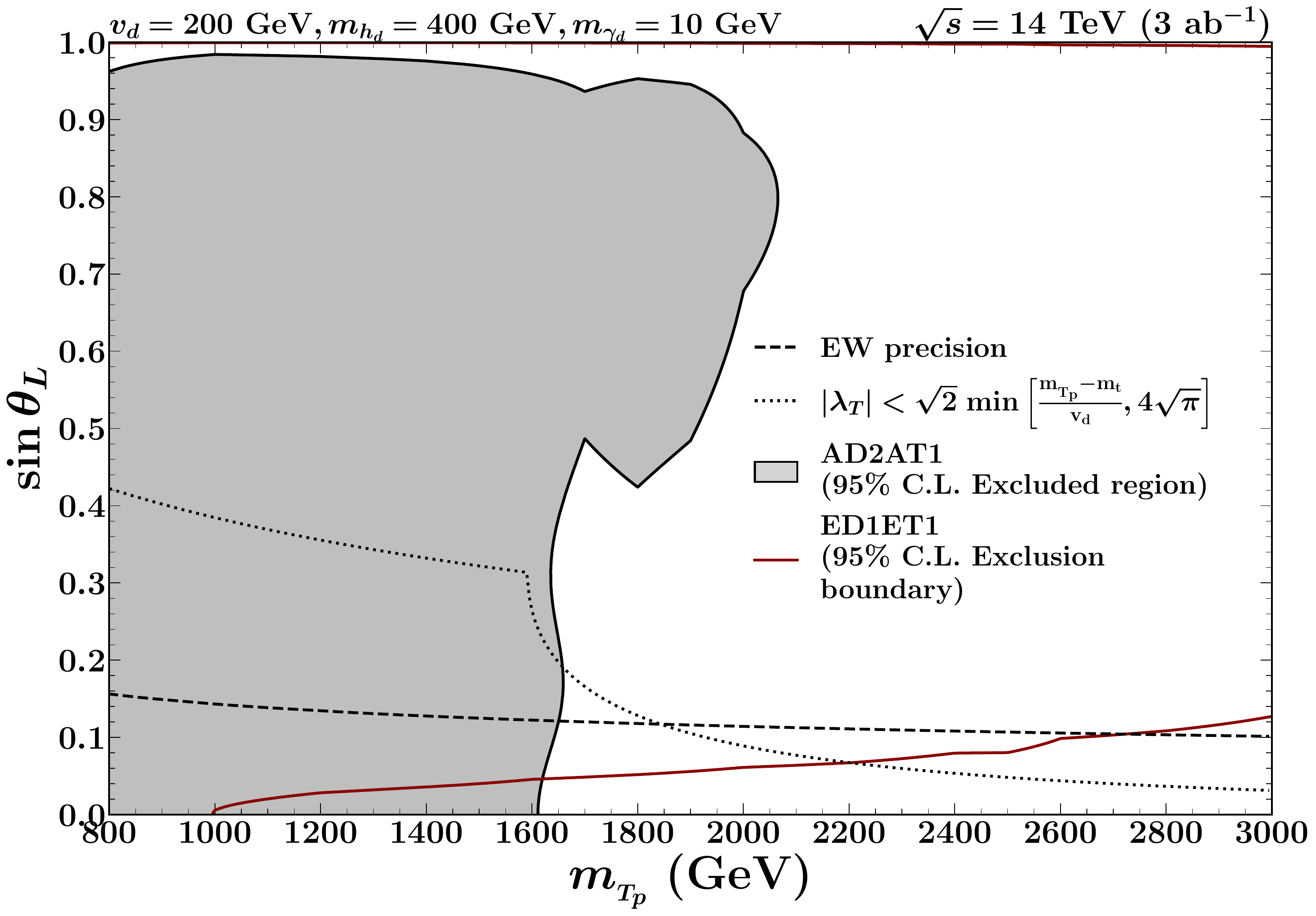}}}
  \caption{{ Various exclusion limits in the $\sin\theta_L - m_{_{T_p}}$ plane. The constraints coming from the Eq.~\eqref{eq:bound_lamT} and the electro-weak precision measurements are depicted, respectively by dotted (black) and dashed (black) lines. The region bounded between two solid (red) lines and the gray colored region are {disfavored} with 95\% {CL} using our analyses in the ED1ET1 and AD2AT1 final states, respectively, assuming actual BR($T_p \to t \gamma_d$) at $\sqrt{s}=14$ TeV with 3 ab$^{-1}$ of integrated luminosity. The two illustrative {subfigures} correspond to (a) ${ v_d}$ =100 GeV, ${ m_{h_d}}$ = 200 GeV, ${ m_{\gamma_d}}$ = 10 GeV and (b) ${\rm v_d}$ =200 GeV, $m_{h_d}$ = 400 GeV, ${ m_{\gamma_d}}$ = 10 GeV. 
  }}
  \label{fig:exclusion_plot_14_3000}
\end{figure}

{We use above CMS data to set limit in the $\sin\theta_L - m_{_{T_p}}$ plane and corresponding exclusion limit is depicted in Fig.~\ref{fig:exclusion_plot_13}. The gray filled exclusion region and solid (red) exclusion boundary in Fig.~\ref{fig:exclusion_plot_13} corresponds to expected 95\% {confidence level} (CL) obtained assuming $p-p$ collision center of mass energy to be 13 TeV at an integrated luminosity of 139 fb$^{-1}$ in the AD2AT1 and ED1ET1 final states, respectively.  In Fig.~\ref{fig:exclusion_plot_14_300} and \ref{fig:exclusion_plot_14_3000}, we also present the expected 95\% CL exclusion limits in the $\sin\theta_L - m_{_{T_p}}$ plane using our analysis assuming $\sqrt{s}= 14$ TeV at an integrated luminosity of 300 fb$^{-1}$ and 3 ab$^{-1}$, respectively. The bounds coming from Eq.~\eqref{eq:bound_lamT} and electroweak precision data are also shown by dotted and dashed line, respectively in these plots. One can see that a large fraction of the region excluded by CMS data is already {disfavored} by these constraints. However, all these bounds depend considerably on the choices of $v_d,~m_{h_d}$ and $m_{\gamma_d}$. We illustrate this dependency for two different sets of these parameters. The Figs.~\ref{subfig:exclusion_13_1}, \ref{subfig:exclusion_14_300fb_1} and \ref{subfig:exclusion_14_3ab_1}, correspond to $v_d = 100 {~\rm GeV}, ~m_{h_d} = 200{~\rm GeV}, m_{\gamma_d} = 10 {~\rm GeV}$ whereas the Figs.~\ref{subfig:exclusion_13_2}, \ref{subfig:exclusion_14_300fb_2} and \ref{subfig:exclusion_14_3ab_2} correspond to $v_d = 200 {~\rm GeV}, ~m_{h_d} = 400{~\rm GeV}, m_{\gamma_d} = 10 {~\rm GeV}$.

If one compares the Figs.~\ref{subfig:exclusion_13_1} and \ref{subfig:exclusion_13_2}, it can be clearly seen how the bound obtained by the CMS data is relaxed if we decrease $v_d$ and $m_{h_d}$. This is due to the reduction of branching ratio in the SM decay modes of the top partner.
The complementary behavior can be noted in the exclusion regions obtained by our analyses in these {subfigures} (and also in the {subfigures} of Fig.~\ref{fig:exclusion_plot_14_300} and \ref{fig:exclusion_plot_14_3000}).


In Figs.~\ref{fig:exclusion_plot_13}, \ref{fig:exclusion_plot_14_300} and \ref{fig:exclusion_plot_14_3000}, one can see that the exclusion regions in both the AD2AT1 and ED1ET1 final states display a significant dependency on $\sin\theta_L$. This is due to the fact that, both of these final states comprises significant admixture of single and pair production of top partner processes. The $\sin\theta_L$ dependency in the contribution from pair production enters through branching ratio of the top partner into top and dark photon mode, while for the contribution coming from the single production, $\sin\theta_L$ dependency enters through both production {cross section}  and branching ratio.
The exclusion regions in the AD2AT1 final state can be understood as follows: At low $\stl$, the single production {cross section}  is much smaller than the pair production cross-section. At the same time, $\ppxssecbr$ is enhanced due to the larger branching ratio compared to higher $\stl$ values (see Fig.~\ref{fig:br_tp}). As a result, the signal significance is primarily driven by pair production at low $\stl$. On the other hand, at higher values of $\stl$ the contribution from single production becomes substantial and can dominate over the pair production {cross section}  for higher $\mtp$ values. However, both of these contributions suffer branching ratio suppression at higher $\stl$ values, leading to an intricate interplay between them in the estimation of signal significance. This is illustrated in Fig.~\ref{fig:excl_explain_ad2at1} for four different choices of benchmark points.
The exclusion region in the ED1ET1 final state on the other hand can be discerned as follows: As can be seen in Fig.~\ref{fig:excl_explain_ed1et1}, the contribution to this final state is dominated by single production of top partner for a wide range of $\stl$ values. {Therefore, the exclusion limit is predominantly driven by the $\spxssecbr$ as a function of $\stl$ and has a much simpler behavior unlike AD2AT1 case.} The exclusion regions  bounded by the two solid (dark red) lines corresponding to this final state are also depicted in Figs.~\ref{fig:exclusion_plot_13}, \ref{fig:exclusion_plot_14_300} and \ref{fig:exclusion_plot_14_3000}. 

{From Fig.~\ref{fig:exclusion_plot_13}, we can see that top partner mass approximately up to 1.17 TeV can be ruled out using the 13 TeV (139 fb$^{-1}$) LHC data in the AD2AT1 final state for $\stl\leq$ 0.96 assuming $v_d = 100$ GeV and $m_{h_d}=200$ GeV. 
For $v_d = 200$ GeV ( and $m_{h_d}=400$ GeV ) with other parameters kept fixed, in the region $\stl \leq$ 0.22, $\mtp$ approximately up to $1.19$ TeV can be excluded with $\geq 2\sigma$ significance. The same plot also depicts that the region corresponding to $\stl\geq 0.23 (0.32)$ and $\mtp$ up to 2.6 TeV is disallowed by the 13 TeV (139 fb$^{-1}$) LHC data in the ED1ET1 final state assuming $v_d = 100~ (200)$ GeV. There is a small region in the top right corner of Fig.~\ref{subfig:exclusion_13_2} for $v_d =200$ GeV which will still be allowed by the same data in the ED1ET1 final state.

Similar exclusion limits can be obtained at 14 TeV LHC center of mass energy corresponding to two different integrated luminosities, {\it i.e.}~300 fb$^{-1}$ and 3 ab$^{-1}$, and these are highlighted in Fig.~\ref{fig:exclusion_plot_14_300} and Fig.~\ref{fig:exclusion_plot_14_3000}, respectively.
For example, in the AD2AT1 final state, we illustrate that the top partner masses approximately up to $1.6$ TeV is disfavored for $\stl \leq 0.99 (0.98)$ assuming $v_d = 100~(200)$ GeV. In addition, the exclusion limit can reach up to $\mtp\sim$ 2.8 (2.35) TeV at $\stl = 0.85~(0.8)$ assuming $v_d = 100~(200)$ GeV. In contrast, with 3 ab$^{-1}$ of integrated luminosity we have obtained a $2\sigma$ exclusion limit on $\sin\theta_L$ as low as 0.1 up to $m_{_{T_p}} = $ 3.0 (2.8) TeV assuming $v_d = 100~(200)$ GeV in the ED1ET1 final state.
}

}

\begin{table}[htbp]
  \centering
  \resizebox{0.75\columnwidth}{!}{
  \begin{tabular}{cccccccc}
    \toprule
    \multirow{3}{*}{$\sqrt{s}$} & \multirow{3}{*}{$m_{_{T_p}}$}& 
    \multicolumn{2}{c}{Final state  } & \multicolumn{2}{c}{$M_{\gamma_d  t}$} & \multicolumn{2}{c}{\multirow{2}{*}{Significance}} \\
     &  & \multicolumn{2}{c}{cross-section}  & \multicolumn{2}{c}{$\in [\alpha,\beta]$} & \multirow{3}{*}{{BR$(T_p)_{100}$} }& 
     \multirow{3}{*}{{BR$(T_p)_{a}$}} \\
     \multirow{2}{*}{(TeV)}& \multirow{2}{*}{(TeV)} &  \multicolumn{2}{c}{(fb)}  & \multicolumn{2}{c}{(fb)}  & & \\
     &  &  $T_p j$ & $T_p \bar{T}_p$ & $T_p j$ & $T_p \bar{T}_p$ & & \\
    \midrule
    \midrule

       \multirow{16}{*}{13}
       
       &\multirow{2}{*}{0.8} & 0.007 & 3.553 & 0.007& 3.184 & \multirow{2}{*}{12.9}& \multirow{2}{*}{5.8}\\
       &  & \multicolumn{2}{c}{\it 1.086} & \multicolumn{2}{c}{\it 0.422}&& \\
       &\multirow{2}{*}{1.0} & 0.011 & 1.734 &  0.008 & 1.530 & \multirow{2}{*}{8.6}& \multirow{2}{*}{3.5}\\
       & & \multicolumn{2}{c}{\it 1.086} & \multicolumn{2}{c}{\it 0.239}&& \\
       &\multirow{2}{*}{1.2} &  0.013 & 0.813 & 0.009 & 0.697 & \multirow{2}{*}{5.2}& \multirow{2}{*}{1.9}\\
       & & \multicolumn{2}{c}{\it 1.086} & \multicolumn{2}{c}{\it 0.174}&& \\
       &\multirow{2}{*}{1.4} & 0.011& 0.348 & 0.007 & 0.313& \multirow{2}{*}{3.4}& \multirow{2}{*}{1.2} \\
       & & \multicolumn{2}{c}{\it 1.086} & \multicolumn{2}{c}{\it 0.091} && \\
       &\multirow{2}{*}{1.5} & 0.008& 0.231 & 0.006 & 0.204 &\multirow{2}{*}{2.5} & \multirow{2}{*}{0.8}\\
       & & \multicolumn{2}{c}{\it 1.086} & \multicolumn{2}{c}{\it 0.089} &&\\
       &\multirow{2}{*}{1.6} & 0.007& 0.146 & 0.004 & 0.126 &\multirow{2}{*}{1.7} & \multirow{2}{*}{0.5}\\
       & & \multicolumn{2}{c}{\it 1.086} & \multicolumn{2}{c}{\it 0.083} &&\\
       &\multirow{2}{*}{1.8} &  0.004 & 0.058 & 0.003 & 0.049 & \multirow{2}{*}{0.7}& \multirow{2}{*}{0.2}\\
       & & \multicolumn{2}{c}{\it 1.086} & \multicolumn{2}{c}{\it 0.082} &&\\
       &\multirow{2}{*}{2.0} & 0.003 & 0.023 & 0.002& 0.018& \multirow{2}{*}{0.3}& \multirow{2}{*}{0.1} \\
       & & \multicolumn{2}{c}{\it 1.086} & \multicolumn{2}{c}{\it 0.082}  &&\\

    \cmidrule{2-8}

    \multirow{16}{*}{14}

    &\multirow{2}{*}{0.8} & 0.011 & 4.553 & 0.009 & 4.129 & \multirow{2}{*}{22.5} & \multirow{2}{*}{10.4} \\
    && \multicolumn{2}{c}{\it 1.764} & \multicolumn{2}{c}{\it 0.435} &&\\

    &\multirow{2}{*}{1.0} & 0.025 & 2.367 & 0.020& 2.101 & \multirow{2}{*}{14.1} & \multirow{2}{*}{5.6} \\
    && \multicolumn{2}{c}{\it 1.764} & \multicolumn{2}{c}{\it 0.417} &&\\
    
   &\multirow{2}{*}{1.2} & 0.019 & 1.150 & 0.015 & 0.988 & \multirow{2}{*}{8.3} & \multirow{2}{*}{2.9} \\
    && \multicolumn{2}{c}{\it 1.764} & \multicolumn{2}{c}{\it 0.349} &&\\

   &\multirow{2}{*}{1.4} & 0.014 & 0.512 & 0.010 & 0.421 &\multirow{2}{*}{4.6} & \multirow{2}{*}{1.5} \\
    && \multicolumn{2}{c}{\it 1.764} &  \multicolumn{2}{c}{\it 0.308} &&\\ 

   &\multirow{2}{*}{1.6} &  0.009 & 0.229 & 0.007 & 0.198  & \multirow{2}{*}{2.4} & \multirow{2}{*}{0.7} \\
    && \multicolumn{2}{c}{\it 1.764}  & \multicolumn{2}{c}{\it 0.257} &&\\
  
    &\multirow{2}{*}{1.7} &  0.007 & 0.145 & 0.004 & 0.124  & \multirow{2}{*}{1.7} & \multirow{2}{*}{0.5} \\
    && \multicolumn{2}{c}{\it 1.764}  & \multicolumn{2}{c}{\it 0.195} &&\\
    
   &\multirow{2}{*}{1.8} & 0.006 & 0.094 & 0.004 & 0.078  & \multirow{2}{*}{1.5} & \multirow{2}{*}{0.4}\\
    && \multicolumn{2}{c}{\it 1.764}  & \multicolumn{2}{c}{\it 0.111} &&\\
    
   &\multirow{2}{*}{2.0} & 0.004 & 0.039 & 0.002 & 0.031  &\multirow{2}{*}{0.7} & \multirow{2}{*}{0.2} \\
    && \multicolumn{2}{c}{\it 1.764}  & \multicolumn{2}{c}{\it 0.104} &&\\
    
    \bottomrule
 \end{tabular}
 }
 \caption{ {The effects of cut flow on {cross sections}   for both the signal and 
 the total SM background along with the signal significance in the {\bf AD2AT1} ($\geq 2\gamma_d + \geq t$) final state at $\sqrt{s}=13$ TeV (139 fb$^{-1}$) and at $\sqrt{s}=14$ TeV (300 fb$^{-1}$). {The {cross sections}   presented above after the final state requirement and the invariant mass cut assume BR$(T_{p}\to t \gamma_{d}) = 100\%$ and BR$(\gamma_{d}\to  q\bar{q}) = 100\%$. The signal {cross sections}   are quoted separately for two different {subprocesses} for various choices of $\mtp$ and the corresponding total SM background {cross sections}   are highlighted in {\it italics}. We have assumed $\sin \theta_{L} = 0.1$, $v_d = 200$ GeV, $m_{\gamma_d}=10$ GeV, and $m_{h_d}= 400$ GeV to estimate the {cross sections}   and the signal significance. The estimated signal significances presented in the last two columns assume $100\%$ and \textit{actual} BR($T_p \to t \gamma_d$), respectively, along with ${\rm BR}(\gamma_d\to q\bar{q}) =$ 60\%.}} }
 \label{table:res_ad2at1}
\end{table}

\begin{table}[htbp]
  \centering
  \resizebox{0.95\columnwidth}{!}{
    \begin{tabular}{ccccccccccc}
      \toprule
      \multirow{3}{*}{$\sqrt{s}$} & \multirow{3}{*}{$m_{_{T_p}}$}& 
      \multicolumn{2}{c}{Final state  } & \multicolumn{2}{c}{$p_{_T}\geq p_{_{T0}}$}&\multicolumn{2}{c}{$M_{\gamma_d  t}$} & \multicolumn{2}{c}{\multirow{3}{*}{Significance}}\\
       &  & \multicolumn{2}{c}{cross-section}  &  &&\multicolumn{2}{c}{$\in [\alpha,\beta]$}  & & \\
       \multirow{2}{*}{(TeV)}&  \multirow{2}{*}{(TeV)}&  \multicolumn{2}{c}{(fb) }& \multicolumn{2}{c}{(fb) }&\multicolumn{2}{c}{(fb) } & \multirow{2}{*}{{BR$(T_p)_{100}$} }& 
       \multirow{2}{*}{{BR$(T_p)_{a}$}} \\
       & &  $T_p j$ & $T_p \bar{T}_p$ &  $T_p j$ & $T_p \bar{T}_p$ & $T_p j$ & $T_p \bar{T}_p$ & & \\

      \midrule
      \midrule
         \multirow{16}{*}{13}
         & \multirow{2}{*}{0.8} &   2.854 & 10.245 &  2.782 & 9.979 & 2.514 & 2.272 & \multirow{2}{*}{ 4.1 }& \multirow{2}{*}{2.1}\\
         &   & \multicolumn{2}{c}{\it 254.52	} &  \multicolumn{2}{c}{\it 181.341} & \multicolumn{2}{c}{\it 42.826}  &&\\
        
         & \multirow{2}{*}{1.0} &    3.275 & 3.245 & 3.106 & 3.071 & 2.772 & 1.021 & \multirow{2}{*}{ 5.1 }& \multirow{2}{*}{2.7}\\
         &   & \multicolumn{2}{c}{\it 254.52	} &  \multicolumn{2}{c}{\it 112.37} & \multicolumn{2}{c}{\it 20.964}  &&\\
        
         & \multirow{2}{*}{1.2} &   2.797 & 0.982 &2.562 & 0.879 &2.197 & 0.321 & \multirow{2}{*}{ 4.8}& \multirow{2}{*}{2.6}\\
         &   & \multicolumn{2}{c}{\it 254.52	} & \multicolumn{2}{c}{\it  71.612} & \multicolumn{2}{c}{\it 11.248}  &&\\        

         & \multirow{2}{*}{1.4} &   2.006 & 0.287 & 1.780 & 0.246 & 1.441 & 0.088 &\multirow{2}{*}{4.1}& \multirow{2}{*}{2.2}\\
         &   & \multicolumn{2}{c}{\it 254.52	} & \multicolumn{2}{c}{\it  47.533} & \multicolumn{2}{c}{\it 5.925} &&\\
     
         & \multirow{2}{*}{1.5} &   1.602 & 0.159 & 1.379 & 0.133 & 1.089 & 0.048 &\multirow{2}{*}{3.3}& \multirow{2}{*}{1.8}\\
         &   & \multicolumn{2}{c}{\it 254.52	}  &  \multicolumn{2}{c}{\it  32.470} &     \multicolumn{2}{c}{\it 3.998}  &&\\
         
         & \multirow{2}{*}{1.6} &   1.299 & 0.088 & 1.106 & 0.073 & 0.830 & 0.027 &\multirow{2}{*}{2.9}& \multirow{2}{*}{1.5}\\
         &   & \multicolumn{2}{c}{\it 254.52	}  &  \multicolumn{2}{c}{\it  32.470} &     \multicolumn{2}{c}{\it 3.998}  &&\\
        
         & \multirow{2}{*}{1.8} &  0.838 & 0.027 & 0.682 & 0.022 & 0.487 & 0.008  &\multirow{2}{*}{2.3}& \multirow{2}{*}{1.2}\\
         &  & \multicolumn{2}{c}{\it 254.52	} &  \multicolumn{2}{c}{\it  22.360} &     \multicolumn{2}{c}{\it  2.096}  &&\\
        
         & \multirow{2}{*}{1.9} &  0.667 & 0.015 & 0.536 & 0.012 & 0.371 & 0.004 &\multirow{2}{*}{1.8}& \multirow{2}{*}{1.0}\\
         & & \multicolumn{2}{c}{\it 254.52	}  &    \multicolumn{2}{c}{\it 18.760} & \multicolumn{2}{c}{\it 1.946} &&\\
        
         & \multirow{2}{*}{2.0} &  0.525 & 0.008 & 0.414 & 0.006 & 0.280 & 0.002 &\multirow{2}{*}{1.6}& \multirow{2}{*}{0.9}\\
         & & \multicolumn{2}{c}{\it 254.52	}  &    \multicolumn{2}{c}{\it 15.939} & \multicolumn{2}{c}{\it 1.364} &&\\

        \cmidrule{2-10}

        \multirow{22}{*}{14}
        &\multirow{2}{*}{0.8} & 3.590 & 13.664 & 3.511 & 13.339 & 3.167 & 3.270  &\multirow{2}{*}{7.8}& \multirow{2}{*}{3.9} \\
        & &  \multicolumn{2}{c}{\it 292.505} &  \multicolumn{2}{c}{\it 210.700} &      \multicolumn{2}{c}{  \it 45.233}  &&\\
      
        &\multirow{2}{*}{1.0} &   4.117 & 4.456 & 3.908 & 4.196 & 3.499 & 1.327 &\multirow{2}{*}{8.7}& \multirow{2}{*}{4.6} \\
        & &  \multicolumn{2}{c}{\it 292.505 }& \multicolumn{2}{c}{\it  136.306} &       \multicolumn{2}{c}{\it  24.345}&&\\
                 
        &\multirow{2}{*}{1.2} &   3.476 & 1.365 & 3.189 & 1.229 & 2.736 & 0.434 &\multirow{2}{*}{8.0}& \multirow{2}{*}{4.3} \\
        & &  \multicolumn{2}{c}{\it 292.505}&  \multicolumn{2}{c}{\it 89.607} &       \multicolumn{2}{c}{\it  13.861}  &&\\
       
        &\multirow{2}{*}{1.4} &  2.559 & 0.433 & 2.264 & 0.364 & 1.814 & 0.133 &\multirow{2}{*}{6.8}& \multirow{2}{*}{3.6}\\
        & & \multicolumn{2}{c}{\it 292.505} &  \multicolumn{2}{c}{\it 61.513} &        \multicolumn{2}{c}{\it  7.663}  &&\\
          
        &\multirow{2}{*}{1.6} & 1.708 & 0.134 & 1.450 & 0.109 & 1.100 & 0.040&\multirow{2}{*}{4.8}& \multirow{2}{*}{2.5}\\
        & &  \multicolumn{2}{c}{\it 292.505}&  \multicolumn{2}{c}{\it 42.318} &        \multicolumn{2}{c}{ \it 5.600} &&\\
        
        &\multirow{2}{*}{1.8} &1.083 & 0.044 & 0.884 & 0.035 & 0.622 & 0.013 &\multirow{2}{*}{3.8}& \multirow{2}{*}{2.0}\\
        &&   \multicolumn{2}{c}{\it 292.505} &  \multicolumn{2}{c}{\it 29.28}&        \multicolumn{2}{c}{\it 2.671}  &&\\         

        &\multirow{2}{*}{2.0} &  0.737 & 0.015 & 0.581 & 0.012 & 0.389 & 0.004&\multirow{2}{*}{2.9}& \multirow{2}{*}{1.5}\\
        & &   \multicolumn{2}{c}{\it 292.505} &  \multicolumn{2}{c}{\it 20.713} &        \multicolumn{2}{c}{\it  1.844} &&\\      

        &\multirow{2}{*}{2.2} & 0.474 & 0.005 & 0.365 & 0.004 & 0.231 & 0.001&\multirow{2}{*}{2.3}& \multirow{2}{*}{1.2}\\
        & &  \multicolumn{2}{c}{ \it 292.505} &  \multicolumn{2}{c}{\it 14.742} &        \multicolumn{2}{c}{\it 0.983} &&\\     

        &\multirow{2}{*}{2.4} &  0.309 & 0.002 & 0.236 & 0.0013 & 0.141 & 4$\times 10^{-4}$  &\multirow{2}{*}{1.7}& \multirow{2}{*}{0.9}\\
        &&   \multicolumn{2}{c}{ \it 292.505}&  \multicolumn{2}{c}{\it 10.355} &         \multicolumn{2}{c}{\it 0.697}&&\\

        &\multirow{2}{*}{2.6} &  0.202 & 6 $\times 10^{-4}$ & 0.150 & 5 $\times 10^{-4}$ & 0.090 & 1 $\times 10^{-4}$&\multirow{2}{*}{1.1}& \multirow{2}{*}{0.6}\\
        & &  \multicolumn{2}{c}{\it 292.505} &  \multicolumn{2}{c}{\it 6.976} &         \multicolumn{2}{c}{\it 0.563} &&\\
       
      \bottomrule
   
   \end{tabular}
   }
 \caption{ {The effects of cut flow on {cross sections}   for both the signal and 
 the total SM background along with the signal significance in the {\bf ED1ET1} ($\gamma_d + t$) final state at $\sqrt{s}=13$ TeV (139 fb$^{-1}$) and at $\sqrt{s}=14$ TeV (300 fb$^{-1}$). {The {cross sections}   presented above after the final state requirement, the minimum transverse momentum threshold and the invariant mass cut assume BR$(T_{p}\to t \gamma_{d}) = 100\%$ and BR$(\gamma_{d}\to  q\bar{q}) = 100\%$. The signal {cross sections}   are quoted separately for two different {subprocesses} for various choices of $\mtp$ and the corresponding total SM background {cross sections}   are highlighted in {\it italics}. We have assumed $\sin \theta_{L} = 0.1$, $v_d = 200$ GeV, $m_{\gamma_d}=10$ GeV, and $m_{h_d}= 400$ GeV to estimate the {cross sections}   and the signal significance. The estimated signal significances presented in the last two columns assume $100\%$ and \textit{actual} BR($T_p \to t \gamma_d$), respectively, along with ${\rm BR}(\gamma_d\to q\bar{q}) =$ 60\%.}}}
  \label{table:res_ed1et1}
\end{table}

\section{Conclusions}\label{conclusions}

We have considered an extension of the SM with a {vectorlike} fermion having same quantum number as that of the right-handed top quark under 
the SM gauge group. In addition, the model also possesses a local $U(1)_d$ symmetry with corresponding gauge boson, {\it i.e.}, the dark photon. The
top partner is charged under the local $U(1)_d$ gauge group while all the SM particles are neutral under it. The $U(1)_d$ symmetry is spontaneously 
broken by the {nonzero} vacuum expectation value of an additional scalar field. The top partner in this model decays to both standard ($T_p \to bW,~tZ,~th$) 
and {nonstandard} modes ($T_p \to t {\gamma_d},~th_d$). The key feature of this model is that the top partner substantially decays to the {nonstandard} modes 
while the branching ratio in the corresponding standard modes are suppressed in a wide range of parameter space.

The dark photon in this scenario is considered unstable and promptly decays to a pair of SM fermions via the gauge kinetic mixing. The dark photon produced in the decay of the heavy top-partner is highly boosted, and corresponding decay products are extremely collimated. 
A crucial part of our analysis is to isolate a dark photon initiated jet in presence of other hadronic activity in a given event. In particular isolating it from the highly boosted top-quark and/or other jets present in the same event while also efficiently reducing the {mistagging} rate of {nondark} photon jets present in the SM background processes. We have used a hybrid deep neural network based algorithm to identify the highly boosted dark photon initiated jet on an event-by-event basis for the low mass dark photon ($\sim 5-20$ GeV). The method proposed in this work to identify a dark photon jet can be applied to other light exotic particles in their hadronic decay mode if highly boosted. However, our method is not sensitive to the spin of the light exotic particle. We have been able to achieve an average tagging efficiency of $30\%$ to $93\%$ for the dark photon jet in the top partner mass range, $0.8 - 2.8$ TeV while the corresponding average {mistagging} rate for {nondark} photon jets present in the signal and in the SM background varies in the range {$1.0\%$ to $2.6\%$} using a HDNN binary classfier trained over a sample corresponding to $m_{_{T_p}} = 1.4$ TeV, $v_d = 200$ GeV, $m_{h_d} = 400$ GeV, and $m_{\gamma_d} = 5$ GeV.

We have considered two different final states for our analysis, namely, (i) {\it at least two tagged dark photon jets along with at least one tagged top quark jet} and, (ii) {\it exactly one tagged dark photon jet along with exactly one tagged top quark jet}. 
We have explored the topology of the signal events, namely, the presence of highly boosted top and dark photon structure along with the fact that the invariant mass of the combined system peaks at around the mass of the top partner. {These immensely help us to reduce the contribution coming from the SM $V+jets~,t\bar{t}+jets,~t\bar{t}W,~t\bar{t}Z,~tW$ and $tj$ backgrounds. 
We have shown that with the help of certain kinematic observables such as the transverse momentum of the dark photon jet and the invariant mass of the top quark-dark photon jet system it is possible to exclude top partner masses up to 1.2 (1.6) TeV and 1.5 (1.9) TeV in the AD2AT1 and ED1ET1 final states respectively, at 13 TeV LHC center of mass energy with 139 fb$^{-1}$ integrated luminosity assuming actual (100\%) branching ratio of the top partner and $\sin\theta_L$ to be $0.1$ with $\sim$ 2$\sigma$ significance}. 

Additionally, we have also presented exclusion limits in the $\sin\theta_L - m_{_{T_p}}$ plane using CMS data corresponding to 137 fb$^{-1}$ integrated luminosity at 13 TeV LHC center of mass energy in the fully hadronic and {mutilepton} (which includes single, same-sign di-lepton and at least three leptons) final states in the context of single and pair production of {vectorlike} top partner, respectively. For {\bf$\sin\theta_L > 0.3$} one can exclude $m_{_{T_{p}}}$ in the range $0.9 - 1.4$ TeV assuming $v_d = 200$ GeV. However, this limit can be relaxed if one assumes $v_d < 200$ GeV.

{Finally, we have presented the $2\sigma$ exclusion limits in the $\stl-\mtp$ plane at $\sqrt{s} = 13$ TeV (139 fb$^{-1}$) and $\sqrt{s} = 14$ TeV (300 fb$^{-1}$ and 3 ab$^{-1}$) assuming the {\it actual} branching ratio of top partner to the top quark and the dark photon ($T_p \to t \gamma_d$) for two different final states. 
{Our analysis indicates that for $\mtp \lesssim 1.6$ TeV, nearly the entire range of $\stl$ ({\it i.e.}~ $\stl \leq 0.99~ (0.96)$ for $v_d= 100 ~(200)$ GeV) can be excluded with 95\% confidence level using $\sqrt{s} =14$ TeV (3 ab$^{-1}$) of data in the AD2AT1 final state. Additionally, the exclusion limit can reach up to $\mtp\sim$ 2.38 (2.07) TeV at $\stl = 0.86~(0.8)$, assuming $v_d = 100~(200)$ GeV.
Similarly, in the ED1ET1 final state, the analysis presented in this work shows that values of $\stl$ as low as 0.03 (0.1) can be excluded with 95\% confidence level for $\mtp \leq 1.2~(2.8)$ TeV at $\sqrt{s} = 14$ TeV and 3 ab$^{-1}$ of integrated luminosity, assuming $v_d = 200$ GeV.}

Our collider analysis is applicable to scenarios where a top partner decays into a top quark and a neutral, light exotic particle that subsequently decays promptly into a pair of Standard Model fermions.}

\section*{Acknowledgement}

We would like to thank Arun Nayak, Satyaki Bhattacharya and Shilpi Jain for useful discussions.
S.V. is thankful for support provided by Council of Scientific $\&$ Industrial Research (CSIR, India) under CSIR-UGC NET Fellowship 
(File No.: 09/934(0017)/2020-EMR-I). 

\appendix

\setcounter{equation}{0}
\numberwithin{equation}{section}

\section[Dark-photon identification]{HDNN identification of dark photon jets}\label{App:dph_tagging}

In this section we will detail the strategy of identifying a highly boosted dark photon jet produced in the decay of the top partner. The method described below can be applied to the dark photon produced in the decay of top partner in a wide range of its mass. Our main purpose is to identify a dark photon initiated jet in an event using hybrid deep neural network. 

\begin{figure}[htbp]
  \centering
  \resizebox{\columnwidth}{!}{
  \subfloat[\label{subfig:jetmass_dist_mgd_10}]{\includegraphics[width=0.5\columnwidth]{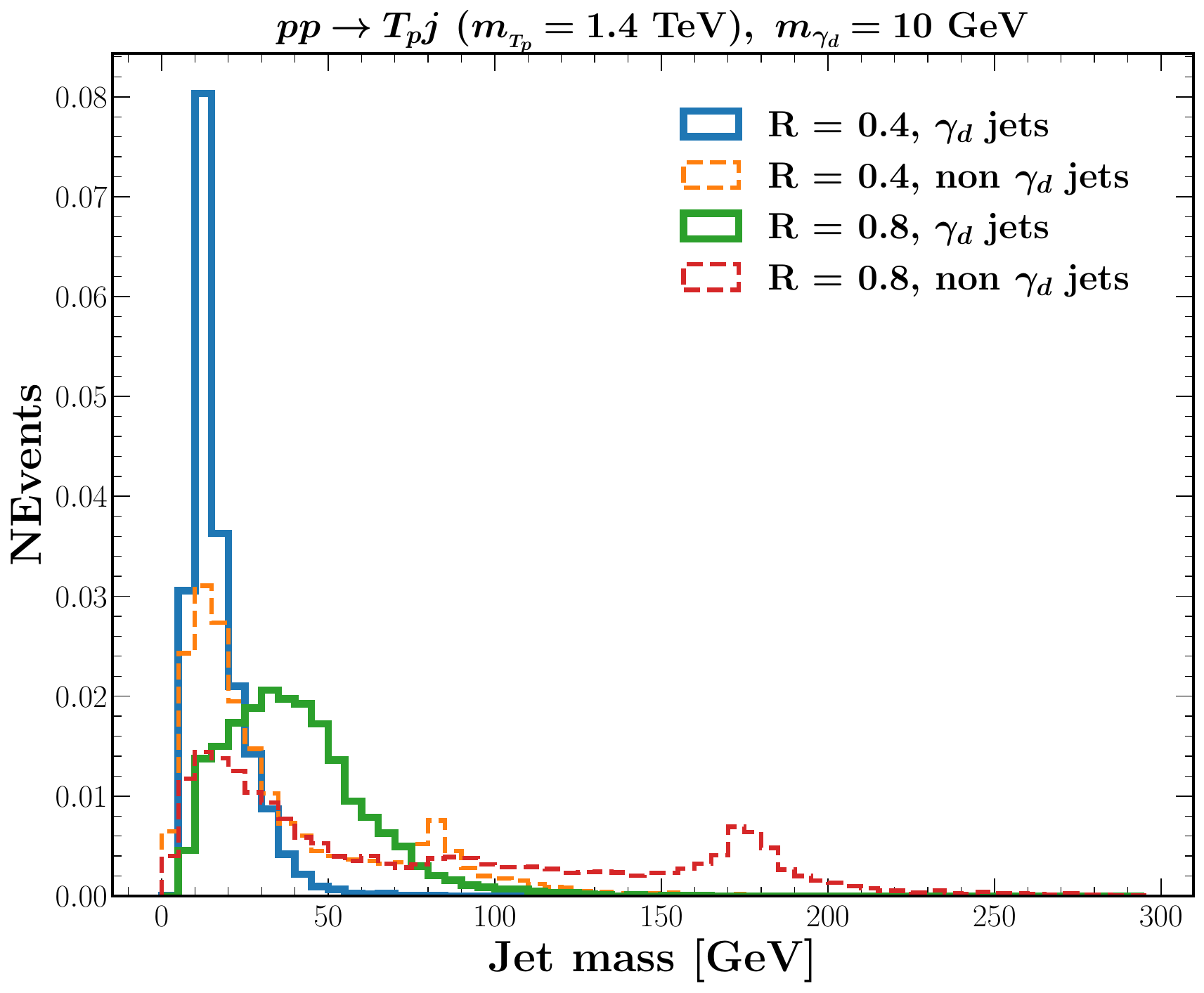}}~
  \subfloat[\label{subfig:jetmass_dist_mgd_50}]{\includegraphics[width=0.5\columnwidth]{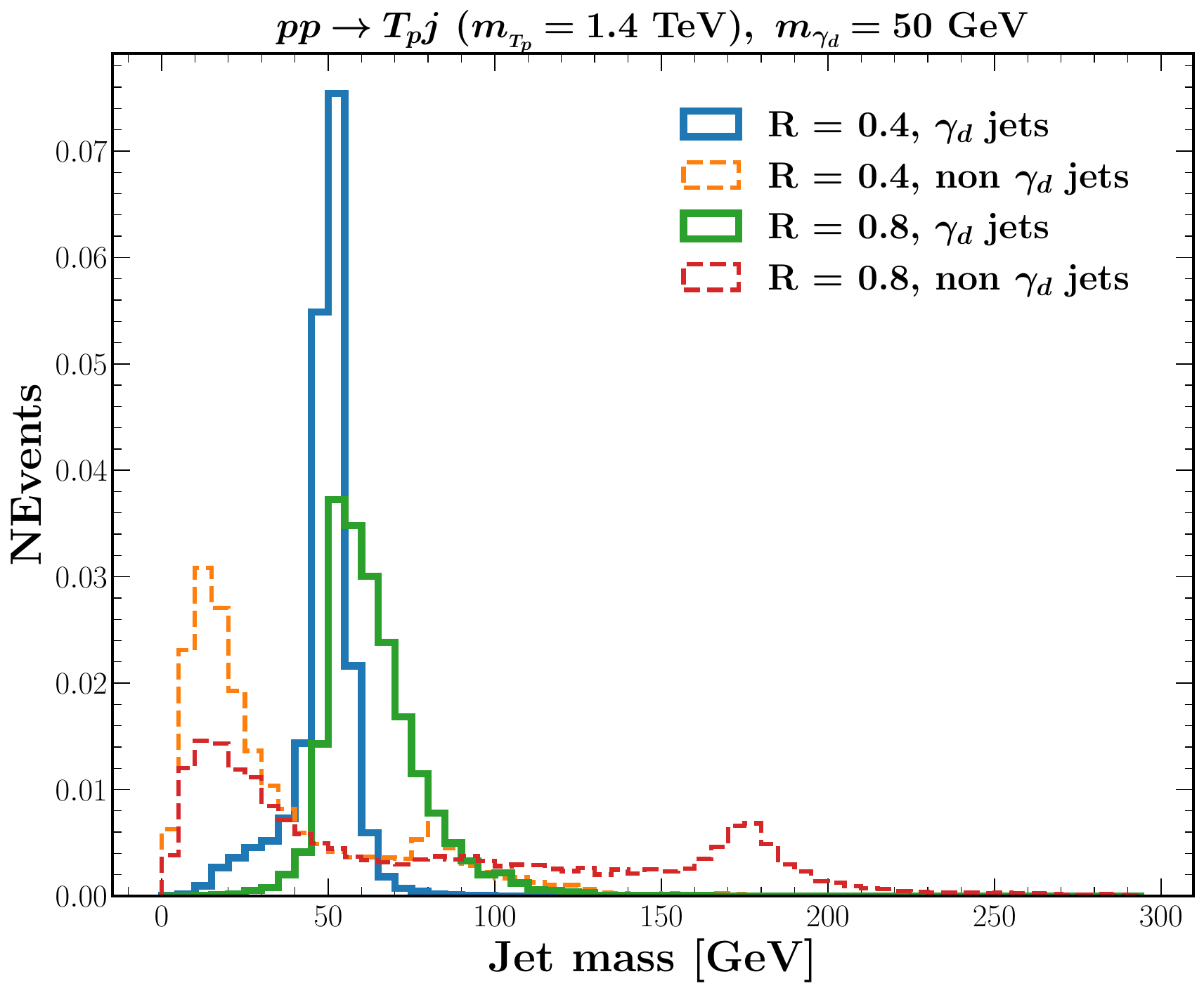}}}
  \caption{{ Jet mass distribution for two clustering radii, 0.4 and 0.8, comparing dark photon jets and {nondark} photon jets. The top partner mass is fixed at 1.4 TeV, while the dark photon mass is varied as follows: (a) 10 GeV, (b) 20 GeV}}
  \label{fig:clus_choice}
\end{figure}

To identify a dark photon initiated jet in an event consisting of hadrons, we cluster the stable hadrons (including photons coming from $\pi^{0}$) using anti-$k_T$ clustering algorithm with a clustering radius ($R$) of 0.4 to form jets. {This choice of clustering radius results in a sharper jet invariant mass peak for the jet associated with the parton-level dark photon\footnote{{For the parton level true dark photon jet tagging, we ensure that the distance in the $\Delta \eta - \Delta \phi$ plane between the jet and the parton-level dark photon is minimum and within 0.02.}} as can be seen in Fig.~\ref{fig:clus_choice}. The jet momentum are reconstructed out of its constituents using $p_{_T}$-recombination scheme. However, one can also use other recombination schemes such as ($E$-recombination) as the method employed below is not sensitive to any particular recombination scheme. }

\subsection{Inputs}\label{subapp:inputs}

The inputs of our hybrid deep neural network, are certain one and two-dimensional
features of the potential dark photon jet candidates having $p_{_T}>200$ GeV. 
we use both the constituent level and jet level information to capture the intricate features of both, the dark photon jets and the {nondark} photon jets in a binary classification framework \cite{CMS:2022prd, Hammad:2022lzo, Ban:2023jfo}.

The constituent level information, represented by jet images, is processed using a convolutional neural network. The CNN is adept at recognizing complex structures, such as localized $p_{_T}$ deposits or distinctive jet shapes. Whereas the jet-level information is processed using a multilayered perceptron. These jet level information include key features like transverse momentum of the jets, fraction of transverse momentum in a smaller cone to that of the whole jet, charge multiplicity, and charge multiplicity fraction in a smaller cone to that of the whole jet. These features offer a summary of the jet’s kinematics and characteristics of the particle content.

\subsection{Jet image preprocessing}\label{subapp:jet_image_preprocessing}

Jet image preprocessing is a crucial step to enhance the interpretability and comparability of the 
underlying structures within jets as well as making it easier for the HDNN model to converge faster as well as for the loss function to get closer to the global minima. To standardize the jet images in $\eta-\phi$ plane we process each jet image through three stages,
namely,

\begin{enumerate}
  \item {\it Centering constituents}: This process begins by centering the jet constituents along the center of the jet maintaining the range of $\phi$ in $[-\pi,\pi]$. This is achieved by transforming the $\eta$ and $\phi$ coordinates into relative distances $\Delta \eta$ and $\Delta \phi$ from the jet's center. This centering allows for a more consistent and clearer visualization of the decay structure within the jet. 
  \item  {\it Rotating the image}: Once centered, the jet image is rotated along its principal axes to align the primary structure in a 
  standardized orientation.
  \item {\it Flipping the image}: The image is flipped to ensure that any additional radiations appear in the first quadrant, 
  facilitating easier comparisons between different jets. 
  \item {\it Pixelating the image}: Finally, the image is discretized the region into a 16 $\times$ 16 grid with
  each pixel weighted by the sum of the transverse momentum in it.
\end{enumerate} 

This preprocessing sequence results in more uniform and comparable jet images, which are useful in helping the HDNN model to learn a standardized and coherent set of data. Moreover, the average jet images as shown in Fig.~\ref{fig:training_ptscheme} give a visual representation of the preprocessed images for the dark photon and {nondark} photon jets in the $\Delta\eta-\Delta\phi$ plane.

\subsection{Constituent level input}\label{subapp:jet_image}
\begin{figure}[htbp]
  \centering
  \resizebox{\columnwidth}{!}{
  \subfloat[\label{subfig:jetimage_larger_sig_pt}]{\includegraphics[width=0.47\columnwidth]{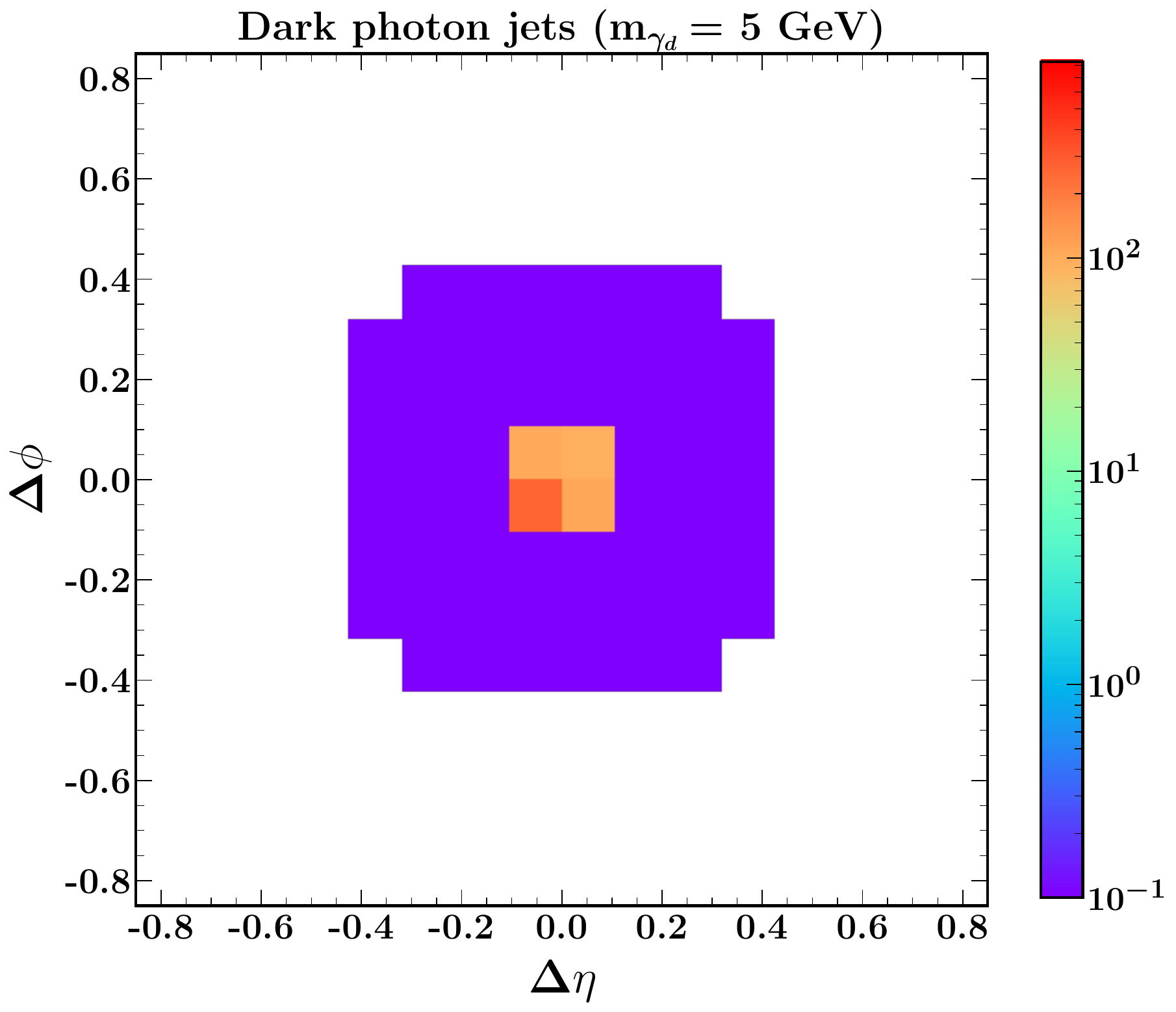}}~
  \subfloat[\label{subfig:jetimage_larger_bkg_pt}]{\includegraphics[width=0.5\columnwidth]{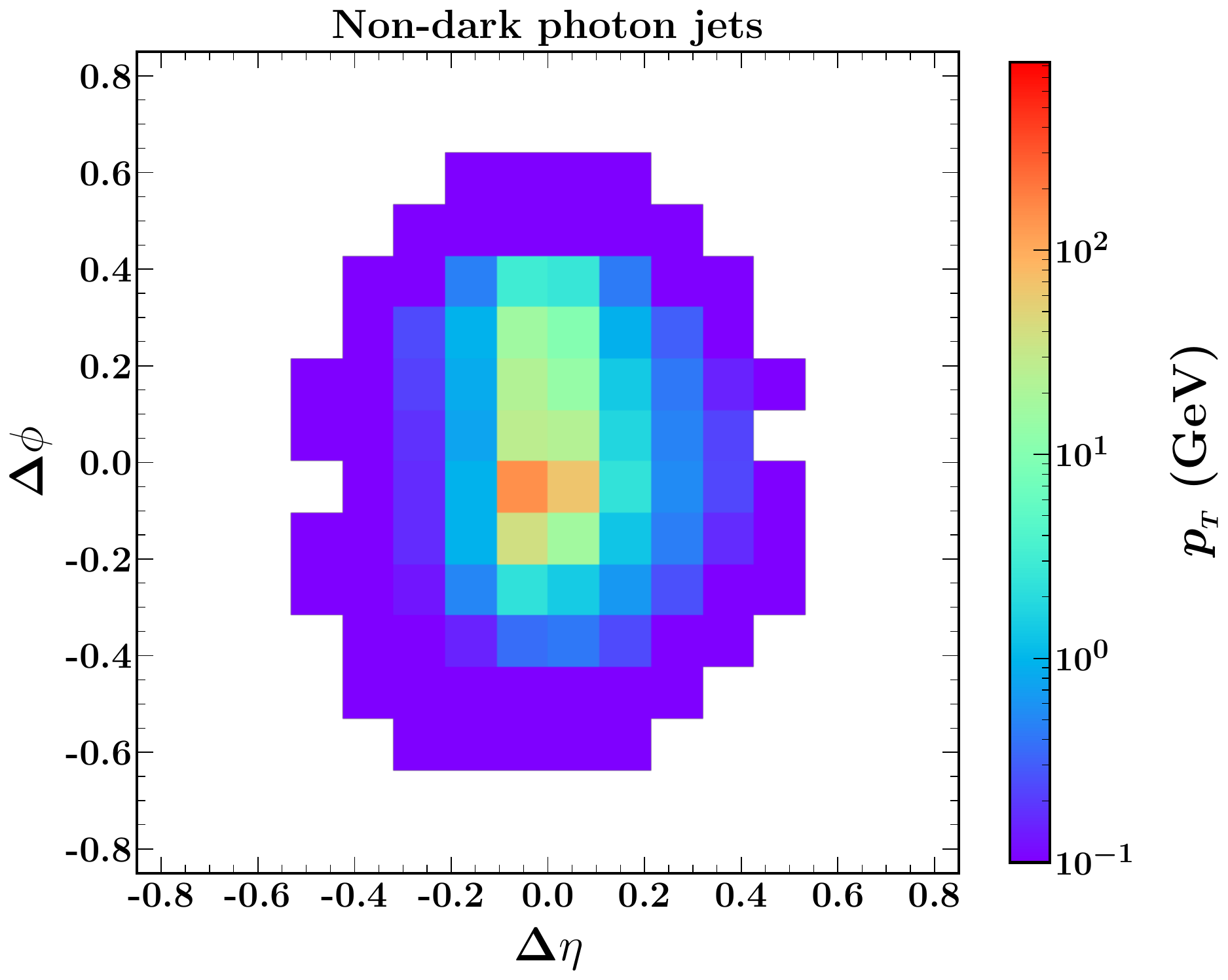}}}

  \resizebox{\columnwidth}{!}{
  \subfloat[\label{subfig:jetimage_smaller_sig_pt}]{\includegraphics[width=0.47\columnwidth]{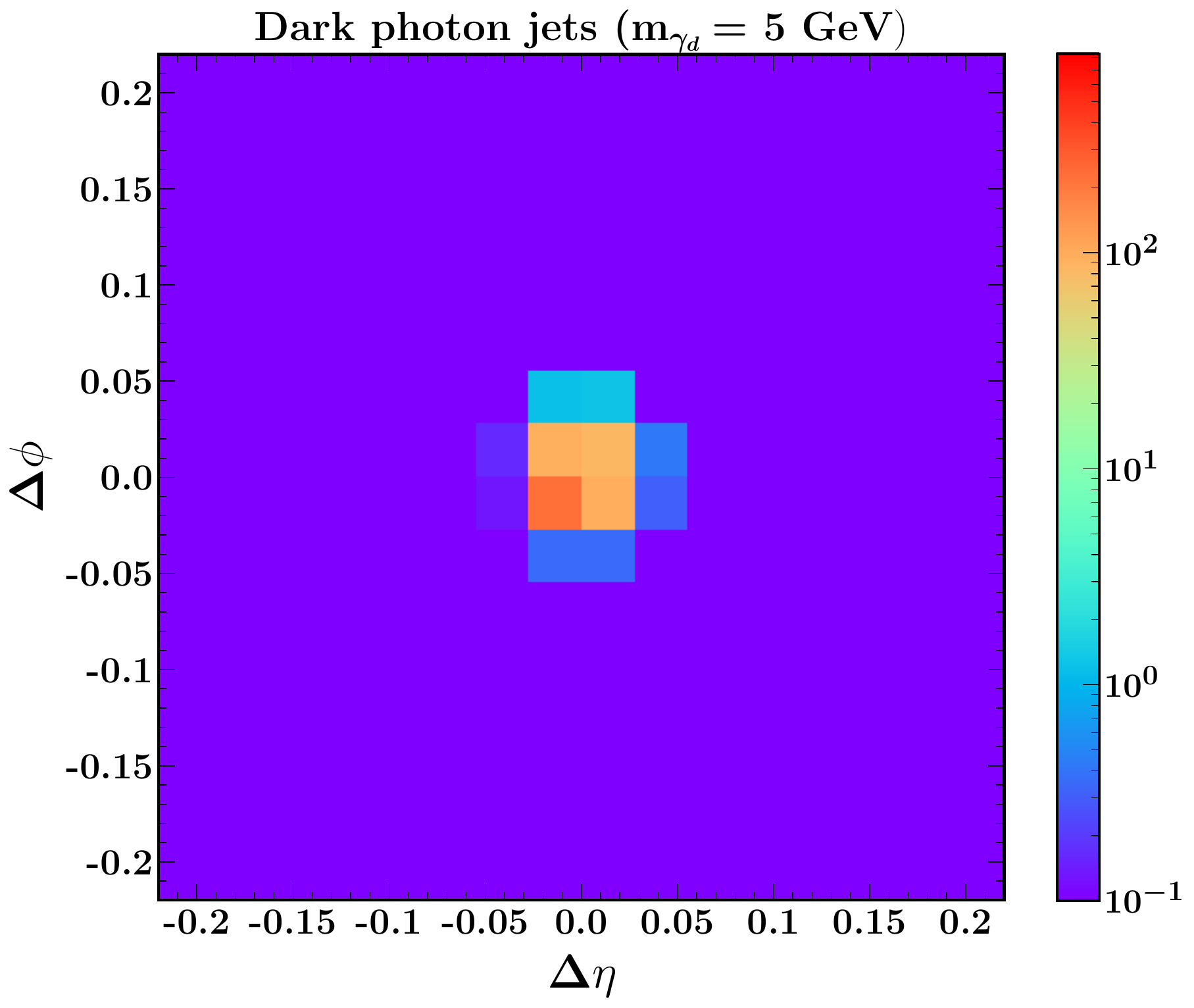}}~
  \subfloat[\label{subfig:jetimage_smaller_bkg_pt}]{\includegraphics[width=0.5\columnwidth]{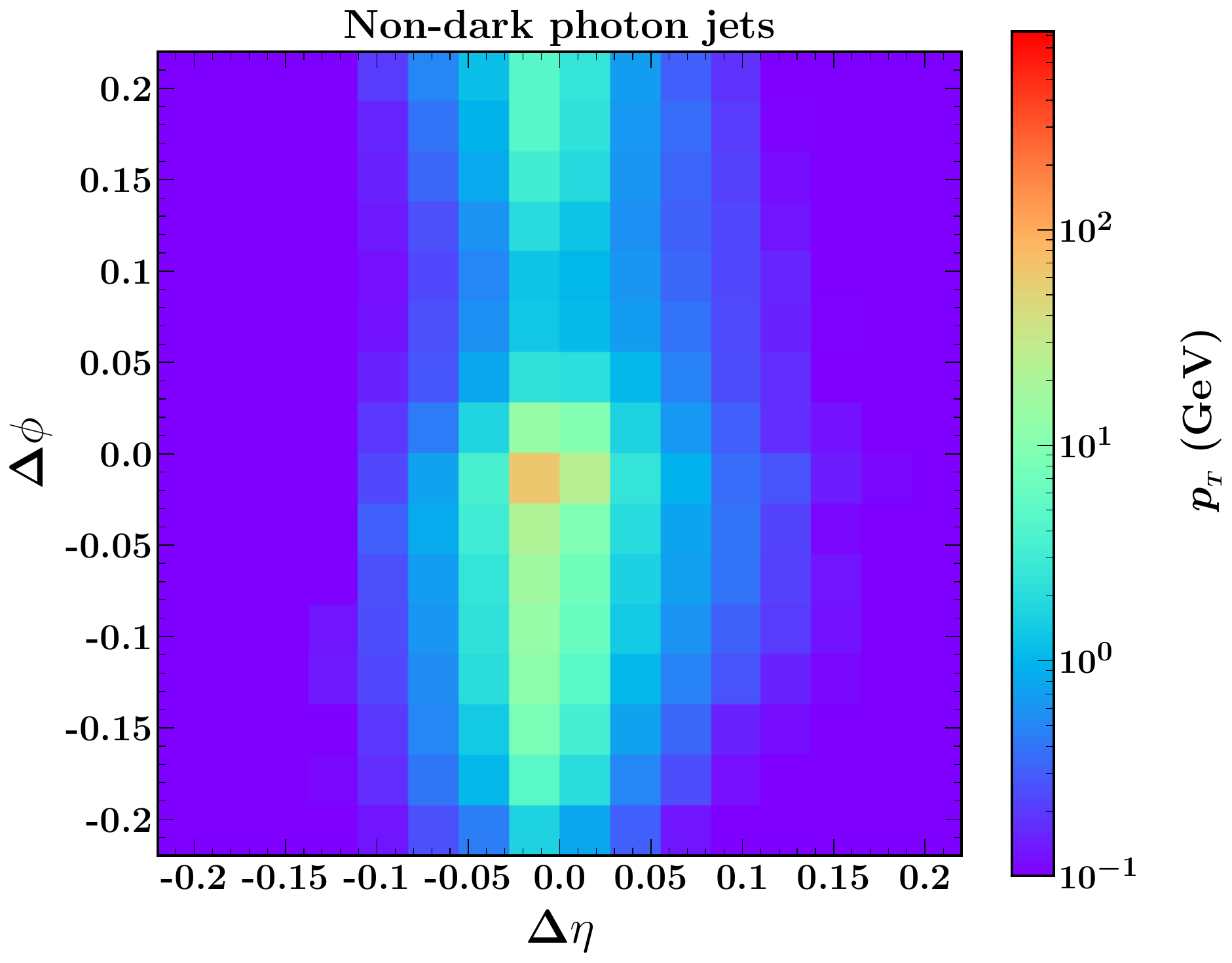}}}
  \caption{{ Average jet images for (a) true dark photon jets ($m_{\gamma_d}=5$ GeV) and (b) {nondark} photon jets in the $\Delta\eta-\Delta\phi$ plane in the range $-0.8$ to $0.8$ with granularity of $0.1 \times 0.1$ assuming $m_{_{T_p}} = 1.4$ TeV.
  (c) and (d) represents the same as in (a) and (b) respectively, but in the $\Delta\eta-\Delta\phi$ range of $-0.2$ to $0.2$ with granularity of $0.025 \times 0.025$. Both the images above are  obtained by using AK4 jets with $ p_{_T}$-scheme recombination.}}
  \label{fig:training_ptscheme}
\end{figure}

The figures in this section display average jet images (averaged over 10,000 individual jet images) that give a visual representation of the $ p_{_T}$ deposition of jet constituents in the $\Delta\eta-\Delta\phi$ plane for both dark photon jets and {nondark} photon jets. The images are prepared with two different granularities, both having $16\times 16$ pixels size.

In Fig.~\ref{subfig:jetimage_larger_sig_pt} and \ref{subfig:jetimage_larger_bkg_pt} we depict the 
jet images with a granularity of $0.1 \times 0.1$ covering a $\Delta\eta-\Delta\phi$ range from $-0.8$ to $0.8$ for the dark photon jets and {nondark} photon jets, respectively. These images include contributions from all the charged, neutral hadrons and photons, capturing the full hadronic content and providing a comprehensive representation of the jet's internal structure. Such an image captures the correlation of transverse momentum of the jet constituents with the {pseudorapidity} and azimuthal angle making it a key component in differentiating between the dark photon jets and {nondark} photon jets.

On the other hand, in Fig.~\ref{subfig:jetimage_smaller_sig_pt} and \ref{subfig:jetimage_smaller_bkg_pt}, we present jet images with a finer granularity of $0.025 \times 0.025$ over a smaller $\Delta\eta-\Delta\phi$ range from $-0.2$ to $0.2$. These images include only the contributions from charged hadrons and photons. Such a finer resolution is possible if one couples the information of the ECAL energy deposition of the photons and HCAL energy deposition of the charged hadrons along with the tracker information associated with the charged hadrons. This higher resolution allows for a more detailed examination of the jet core, where most of the transverse momentum is concentrated in the actual dark photon initiated jet. These two layers will act as two channels of the image input to the CNN part of HDNN architecture.

\subsection{Jet level input}\label{subapp:jet_feats}

\begin{figure}[htbp]
  \centering
  \resizebox{\columnwidth}{!}{
  \subfloat[\label{subfig:jet_feat_pt}]{\includegraphics[width=0.5\columnwidth]{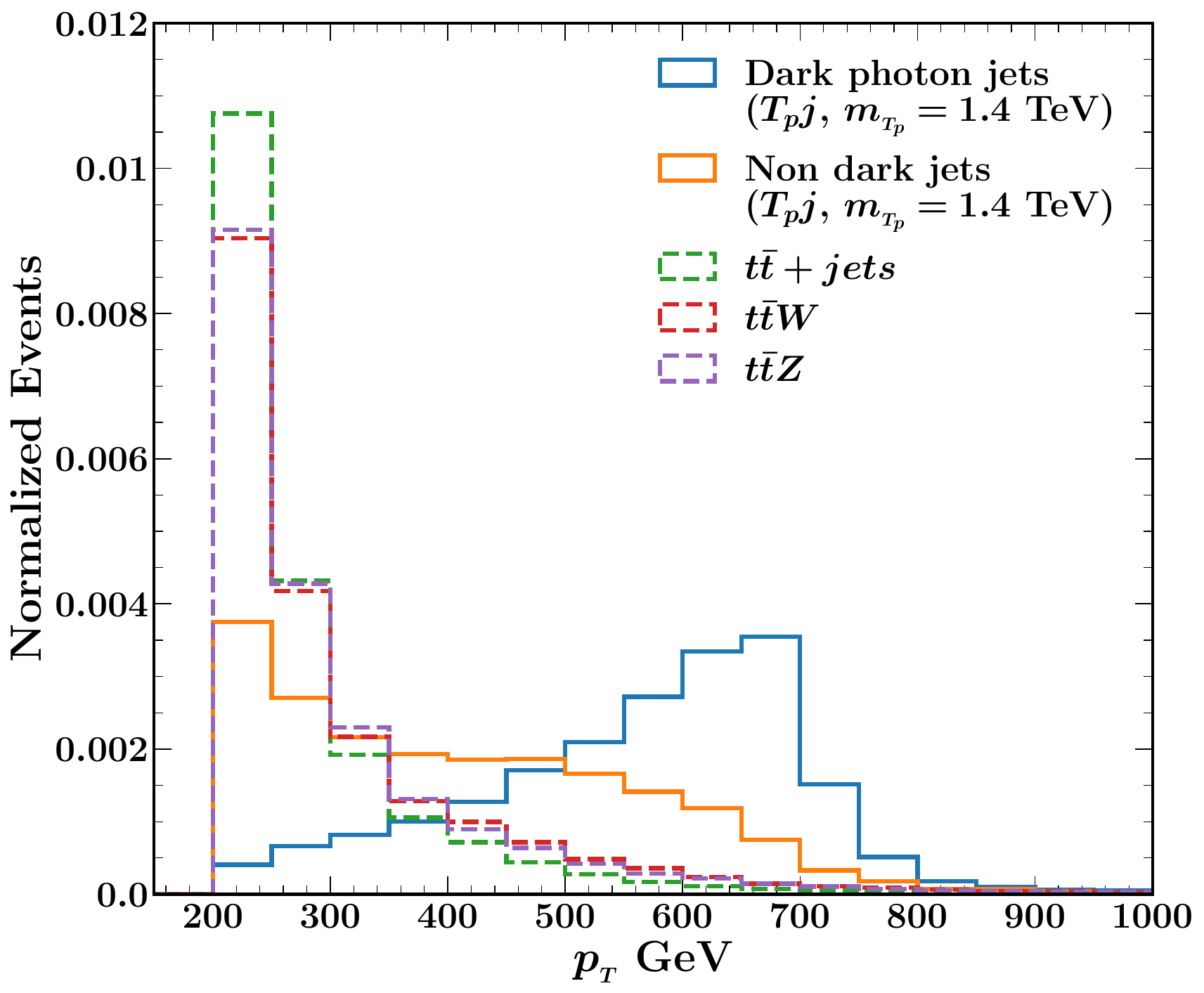}}~
  \subfloat[\label{subfig:jet_feat_pt_ratio}]{\includegraphics[width=0.5\columnwidth]{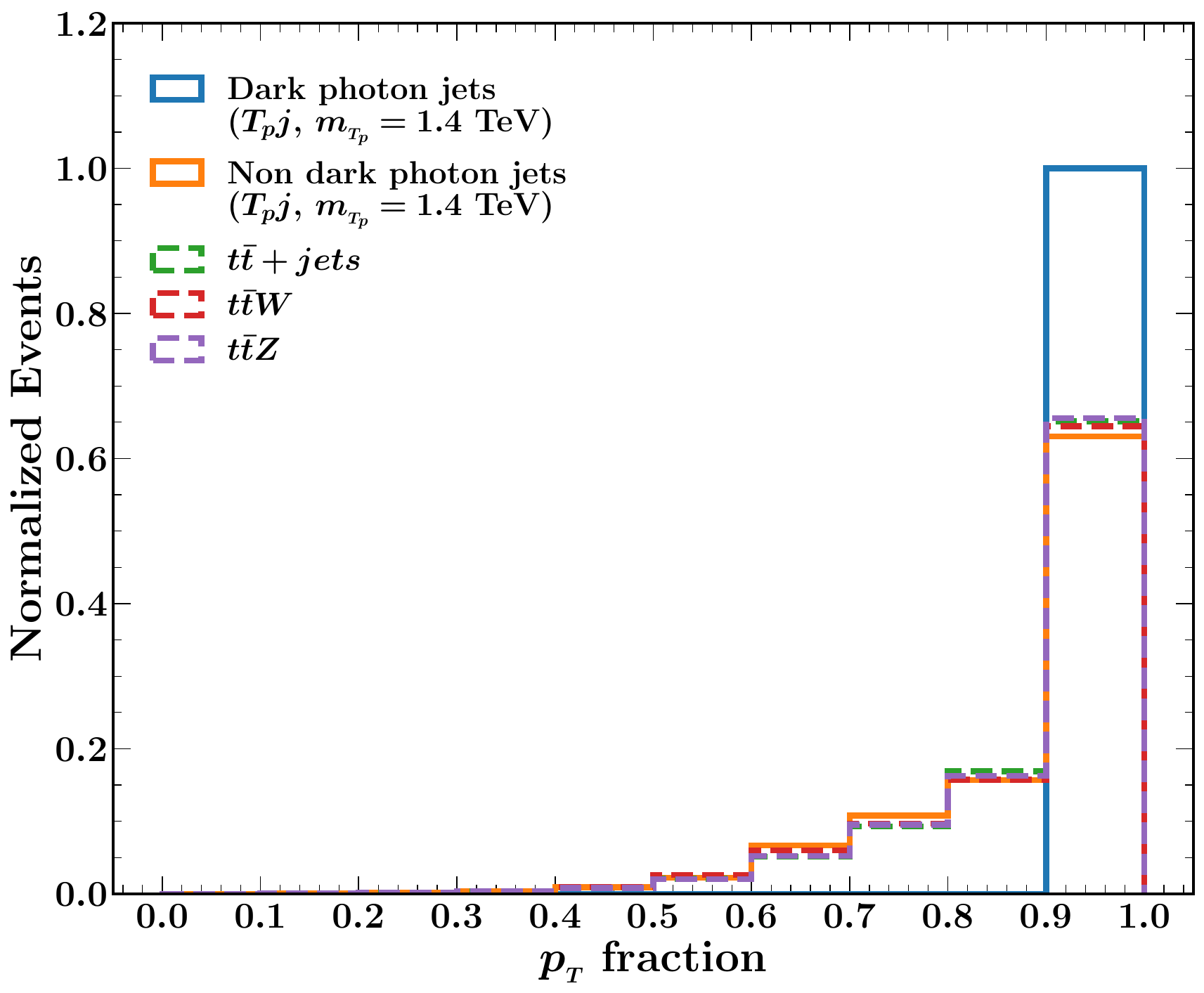}}
  }

  \resizebox{\columnwidth}{!}{
  \subfloat[\label{subfig:jet_feat_cm}]{\includegraphics[width=0.5\columnwidth]{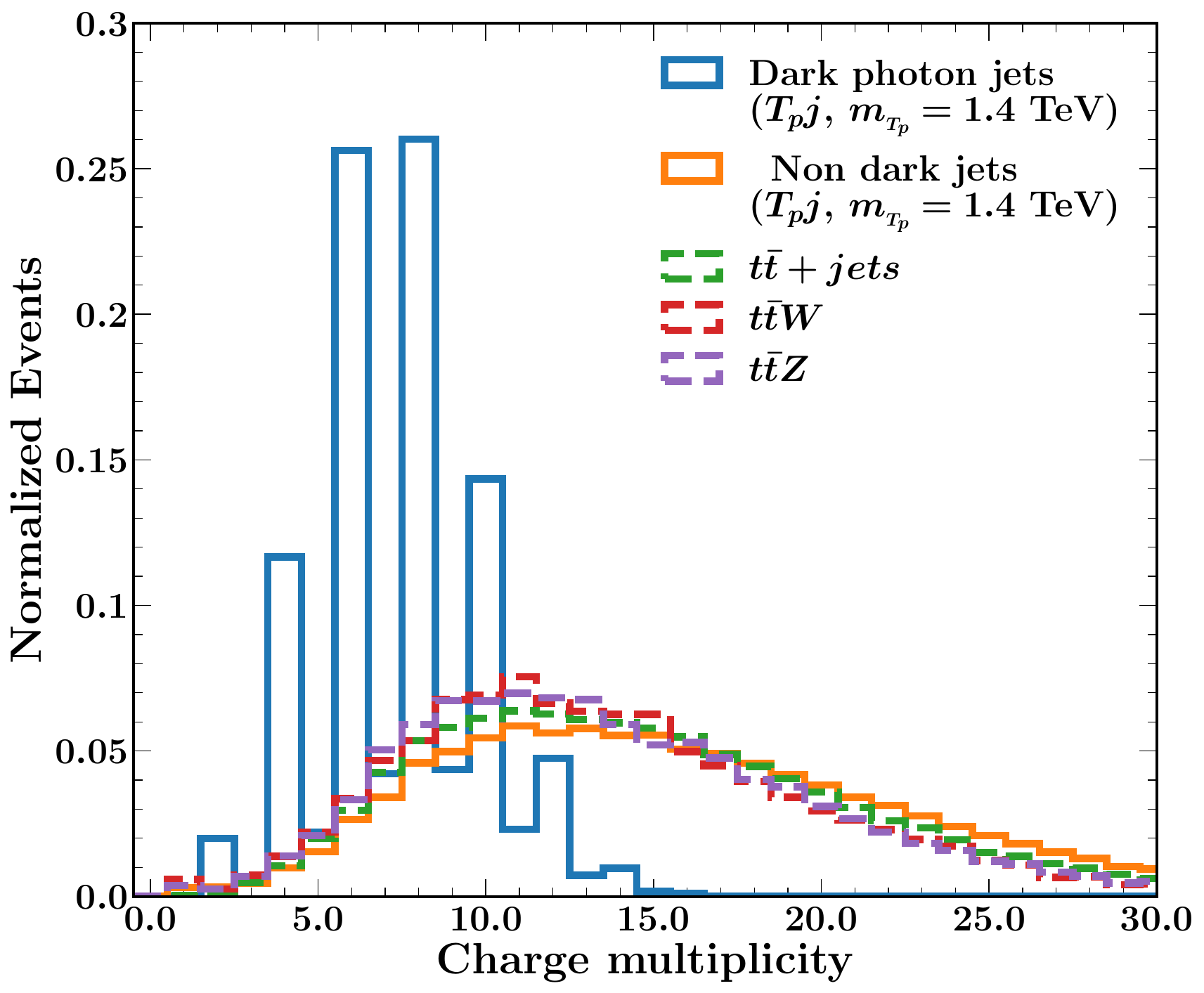}}~
  \subfloat[\label{subfig:jet_feat_cm_ratio}]{\includegraphics[width=0.5\columnwidth]{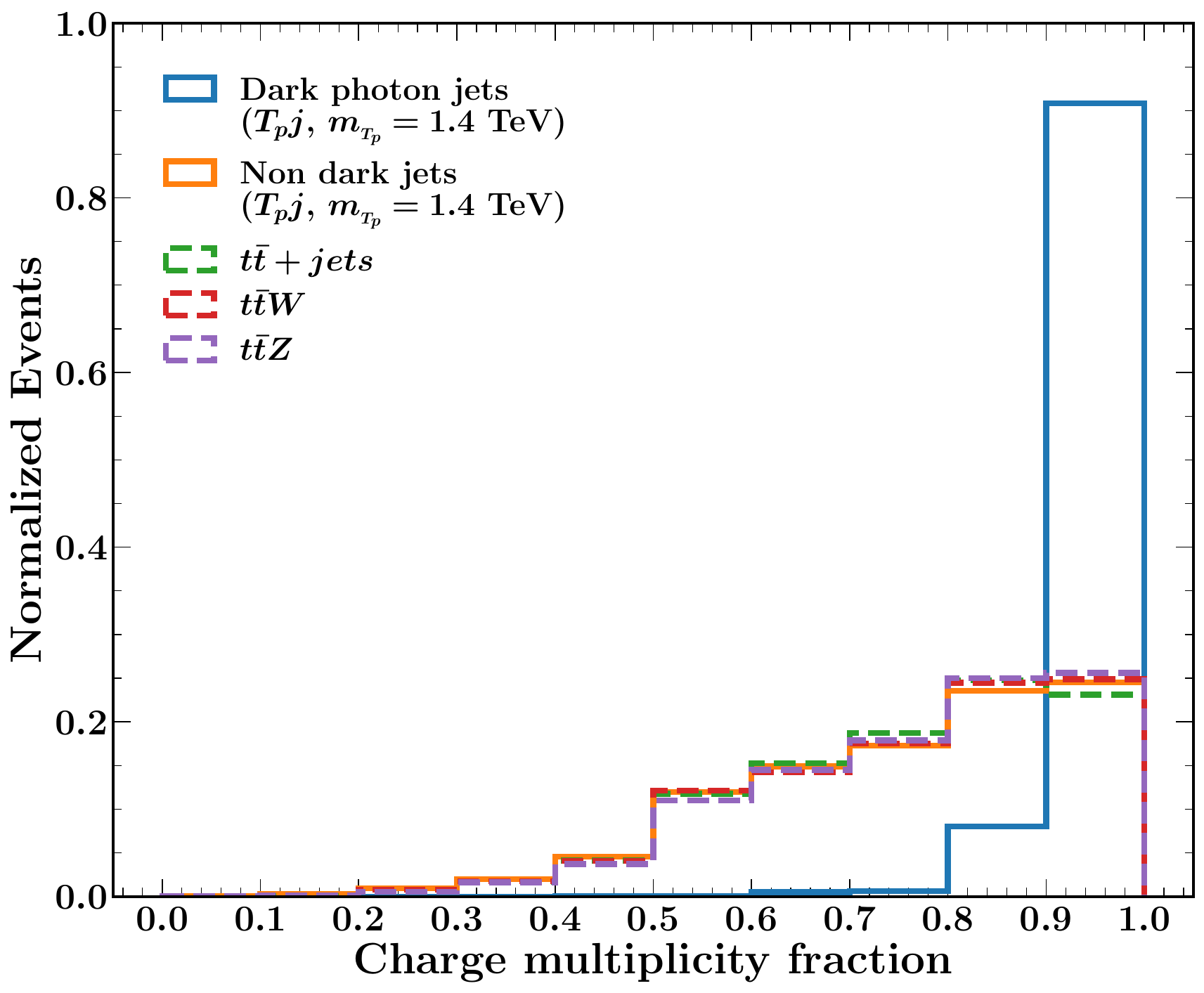}}
  }
  \caption{{ (a) $p_{_T}$ distribution, (b) ratio of total $p_{_T}$ within a cone of radius 0.2 around the jet center to that within the entire AK4 jet, (c) charge multiplicity distribution, and (d) ratio of the number of charged hadrons within a cone of radius 0.2 around the jet center to that within the entire AK4 jet for the parton level tagged dark photon jet and {nondark} photon jets coming from the signal ($T_pj$) itself and other SM background processes.}}
  \label{fig:jetfeatures_distribution}
\end{figure}

The jet level features shown in Fig.~\ref{fig:jetfeatures_distribution} play an important role in differentiating a dark photon jet from jets initiated by particles other than the dark photon. 
Fig.~\ref{subfig:jet_feat_pt} shows the transverse momentum ($p_{_T}$) distribution, 
Fig.~\ref{subfig:jet_feat_pt_ratio} examines the $p_{_T}$ fraction, {\it i.e.} the ratio of total $p_{_T}$ within a cone of radius 0.2 around the jet center to that within the entire AK4 jet for the parton level tagged dark photon jet and {nondark} photon jets coming from the signal ($T_pj$) itself and other SM background processes. The later variable highlights the amassing of transverse momentum for the actual dark photon jet within the jet core and is particularly significant in capturing the more highly collimated $p_{_T}$ depositions.
Fig.~\ref{subfig:jet_feat_cm} focuses on charge multiplicity, {\it i.e.}, the number of charged hadrons within the potential dark photon jet candidate.
Finally, the charge multiplicity fraction, {\it i.e.}, the ratio of the number of charged hadrons within a cone of radius 0.2 around the jet center to that within the entire AK4 jet is shown in Fig.~\ref{subfig:jet_feat_cm_ratio}. 
All these features enhance the ability to differentiate a dark photon jet from other {nondark} photon jets.

\subsection{Network architecture}\label{subapp:DNN_arch}

The model is built to effectively process both jet images and jet-level features, making it well-suited for tasks like jet classification or tagging. The jet images are structured as $16 \times 16 \times 2$ arrays, where the individual $16 \times 16$ array represents each channel as mentioned earlier. The image input undergoes several convolutional layers, starting with a 2D convolutional layer with 32 filters and a $3 \times 3$ kernel. We also use additional padding to preserve the original image size allowing the network to preserve the spatial information throughout the layers.
After this, the dimension of the array becomes $16 \times 16 \times 32$. 
The convolution operation is followed by a parametric rectified linear unit (PReLU) activation which introduces {nonlinearity}, 
and helps the model capture complex patterns in the data. 
Next, another 2D convolutional layer with 64 filters is applied. The array dimension now changes $16 \times 16 \times 64$ at this stage. 
A third 2D convolutional layer with 64 filters is applied before the image is flattened into a 1D vector of shape $16384$, converting the spatial features into a vector that can be combined with the jet-level features.
Simultaneously, the jet-level features, representing kinematic variables like jet $p_{_T}$, charge multiplicity, $p_{_T}$ and charge multiplicity fractions, are processed through a multilayered perceptron with 64 neurons and PReLU activation. This transformation results in a latent space vector with $64$ features, that complements the resultant vector resulting from flattened output from the convolutional neural network part of the HDNN. These two vectors, {\it i.e.}, one from the flattened jet image and the 
other from the jet-level features, are concatenated into a combined feature vector.
At this point, the model performs dimensionality reduction through a series of fully connected layers. The first dense layer, with 64 neurons and PReLU activation. Another dense layer with 32 neurons follows, further compressing the feature vector. Finally, a dense layer with 16 neurons reduces the feature space further.
The final output is generated through a sigmoid activation on a single neuron, providing a probability score between 0 and 1 for binary classification. 

This model is particularly effective because it leverages both the spatial information from the jet images and the {nonspatial}, one dimensional 
jet-level features. 
By applying convolutional layers to the jet images, the model preserves the structure of the transverse momentum distribution across the jet, while the dense layers allow for kinematic variables of the jet. 

\subsection{Training and validation}\label{subapp:model_training}

The model is trained on a dataset of 240,000 samples, which is split into training and validation sets, with 80\% of the data (192,000 samples) used for training and the remaining 20\% (48,000 samples) used for validation.
{The dataset is generated using a benchmark point $m_{_{T_p}} = 1.4$ TeV, $v_d = 200$ GeV, $m_{h_d} = 400$ GeV, and $m_{\gamma_d} = 5$ GeV, in conjunction with the standard model parameters specified in Eq.~\eqref{eq:sm_param_choice}. 
The training dataset includes both signal and background samples. The signal component consists of dark photon-initiated jets, as well as high-$p_{_T}$ ($p_{_T}>200$ GeV) {nondark} photon jets, both originating from the $T_p j$ process. In addition, background samples are included from $t\bar{t} + \text{jets}$, $t\bar{t}W$, and $t\bar{t}Z$ SM processes. Each dataset is constructed with an equal admixture of dark photon jets ({labeled} as 1) and {nondark} photon jets ({labeled} as 0), ensuring a balanced training sample.

Additionally, an independent test dataset comprising 120,000 samples is used to evaluate the model's performance. During training, the model is optimized over multiple epochs, each consisting of forward and backward pass.}

In each epoch, the training data is fed into the model in batches. During the forward pass, the input data is processed through various layers of the network to produce a binary output. The model's predictions are compared with the actual labels using a binary cross-entropy loss function \cite{Mao2023CrossEntropyLF}, defined as

\[
\text{Loss} = - \frac{1}{N} \sum_{i=1}^{N} \left[ y_i \log(p_i) + (1 - y_i) \log(1 - p_i) \right]
\]

Where \(y_i\) represents the true label, \(p_i\) is the predicted probability for the positive class, and \(N\) is the number of samples in the batch. This loss function penalizes the model for incorrect predictions, with the penalty increasing as the prediction diverges from the true label.

We use Adam optimizer \cite{kingma:2017} to update the model's weights based on the gradients of the loss function with respect to the weights, calculated during the backward pass. The gradients are computed using backpropagation, where the error from the loss function is propagated backward through the network, updating the weights to minimize the loss. During training, the model is evaluated on the validation set at the end of each epoch to monitor its performance. We use early stopping, which halts training if the validation loss stops improving for a set number of epochs and learning rate reduction when the validation loss plateaus to prevent overfitting and improve training efficiency. After training, the model's performance is evaluated on a separate test dataset, providing an unbiased estimate of its generalization ability. 

\subsection{Model performance}\label{subapp:model_performance}

\begin{figure}[htbp]
  \centering
  \resizebox{\columnwidth}{!}{
  \subfloat[\label{subfig:roc_curve}]{\includegraphics[width=0.5\columnwidth]{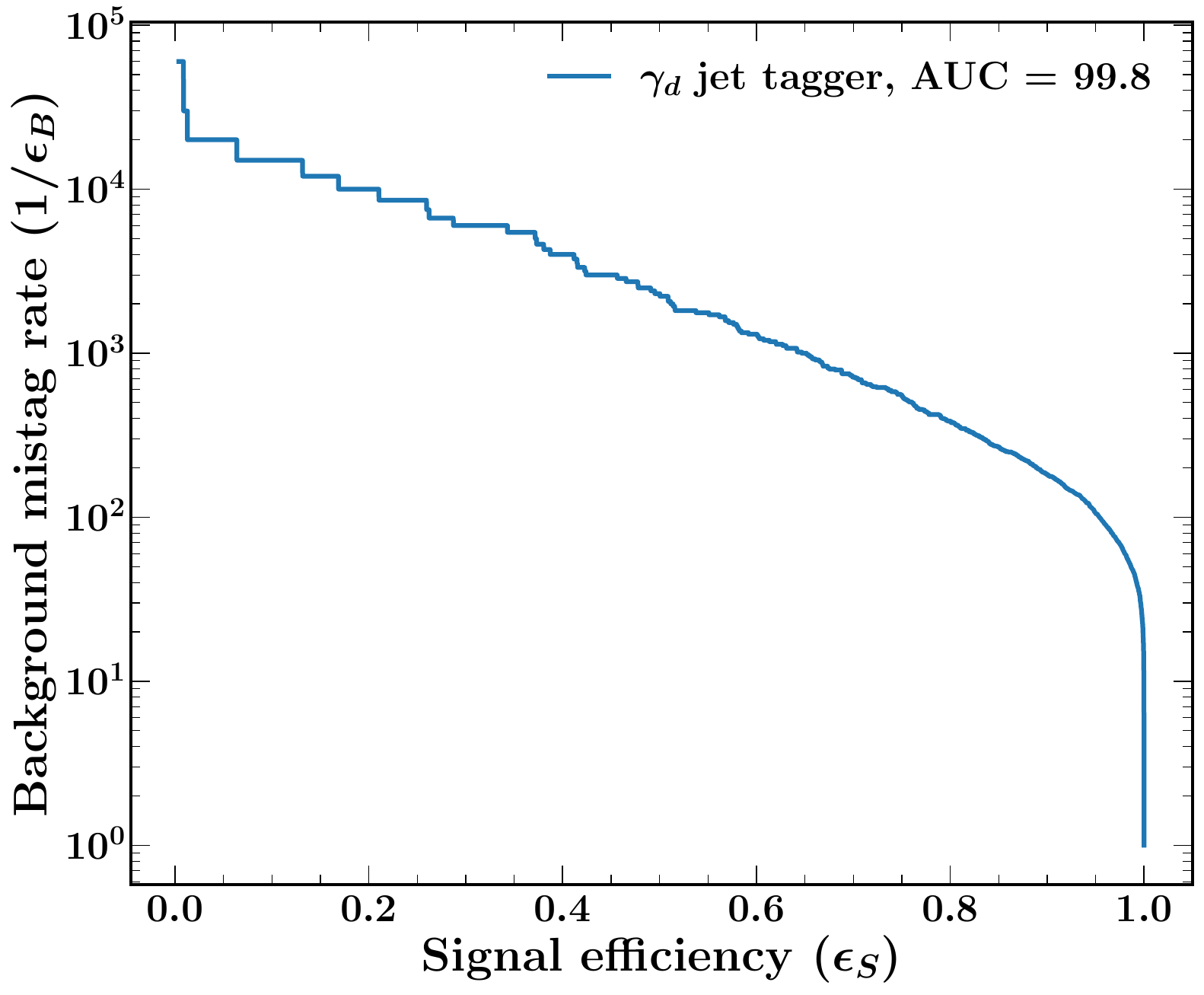}}~
  \subfloat[\label{subfig:score_dis}]{\includegraphics[width=0.5\columnwidth]{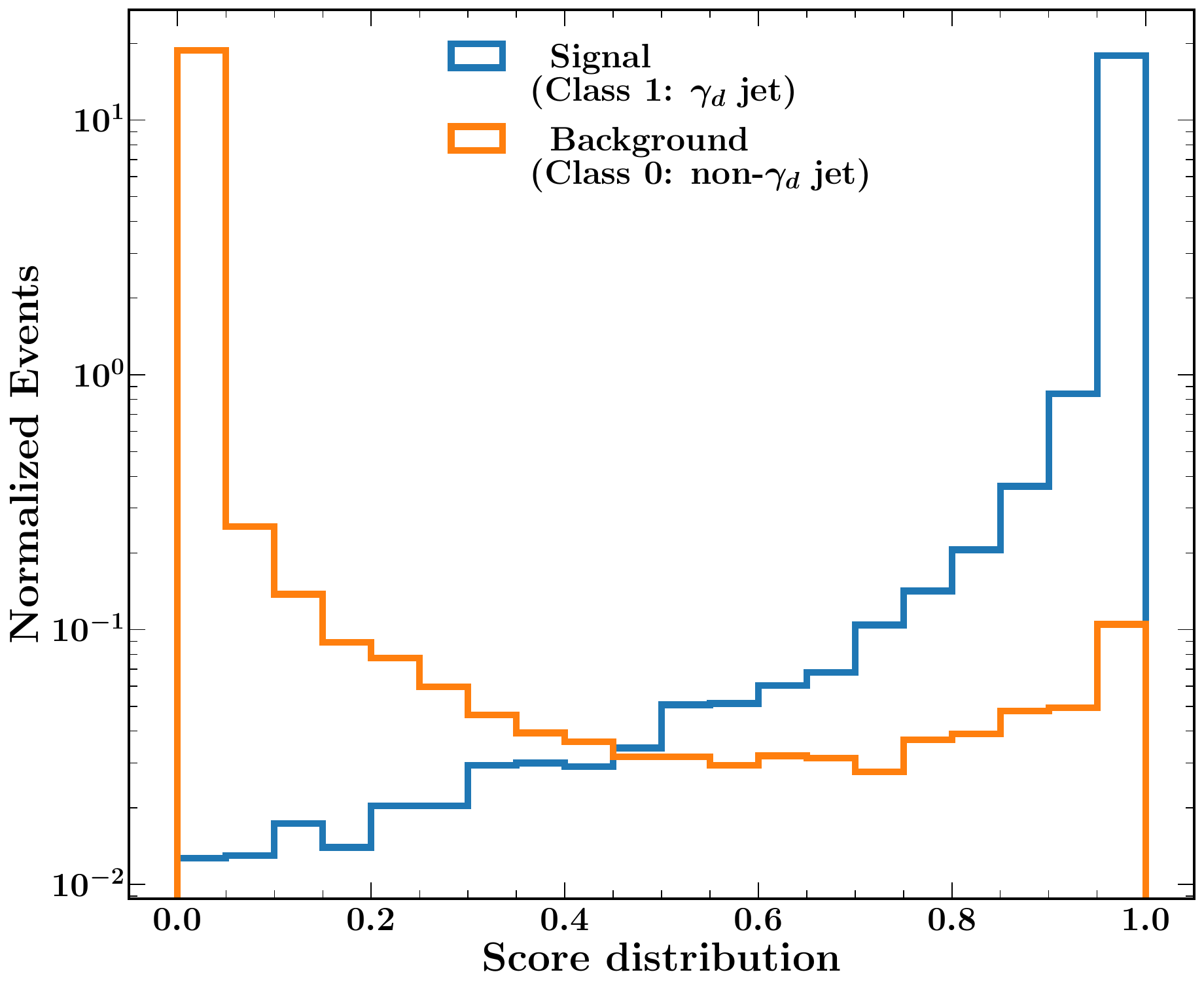}}
  }
  \caption{{ (a) ROC curve and (b) Score distribution for the test sample containing $60,000$ parton level tagged dark photon jets and $60,000$ {nondark} photon jets.}}
  \label{fig:model_performance}
\end{figure}

Our test sample contains $60,000$ parton level tagged $\gamma_d$ jets and $60,000$ non-$\gamma_d$ jets images and corresponding jet level feature that were described in {Appendix} \ref{subapp:jet_feats}. {Other essential parameters used for the test dataset is same as that used for training and validation described in \ref{subapp:model_training}.}
In Fig.~\ref{fig:model_performance}, we illustrate the performance of our binary classification model through two key visualizations: the ROC curve and the score distribution for both signal and background events.
The ROC curve in Fig.~\ref{subfig:roc_curve} provide a comprehensive view of the model's ability to distinguish between $\gamma_d$ and non-$\gamma_d$ jets, with the signal efficiency plotted against the background rejection rate across different decision thresholds. A curve that approaches the top-left corner signifies a better-performing model, as it indicates high signal efficiency and strong background suppression. Additionally, the score distribution in Fig.~\ref{subfig:score_dis} shows the clear separation achieved by the HDNN model to distinguish between a $\gamma_d$ jet and a non-$\gamma_d$ jet. {Based on the performance of the classifier on the test sample, we set the working point at a threshold of 0.4, achieving a true positive rate of $\sim 0.93$ and a false positive rate of $\sim 0.02$.}

\subsection{Efficiency}\label{subapp:pt_binned_eff}
\begin{figure}[htbp]
  \centering
  \resizebox{\columnwidth}{!}{
  \subfloat[\label{subfig:sig_tag}]{\includegraphics[width=0.5\columnwidth]{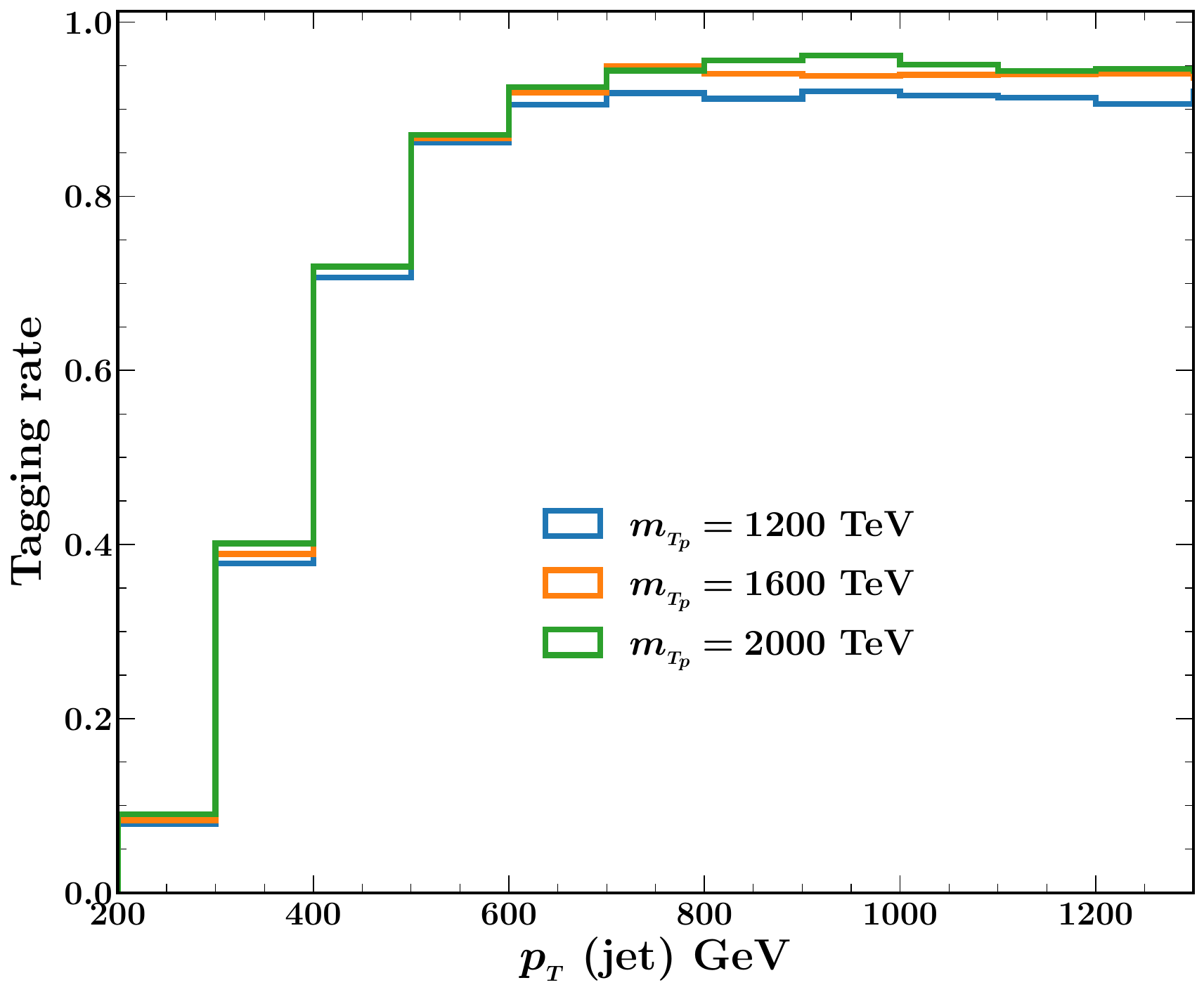}}~
  \subfloat[\label{subfig:bkg_mis-tag}]{\includegraphics[width=0.5\columnwidth]{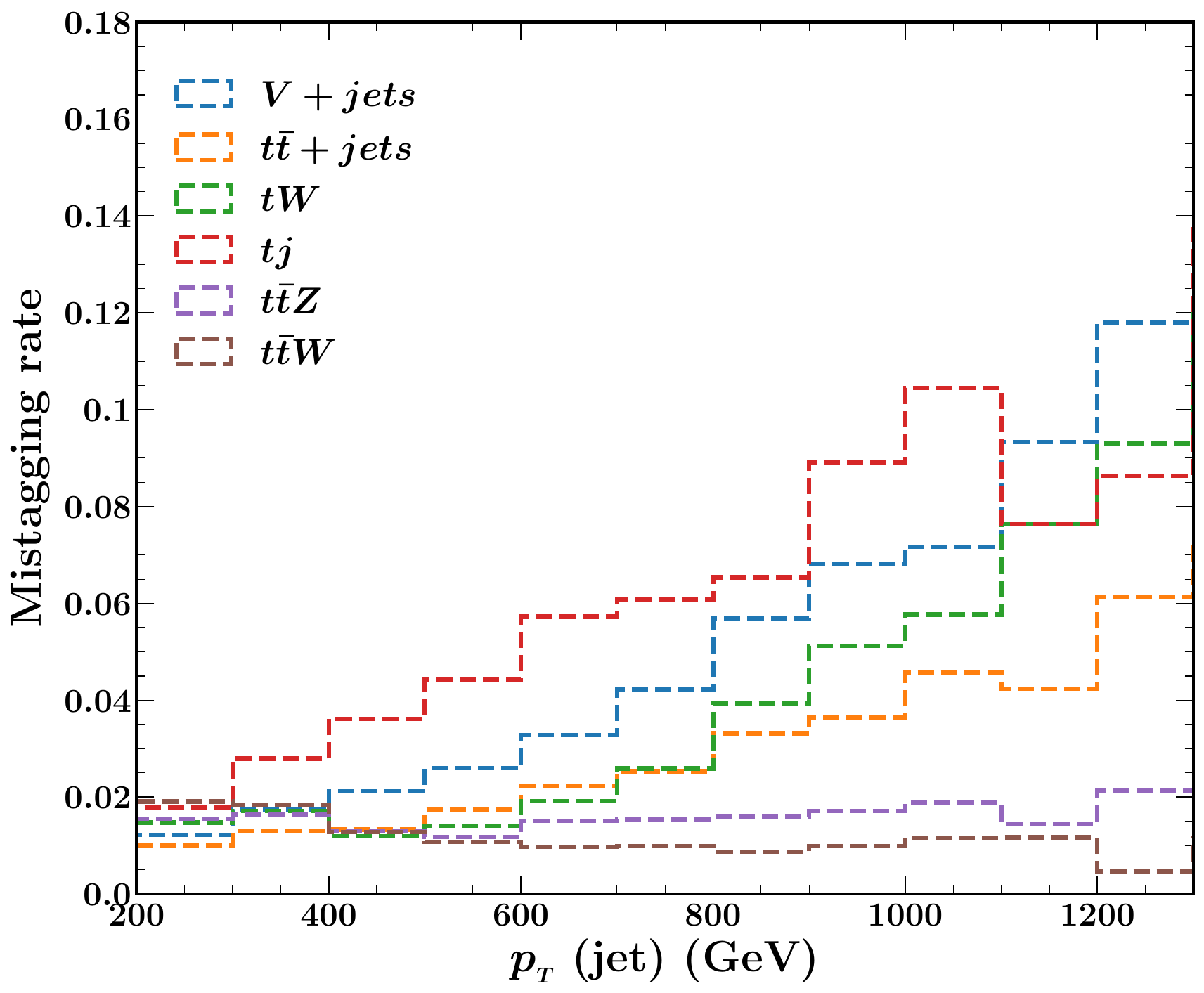}}
  }
  \caption{{ Tagging and {mistagging} rates in various transverse momentum bins for (a) parton level tagged dark photon coming from $T_pj$ process and (b) {nondark} photon jets coming various sources. }}
  \label{fig:eff_pt_binned}
\end{figure}

In Fig.~\ref{fig:eff_pt_binned}, we present the tagging and {mistagging} rate as a function of jet $p_{_T}$.
{We have been able to achieve an average tagging efficiency of $30\%$ to $93\%$ for the true dark photon jet in the top partner mass range, $0.8 - 2.8$ TeV and $m_{\gamma_d} = 10$ GeV for a HDNN classifier that is trained with a sample corresponding to  $m_{_{T_p}} = 1.4$ TeV, and $m_{\gamma_d} = 5$ GeV.} The average {mistagging} rate for the corresponding {nondark} photon jets present in the signal and in the SM background varies in the range $1.0\%$ to $2.6\%$. For the present work it is more than sufficient to distinguish a high $p_{_T}$ true dark photon initiated jet having some characteristic features described above significantly different from jets initiated by other particles. {In this work, we choose this setup to obtain the results quoted in Sec.~\ref{sec:collider_analysis} and Sec.~\ref{sec:results}.}

\begin{figure}[htbp]
  \centering
  \resizebox{\columnwidth}{!}{
  \subfloat[\label{subfig:jetimage_5_pt1_0pt8}]{\includegraphics[width=0.5\columnwidth]{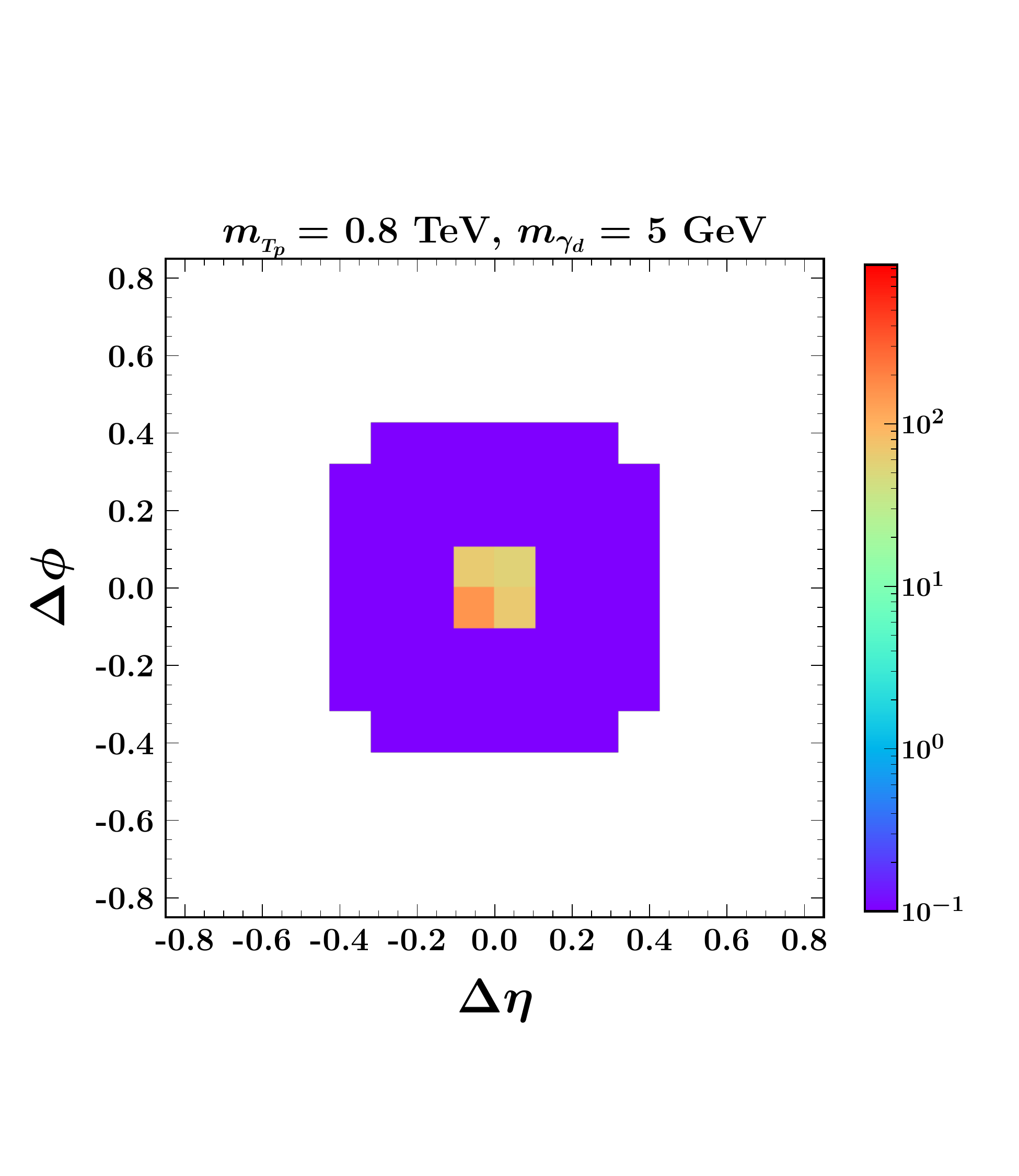}}~
  \subfloat[\label{subfig:jetimage_5_pt1_1pt4}]{\includegraphics[width=0.5\columnwidth]{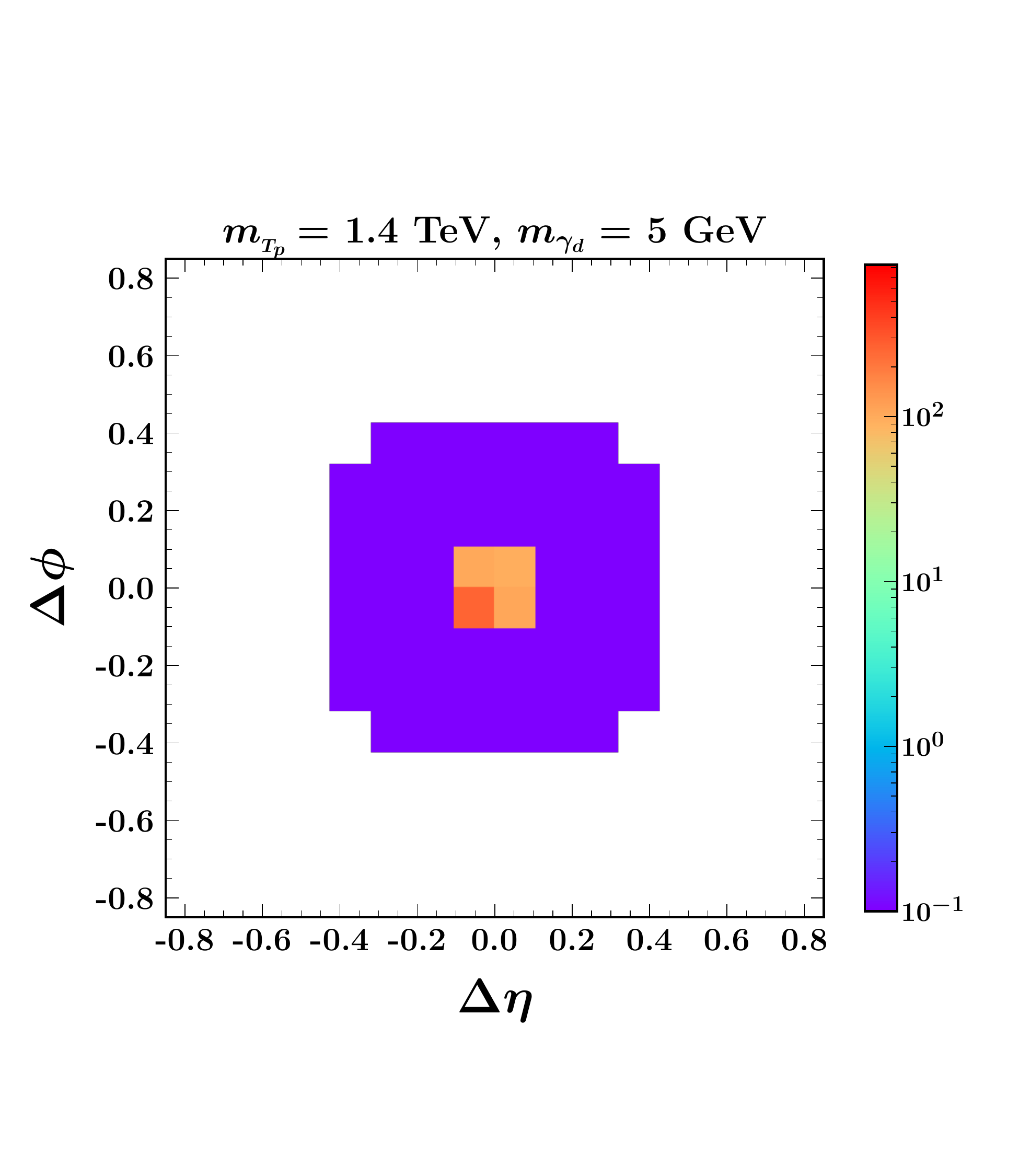}}~
  \subfloat[\label{subfig:jetimage_5_pt1_2pt6}]{\includegraphics[width=0.5\columnwidth]{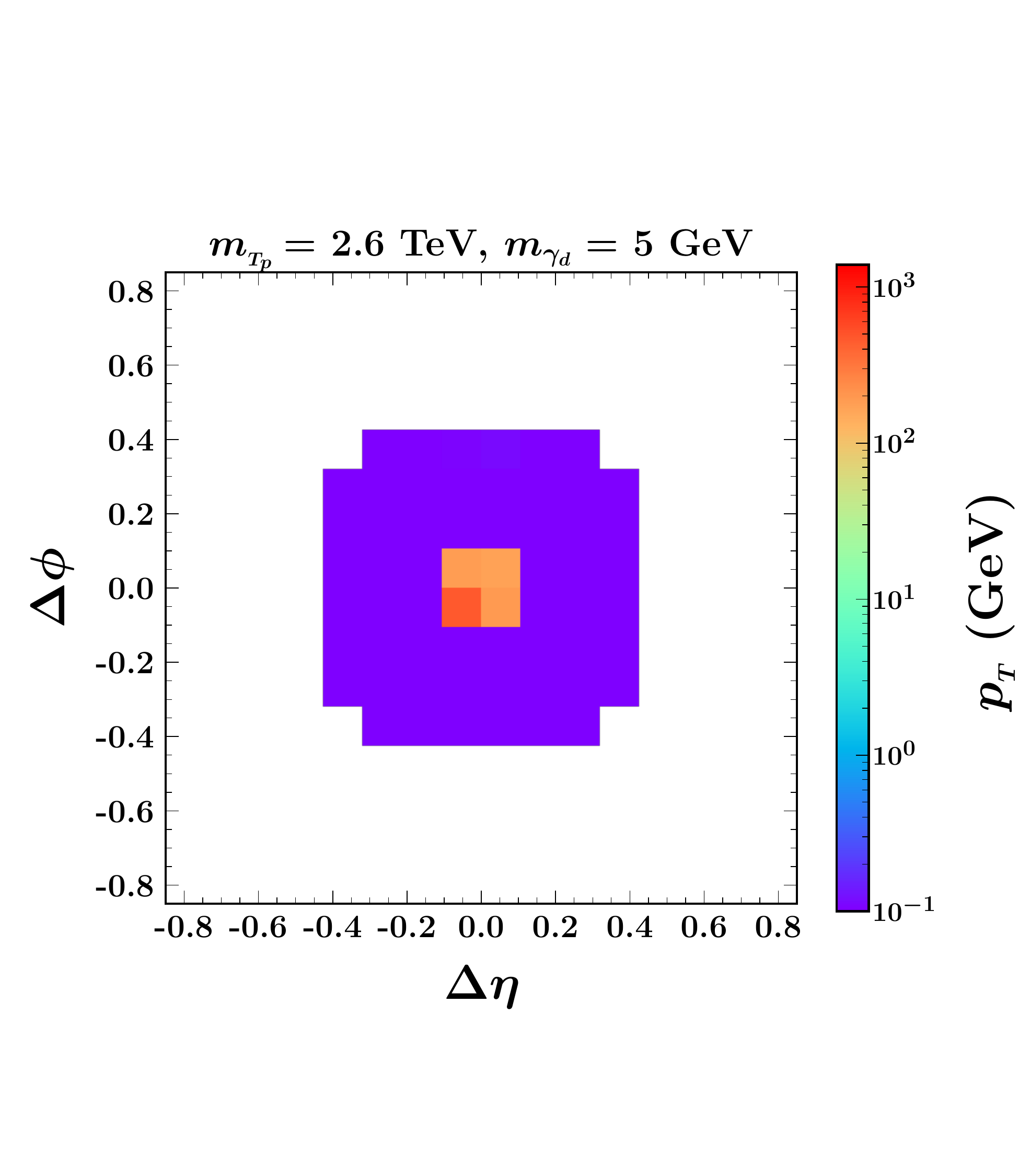}}
  }

  \resizebox{\columnwidth}{!}{
  \subfloat[\label{subfig:jetimage_10_pt1_0pt8}]{\includegraphics[width=0.5\columnwidth]{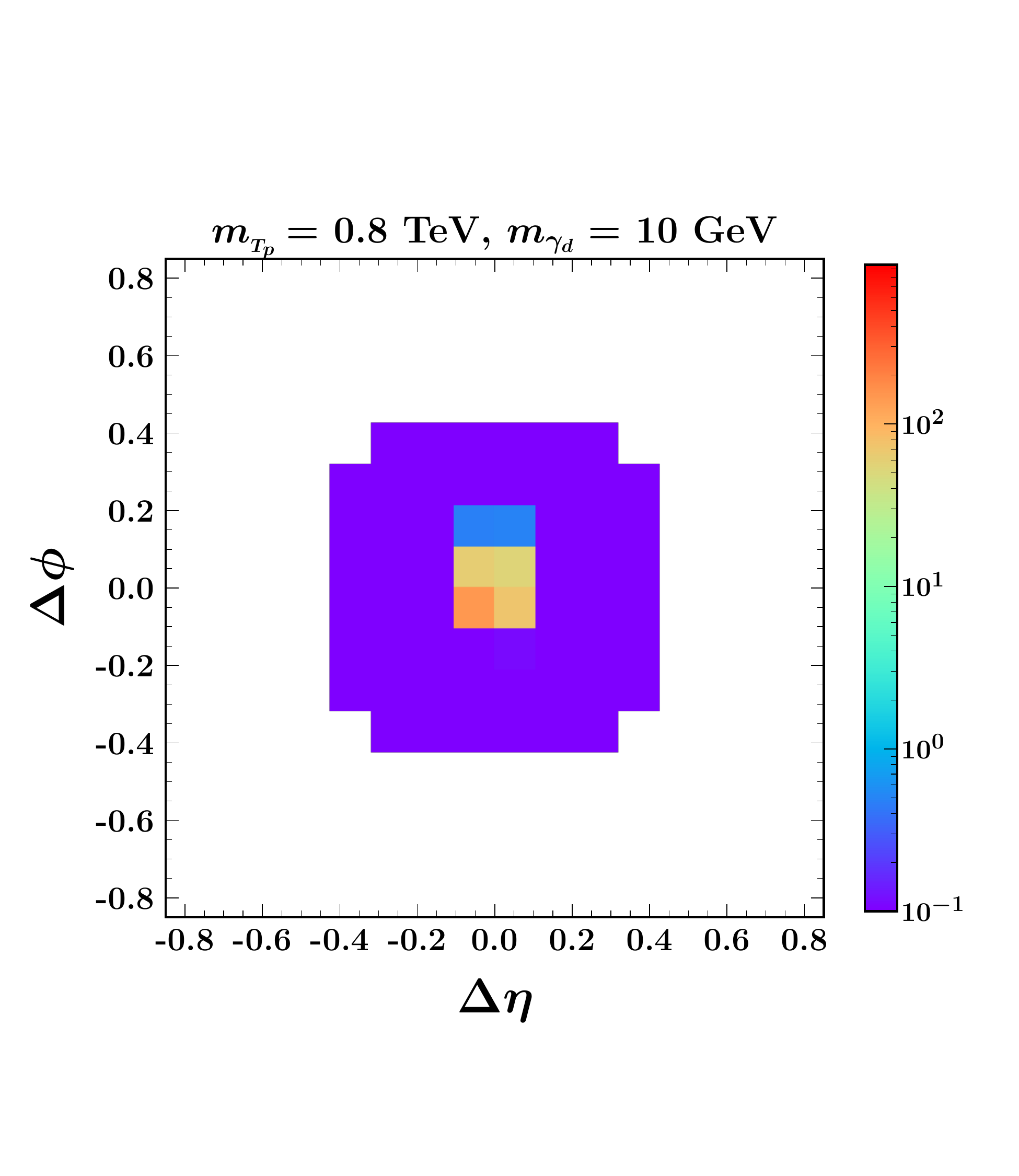}}~
  \subfloat[\label{subfig:jetimage_10_pt1_1pt4}]{\includegraphics[width=0.5\columnwidth]{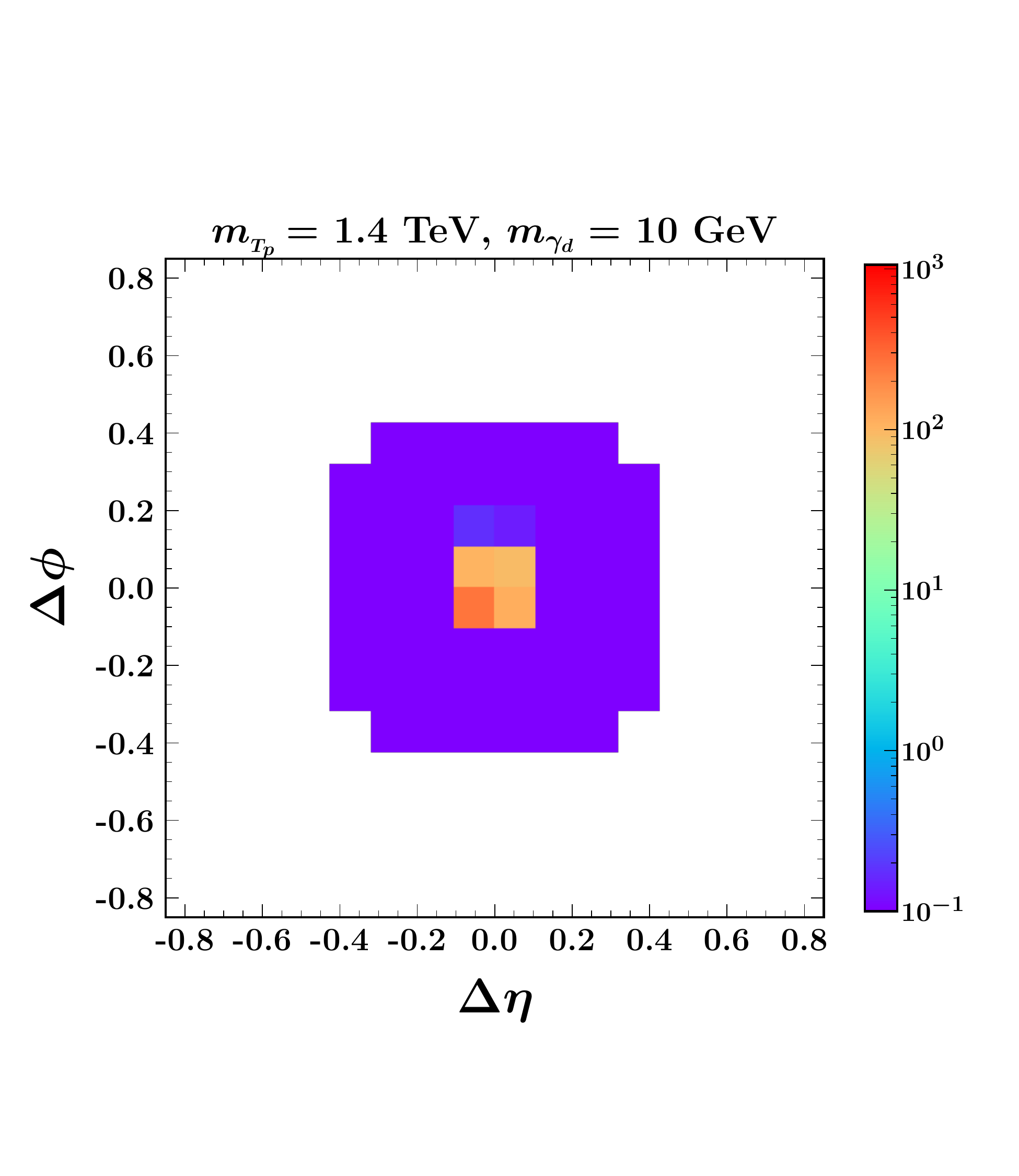}}~
  \subfloat[\label{subfig:jetimage_10_pt1_2pt6}]{\includegraphics[width=0.5\columnwidth]{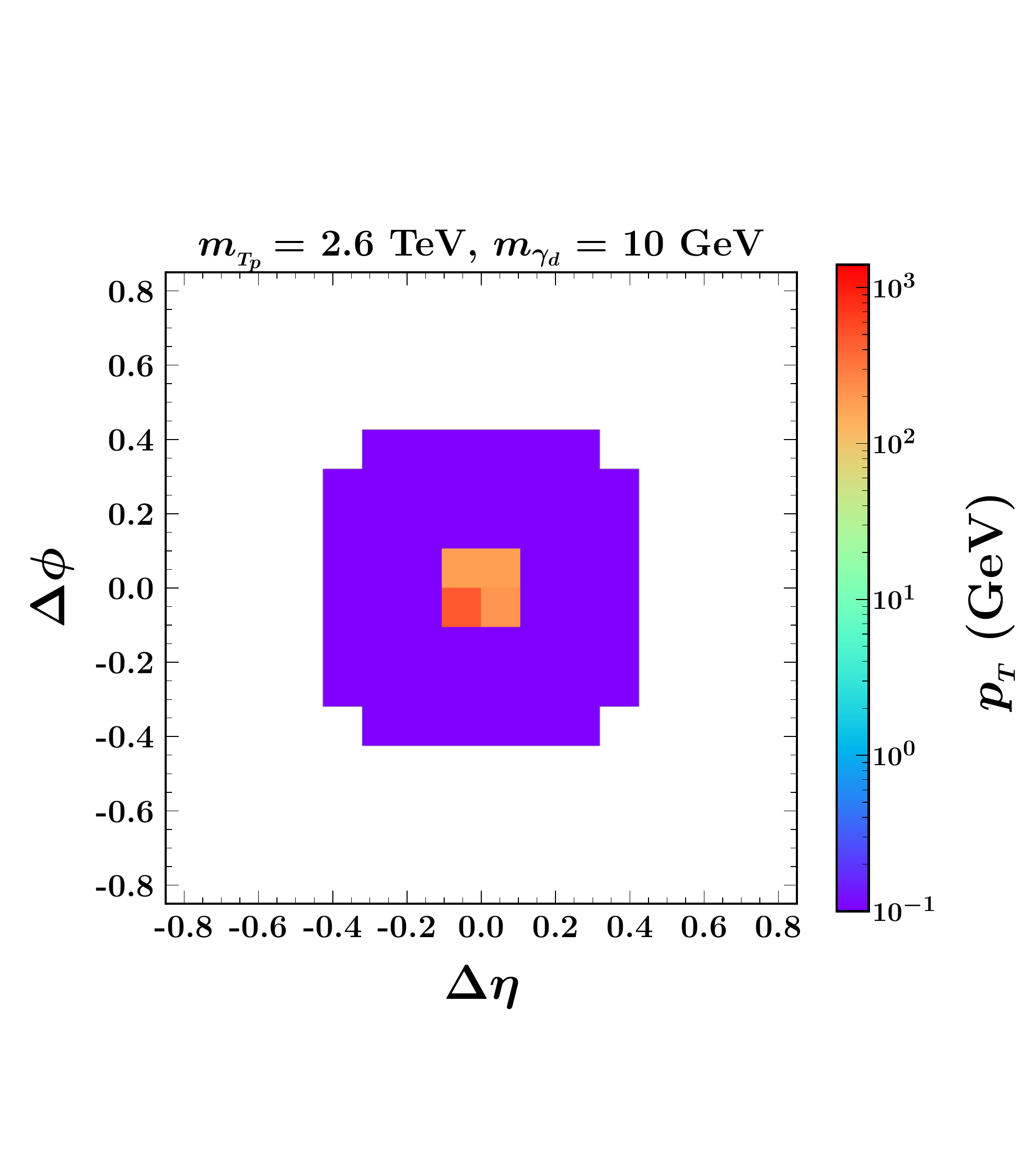}}
  }

  \resizebox{\columnwidth}{!}{
  \subfloat[\label{subfig:jetimage_20_pt1_0pt8}]{\includegraphics[width=0.5\columnwidth]{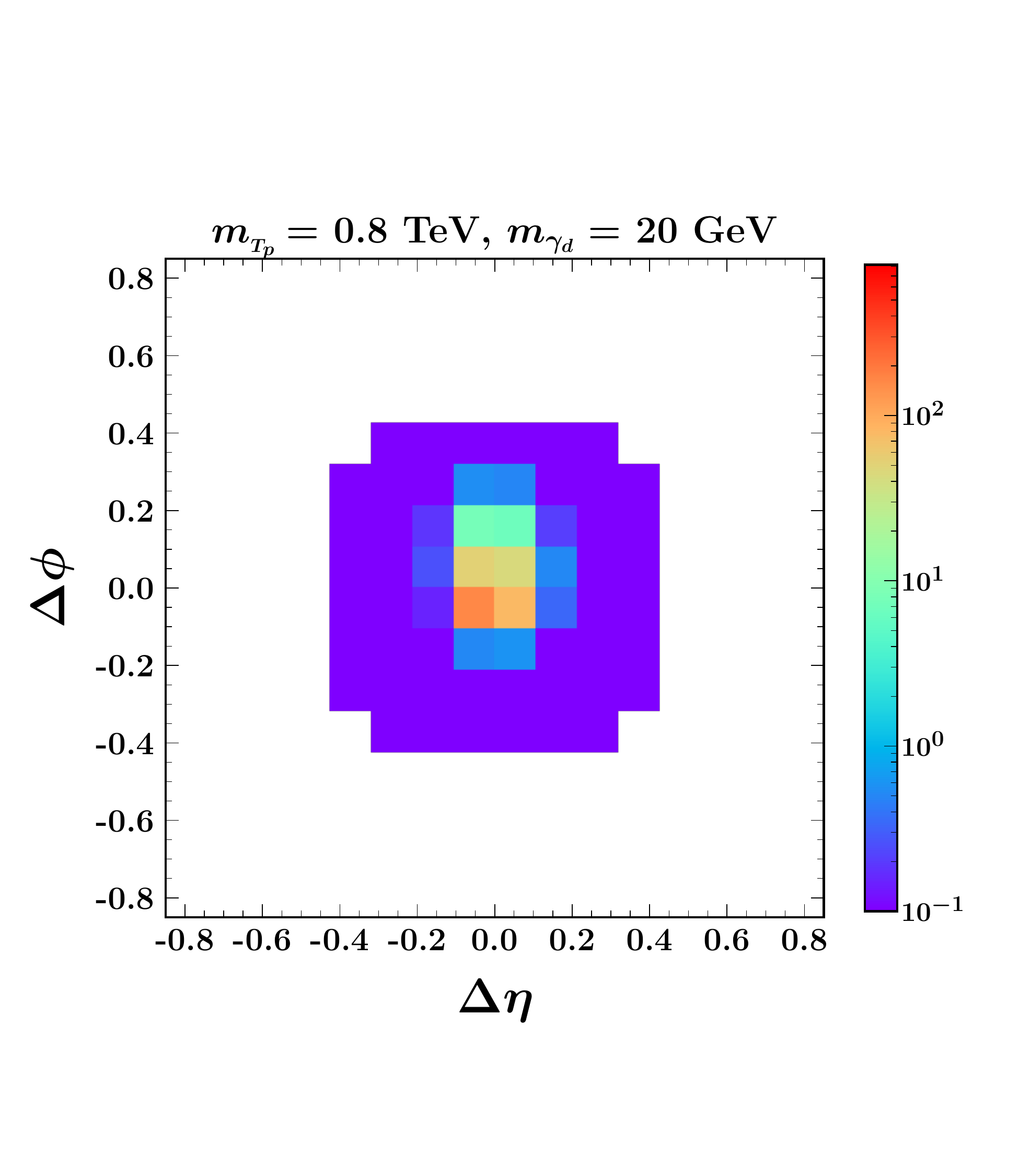}}~
  \subfloat[\label{subfig:jetimage_20_pt1_1pt4}]{\includegraphics[width=0.5\columnwidth]{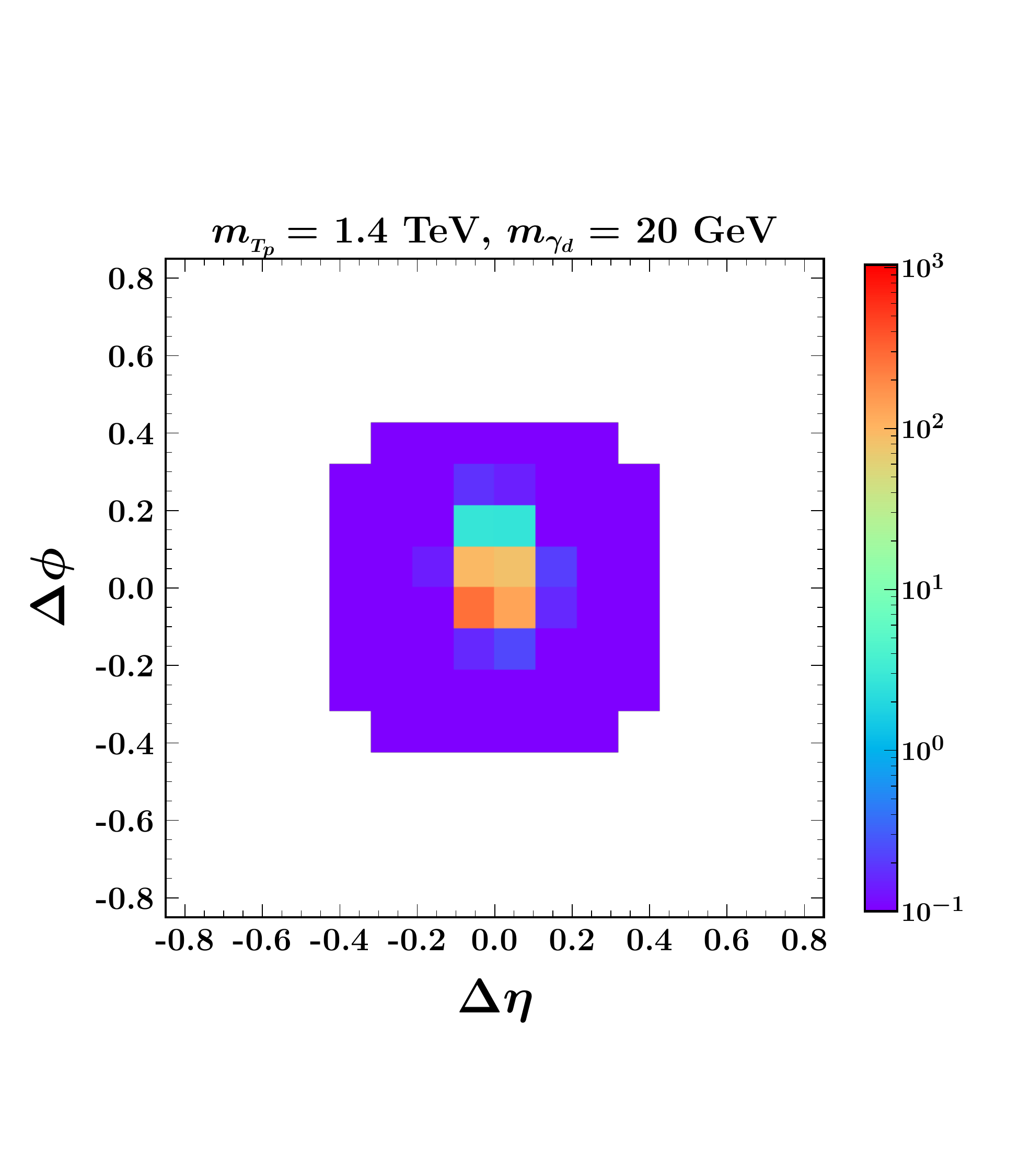}}~
  \subfloat[\label{subfig:jetimage_20_pt1_2pt6}]{\includegraphics[width=0.5\columnwidth]{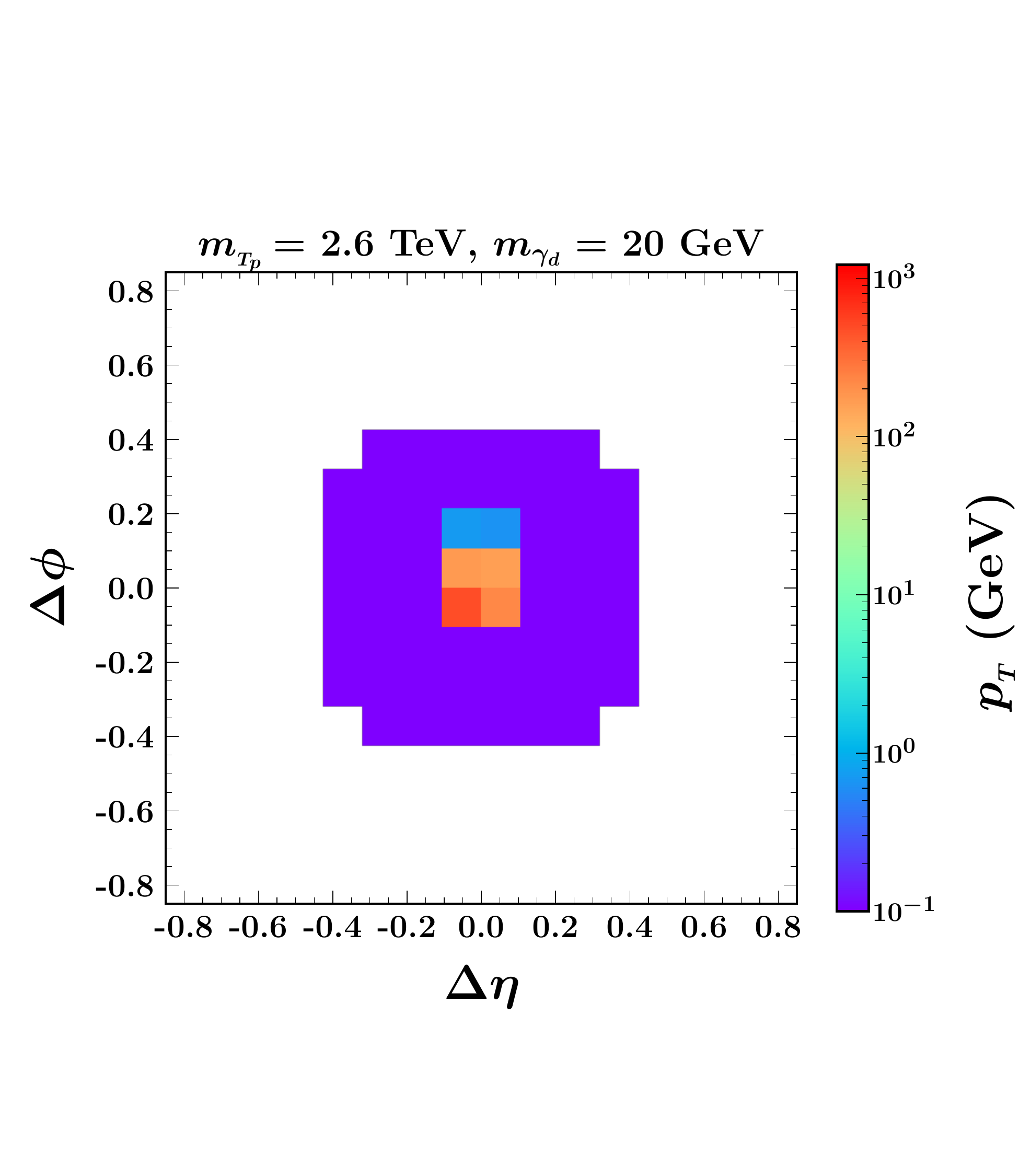}}
  }
  \caption{{ Average jet images of true dark photon jets in the $\Delta\eta-\Delta\phi$ plane, spanning the range $-0.4$ to $0.4$ with a granularity of $0.1 \times 0.1$. The individual sub-plots in this figure correspond to different combinations of $[\mgd, \mtp]$ as follows: (a) [5 GeV, 0.8 TeV], (b) [5 GeV, 1.4 TeV], (c) [5 GeV, 2.6 TeV], (d) [10 GeV, 0.8 TeV], (e) [10 GeV, 1.4 TeV], (f) [10 GeV, 2.6 TeV], (g) [20 GeV, 0.8 TeV], (h) [20 GeV, 1.4 TeV], (i) [20 GeV, 2.6 TeV]. The above jet images are obtained by using AK4 jet clustering algorithm with $ p_{_T}$-scheme recombination.}}
  \label{fig:comp_jetimages_pt1}
\end{figure}

\begin{figure}[htbp]
  \centering
  \resizebox{\columnwidth}{!}{
  \subfloat[\label{subfig:jetimage_5_pt025_0pt8}]{\includegraphics[width=0.33\columnwidth]{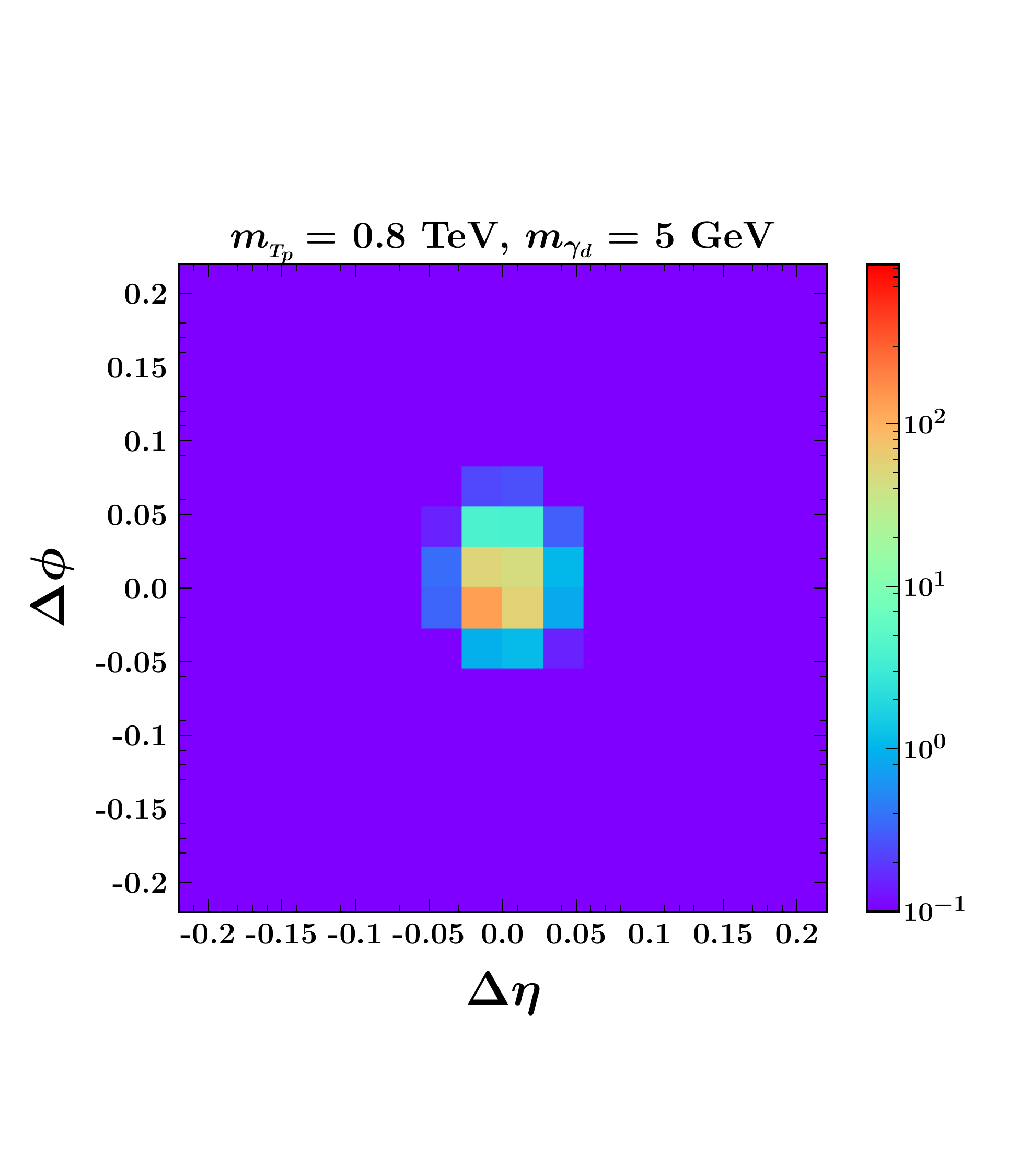}}~
  \subfloat[\label{subfig:jetimage_5_pt025_1pt4}]{\includegraphics[width=0.33\columnwidth]{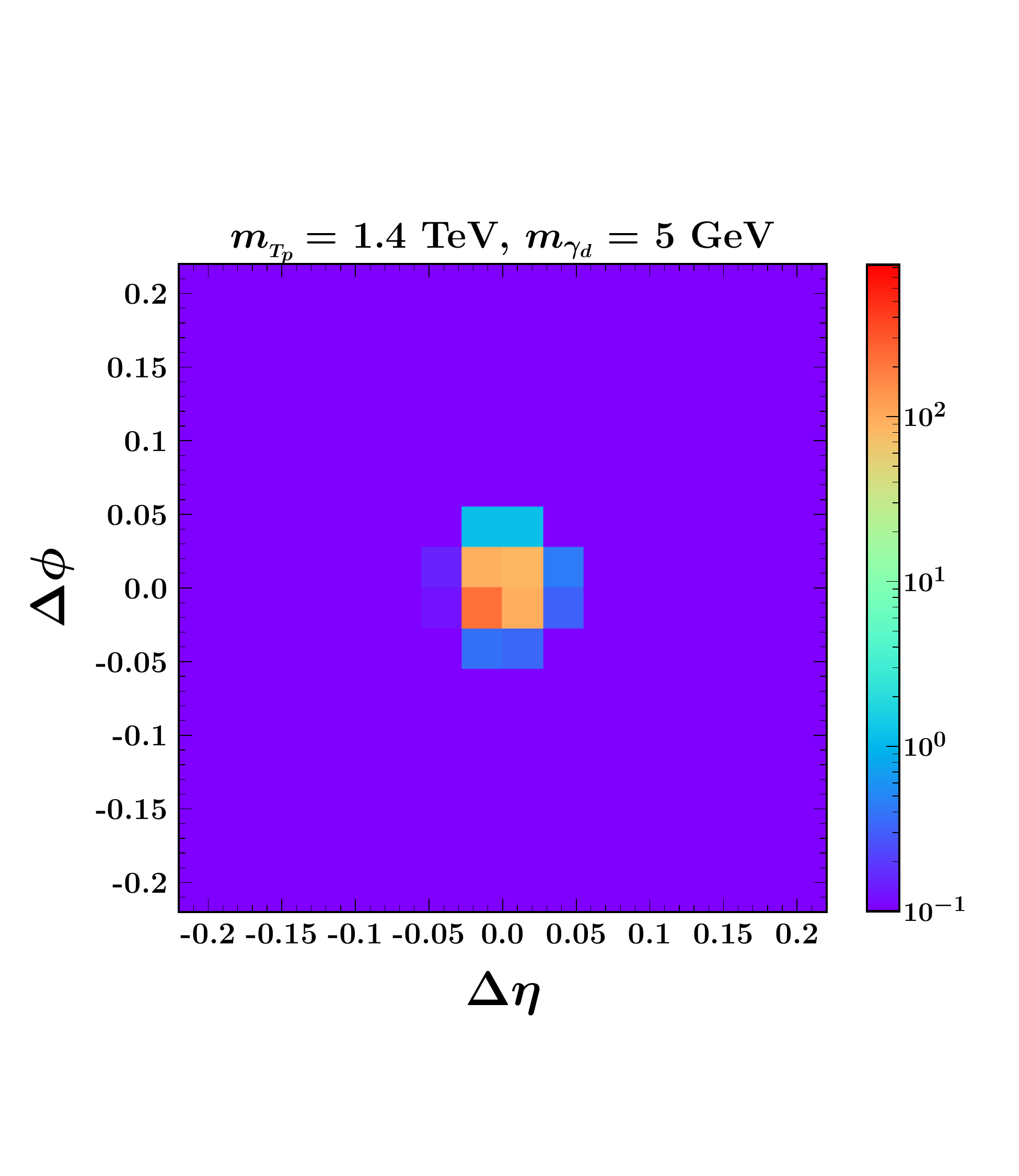}}~
  \subfloat[\label{subfig:jetimage_5_pt025_2pt6}]{\includegraphics[width=0.33\columnwidth]{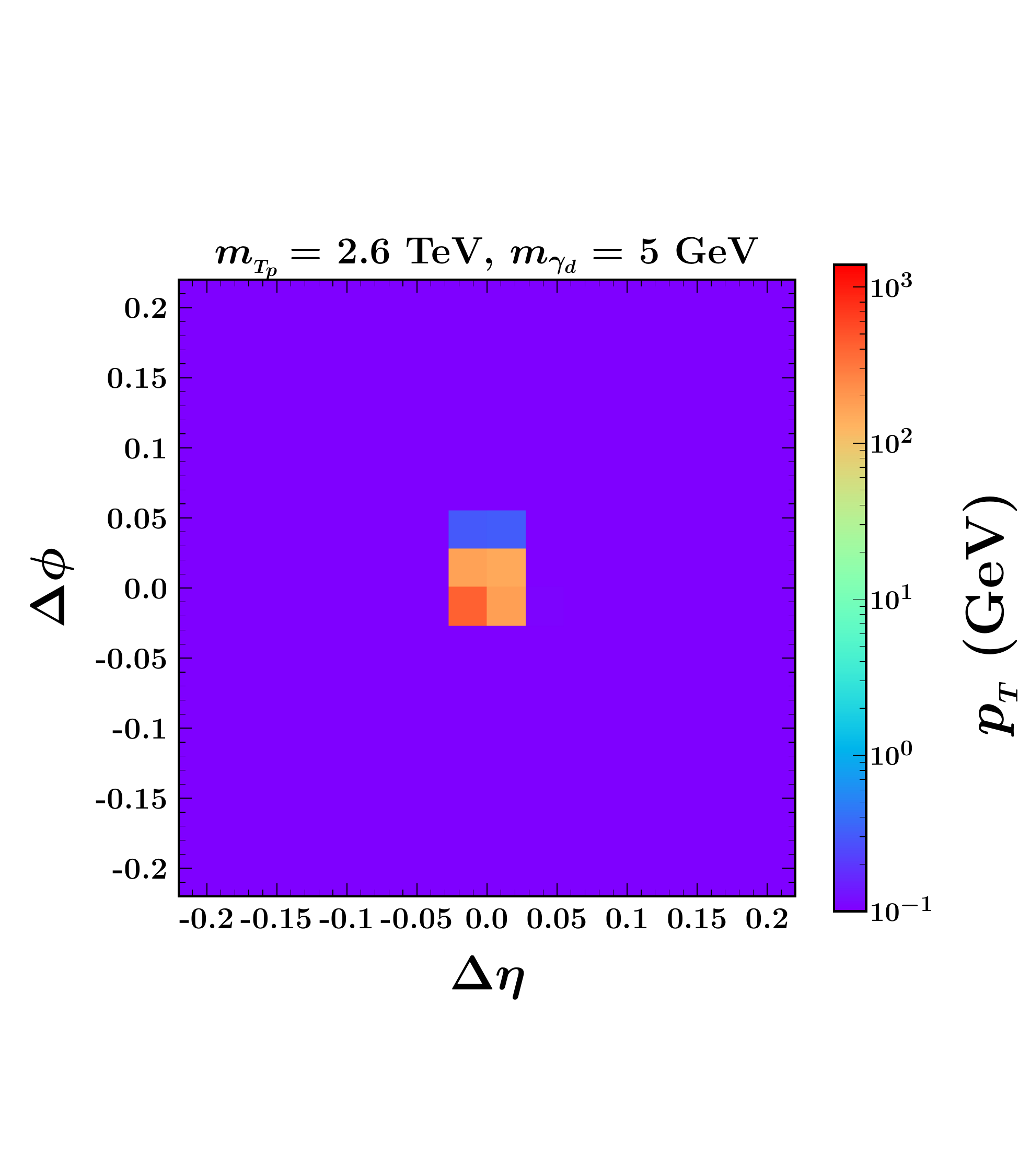}}
  }

  \resizebox{\columnwidth}{!}{
  \subfloat[\label{subfig:jetimage_10_pt025_0pt8}]{\includegraphics[width=0.33\columnwidth]{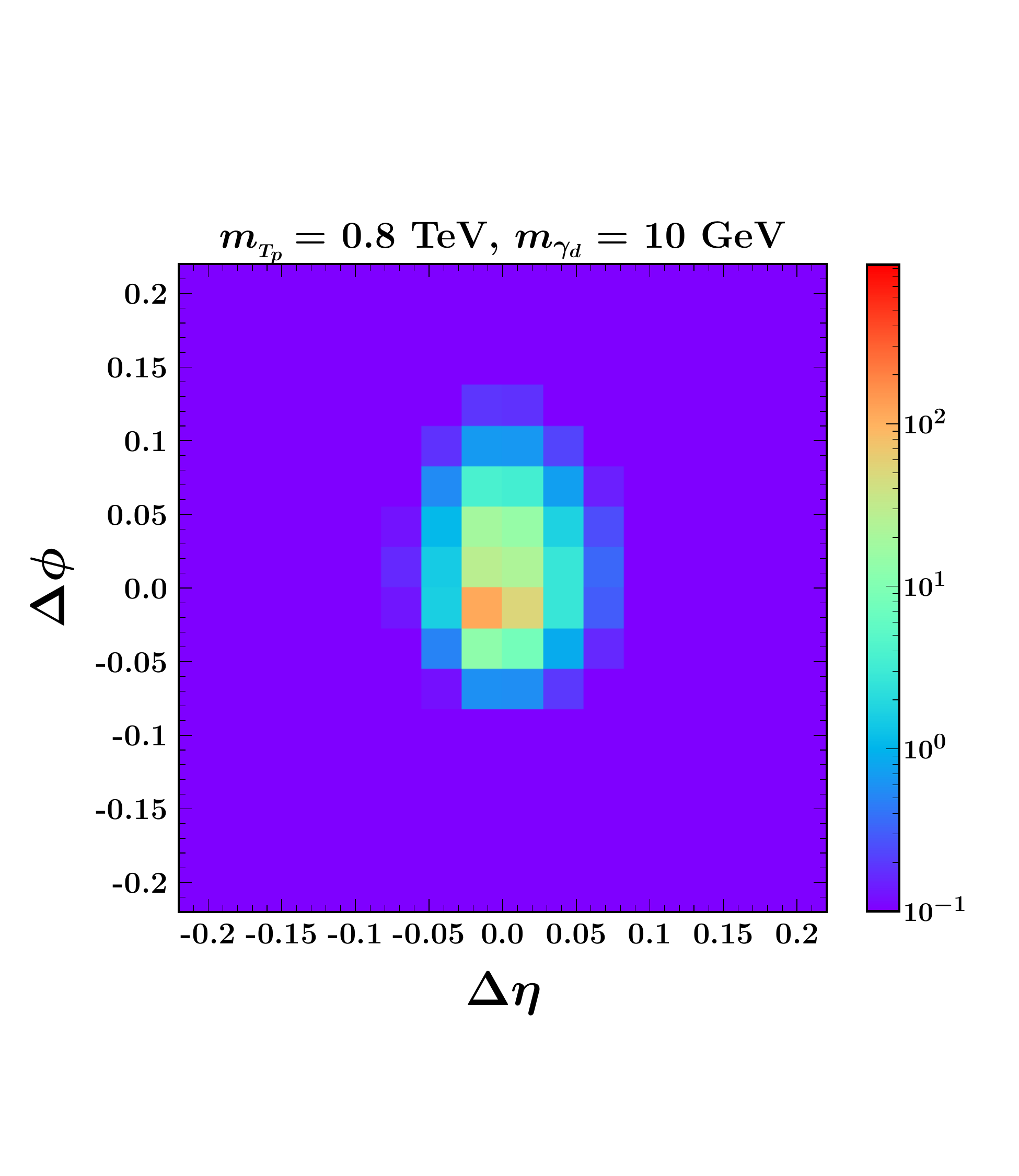}}~
  \subfloat[\label{subfig:jetimage_10_pt025_1pt4}]{\includegraphics[width=0.33\columnwidth]{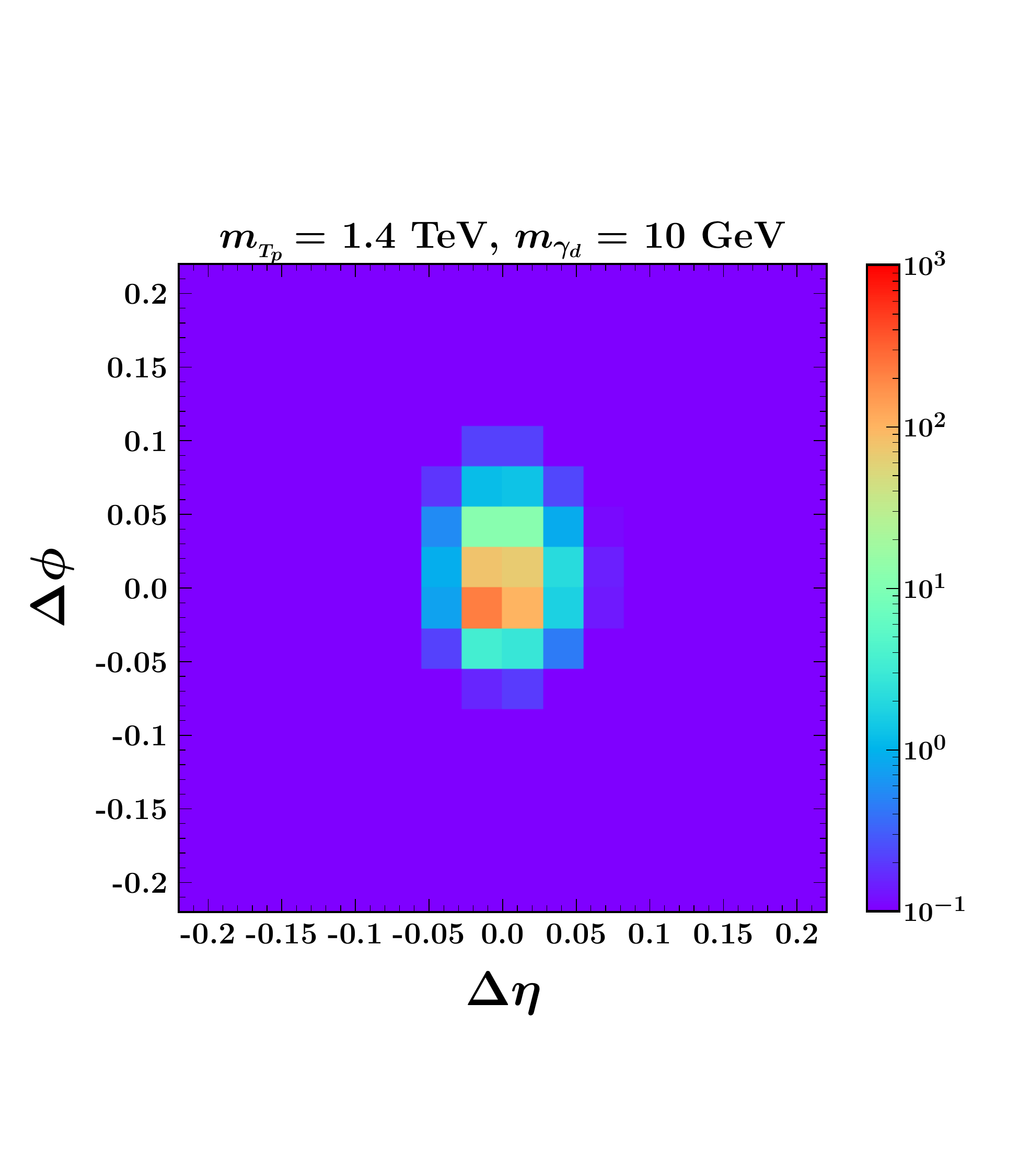}}~
  \subfloat[\label{subfig:jetimage_10_pt025_2pt6}]{\includegraphics[width=0.33\columnwidth]{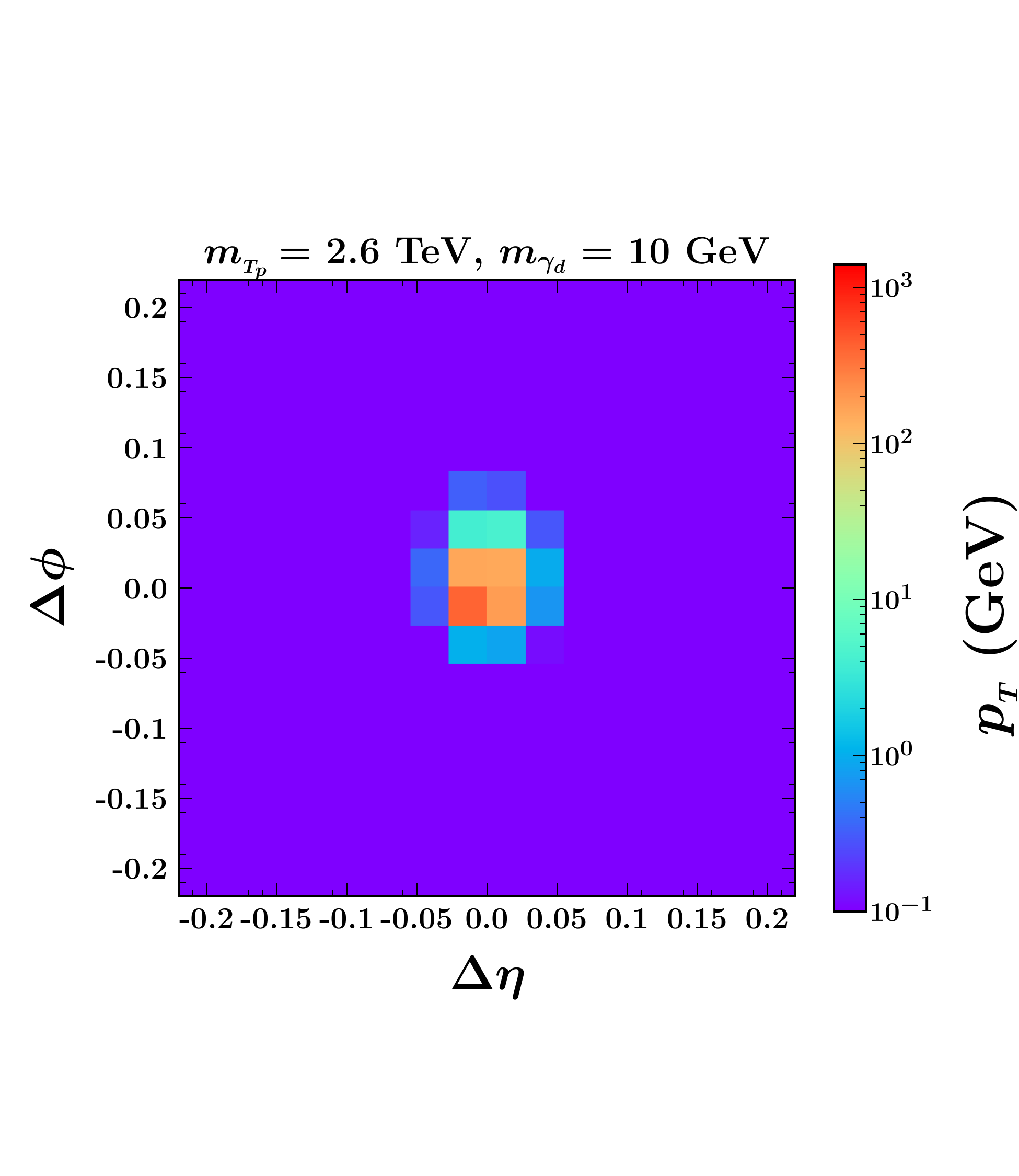}}
  }

  \resizebox{\columnwidth}{!}{
  \subfloat[\label{subfig:jetimage_20_pt025_0pt8}]{\includegraphics[width=0.33\columnwidth]{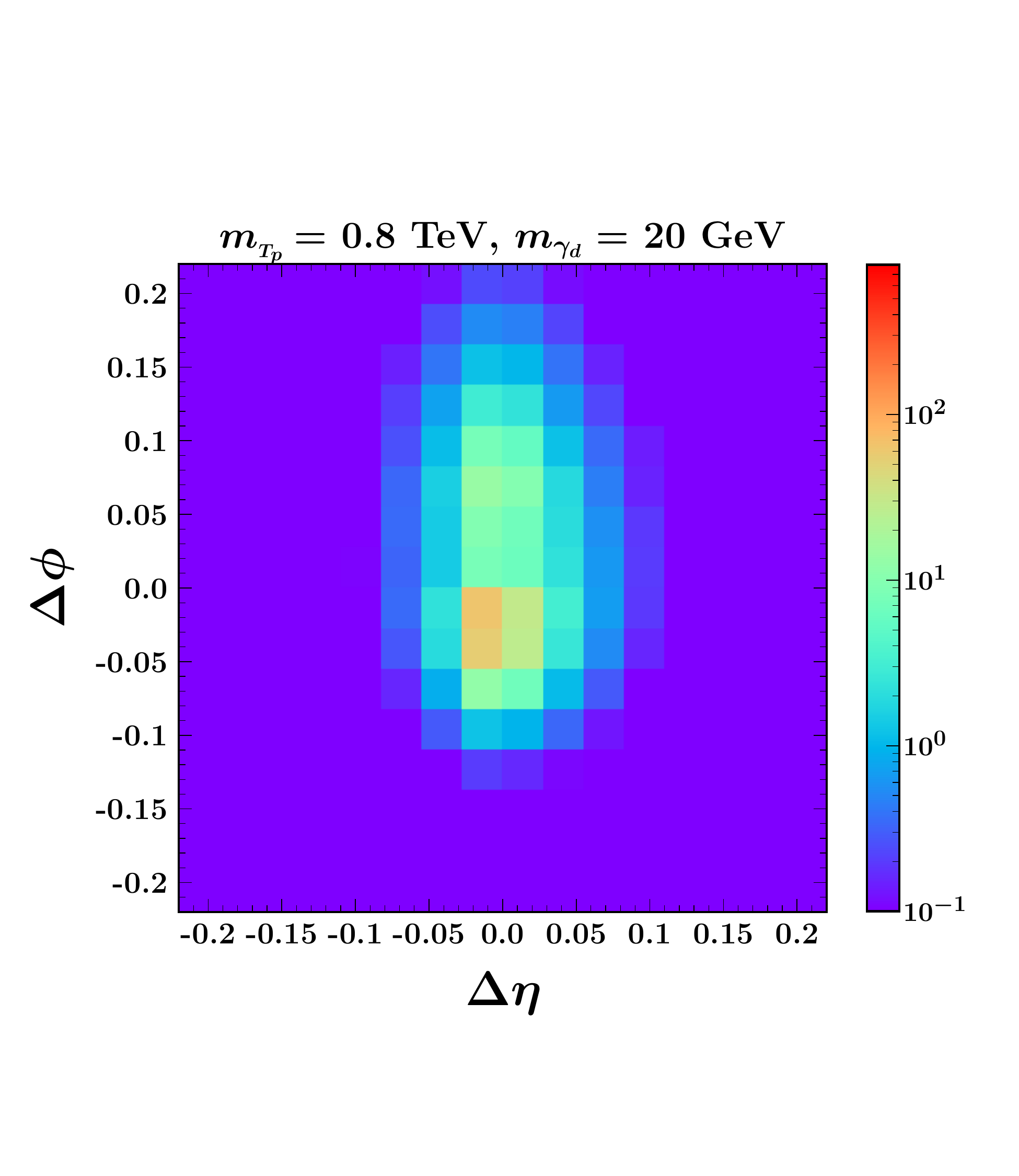}}~
  \subfloat[\label{subfig:jetimage_20_pt025_1pt4}]{\includegraphics[width=0.33\columnwidth]{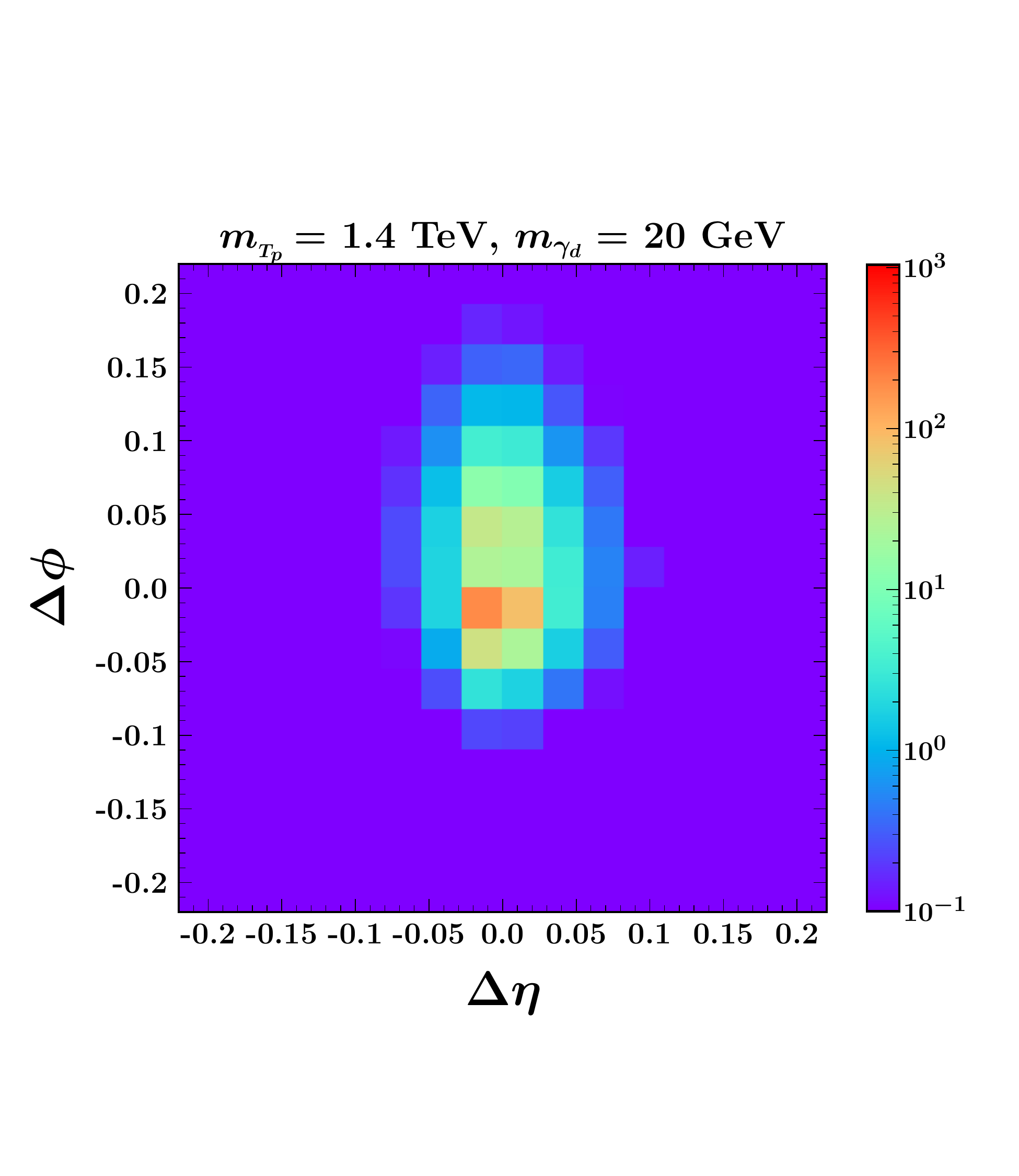}}~
  \subfloat[\label{subfig:jetimage_20_pt025_2pt6}]{\includegraphics[width=0.33\columnwidth]{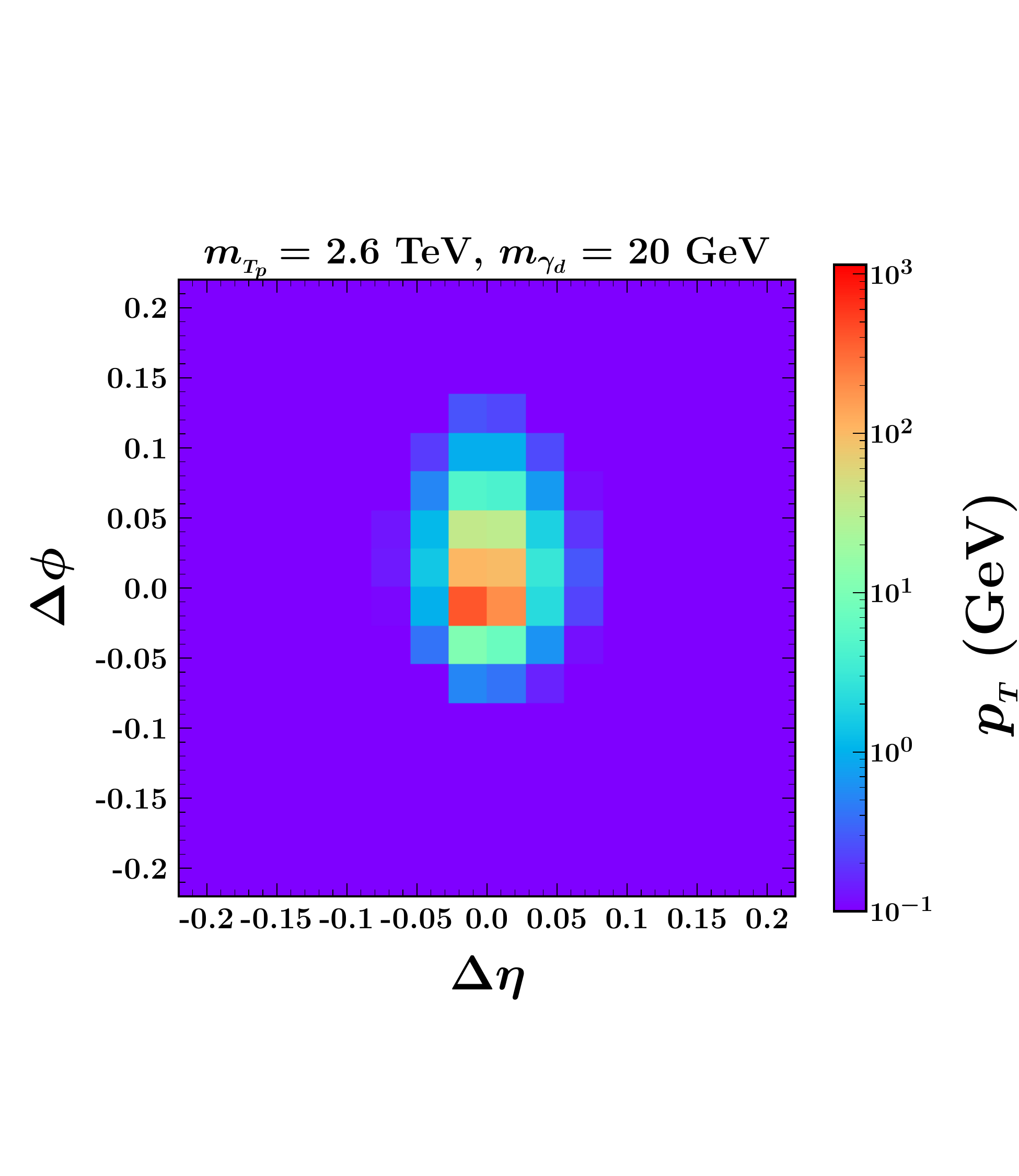}}
  }
  \caption{{ Average jet images of true dark photon jets in the $\Delta\eta-\Delta\phi$ plane, spanning the range $-0.2$ to $0.2$ with a granularity of $0.025 \times 0.025$. The individual sub-plots in this figure correspond to different combinations of $[\mgd, \mtp]$ as follows: (a) [5 GeV, 0.8 TeV], (b) [5 GeV, 1.4 TeV], (c) [5 GeV, 2.6 TeV], (d) [10 GeV, 0.8 TeV], (e) [10 GeV, 1.4 TeV], (f) [10 GeV, 2.6 TeV], (g) [20 GeV, 0.8 TeV], (h) [20 GeV, 1.4 TeV], (i) [20 GeV, 2.6 TeV]. The above jet images are obtained by using AK4 jet clustering algorithm with $ p_{_T}$-scheme recombination.}}
  \label{fig:comp_jetimages_pt025}
\end{figure}

{In Fig.~\ref{fig:comp_jetimages_pt1} and \ref{fig:comp_jetimages_pt025}, we present a comparative study of average jet images for the true dark photon jets for different sets of $\mtp$ and $m_{\gamma_d}$ and two choices of granularity, $0.1 \times 0.1$ and $0.025 \times 0.025$, respectively. A close look at Fig.~\ref{fig:comp_jetimages_pt1} and Fig.~\ref{fig:comp_jetimages_pt025} shows that as the value of $\mtp$ increases, the $p_{_T}$ deposition in the $\Delta\eta$–$\Delta\phi$ plane becomes more collimated around the center. For $\mtp$ values different from 1.4 TeV, the average jet images exhibit significant variations in their $p_{_T}$ distribution in the $\Delta\eta$–$\Delta\phi$ plane. However, for $\mtp > 1.4$ TeV, the difference in jet images compared to $\mtp = 1.4$ TeV is less likely to be captured by the classifier. This is primarily due to the limited resolution of the hadronic calorimeter and the tracker, which makes it harder to distinguish between these more energetic but similarly shaped jets (for a fixed $\mgd$).

In contrast, for $\mtp < 1.4$ TeV, the $p_{_T}$ deposition is more diffuse and spread out, extending beyond just a few central pixels in the jet image. This leads to more noticeable differences in the average jet images compared to the $\mtp = 1.4$ TeV case. Instead, they resemble more closely the features of {nondark} photon jets arising from the signal or Standard Model backgrounds. This qualitative trend is consistent with the estimated tagging efficiencies reported in Tables~\ref{table:tagrate_dph_5} and~\ref{table:tagrate_dph_10}.

We have chosen to train our HDNN classifier using a dataset corresponding to $\mtp = 1.4$~TeV. However, this is not a rigid choice. The classifier also performs well when trained on samples within the range $1.2~\text{TeV} \leq \mtp \leq 1.6~\text{TeV}$, with no significant loss in tagging efficiency for higher $\mtp$ values. The selection of \( \mtp = 1.4 \)~TeV is motivated by the observation that the HDNN classifier maintains stable performance for \( \mtp > 1.4 \)~TeV, while still performing reasonably well at lower values, such as \( \mtp = 1.0 \) and \( 1.2 \)~TeV. This robustness is primarily due to the reasons explained above. As expected, the classifier's performance begins to deteriorate significantly when applied to much lower \( \mtp \) values, such as \( \mtp = 0.8 \)~TeV.

In Table~\ref{table:tagrate_dph_5}, we summarize the tagging  efficiencies of identifying a true dark photon jet over a wide range of parameter space using a HDNN framework that is trained over a sample corresponding to $m_{_{T_p}} = 1.4 ~{\rm TeV},~m_{\gamma_d} = 5 $ GeV. The range of parameter space includes $\mtp$ in the range $\{0.8,2.8\}$ TeV for two different choices of dark photon masses, $m_{\gamma_d} = 5,~10$ GeV. To distinguish dark photon-initiated jets from {nondark} photon jets, we impose a threshold value 0.4 on the binary classifier probability score. Similarly, Table~\ref{table:tagrate_dph_10} highlights the tagging efficiencies obtained with the same HDNN framework for the same range of parameters as in Table~\ref{table:tagrate_dph_5}, however, trained over a sample with $m_{_{T_p}} = 1.4 ~{\rm TeV}$ and $m_{\gamma_d} = 10$ GeV. In this case, we choose a threshold of 0.48. All other model parameters remain consistent with those detailed in \ref{subapp:model_training}. 

We also present the tagging efficiencies for identifying true dark photon jets in Table~\ref{table:tagging_rates_5_10_20}, using a HDNN classifier trained over a wide range of dark photon masses ($m_{\gamma_d} = 5$, 10, and 20 GeV) for a fixed top partner mass ($\mtp = 1.4$ TeV). We use a probability threshold value 0.4 to achieve an optimized tagging efficiency for higher values of $\mtp$ comparable to the dedicated classifiers trained on individual dark photon mass values. However, this approach normally results in a higher background {mistagging} rate. Additionally, we depict the model's performance on intermediate mass points($m_{\gamma_d} = 7$ GeV and 15 GeV) to assess its generalization capability beyond the specific training points in the same table.

The HDNN classifier yields an improved tagging ({mistagging}) efficiencies when compared with that obtained by a simple cut-based selection--where a jet is identified as a dark photon-initiated jet if its $p_{_T} \geq 400$ GeV and both the $p_{_T}$ fraction and charge multiplicity fraction exceed 0.8. For $\mtp = 1.4$ TeV and $\mgd=10$ GeV, the naive cut based analysis yields a tagging efficiency of approximately 63\%, albeit with a significantly higher average {mistagging} rate of $\sim$11\%.}

\begin{table}[htbp]
  \centering
  \resizebox{0.8\columnwidth}{!}{
    \begin{tabular}{c c c  c c c c c c}
      \toprule
      \multirow{2}{*}{ ${ m_{\gamma_d}}$ (GeV)}& \multicolumn{7}{c}{$m_{_{T_p}}$ (TeV)} \\
      \cmidrule{2-9}
      & 0.8 & 1.0 & 1.2 & 1.4 & 1.6 & 2.0 & 2.6 & 2.8 \\
      \midrule
      \midrule
      5 & 86 & 89 & 91 & 93 & 94 & 95 & 95 & 95 \\
      10 & 30 & 51 & 66 & 76 & 82 & 88 & 92 & 93 \\
      \bottomrule
    \end{tabular}
  }
  \caption{{ Identification efficiency of the true dark photon jet for various choices of $\mtp$ in the range, $\{0.8,2.8\}$ TeV using the HDNN framework for two different dark photon masses. The training sample corresponds to $m_{\gamma_d} = 5$ GeV and $\mtp =1.4$ TeV as mentioned in the text (see {Appendix} \ref{subapp:model_training}). }}
  \label{table:tagrate_dph_5}
\end{table}

\begin{table}[htbp]
  \centering
  \resizebox{0.8\columnwidth}{!}{
    \begin{tabular}{c c c c c c c c c}
      \toprule
        \multirow{2}{*}{ ${ m_{\gamma_d}}$ (GeV)}& \multicolumn{7}{c}{$m_{_{T_p}}$ (TeV)} \\
        \cmidrule{2-9}
      & 0.8 & 1.0 & 1.2 & 1.4 & 1.6 & 2.0 & 2.6 & 2.8 \\
      \midrule
      \midrule
      5 & 59 & 74 & 83 & 87 & 89 & 92 & 94 & 94 \\
      10 & 77 & 85 & 88 & 91 & 93 & 94 & 95 & 95 \\
      \bottomrule
    \end{tabular}
  }
  \caption{{ The same as in Table~\ref{table:tagrate_dph_5}, however, the training sample corresponds to $m_{\gamma_d} = 10$ GeV  and $\mtp =1.4$ TeV with all other parameters kept same as mentioned in the text. }}
  \label{table:tagrate_dph_10}
\end{table}

\begin{table}[htbp]
  \centering
  \resizebox{0.8\columnwidth}{!}{
    \begin{tabular}{ c  c c c c c c c c}
      \toprule
      \multirow{2}{*}{ ${m_{\gamma_d}}$ {(GeV)}} & \multicolumn{8}{c}{ ${m_{_{T_p}}}$ { (TeV)}} \\ 
      \cmidrule{2-9}
      & 0.8 & 1.0 & 1.2 & 1.4 & 1.6 & 2.0 & 2.6 & 2.8 \\ 
      \midrule
      \midrule
      {\bf 5}  & 88.1 & 90.5 & 92.4 & 93.3 & 93.9 & 94.5 & 94.9 & 94.9 \\ 
        
        7  & 83.7 & 89.3 & 90.7 & 92.8 & 93.7 & 94.4 & 94.9 & 95.3 \\ 
        
        {\bf 10} & 73.6 & 83.5 & 87.3 & 90.6 & 92.3 & 93.7 & 94.9 & 94.9 \\ 
        
        15 & 63.2 & 77.2 & 83.4 & 87.7 & 89.6 & 92.4 & 94.4 & 94.2 \\ 
        
        {\bf 20} & 53.5 & 69.2 & 79.5 & 85.2 & 88.4 & 91.4 & 93.3 & 93.7 \\
      \bottomrule
  \end{tabular}
  }
  \caption{ Identification efficiency of the true dark photon jet for various choices of $\mtp$ in the range, $\{0.8,2.8\}$ TeV using the HDNN framework for different choices of dark photon masses in the range, $\{5,20\}$ GeV. The training samples correspond to $m_{\gamma_d} = 5,~10,$ and $20$ GeV with $\mtp= 1.4 $ TeV.}
  \label{table:tagging_rates_5_10_20}
\end{table}

{
\section[STL Dependence of Cross Sections and Branching Ratios]{Dependency of $\sigma_c(T_p \bar{T}_p) \cdot {\rm BR}^2$ and $\sigma_c(T_p j) \cdot {\rm BR}$ on $\stl$}\label{app:excl}

In this section, we illustrate the $\stl$ dependency of the signal {cross section}  obtained after imposing the $C_0$ cuts in two different final states, namely, AD2AT1 in Fig.~\ref{fig:excl_explain_ad2at1} and ED1ET1 in Fig.~\ref{fig:excl_explain_ed1et1}.

\begin{figure}[htbp]
  \centering
  \resizebox{\columnwidth}{!}{
  \subfloat[\label{subfig:bp2_ad2at1_1k}]{\includegraphics[width=0.5\columnwidth]{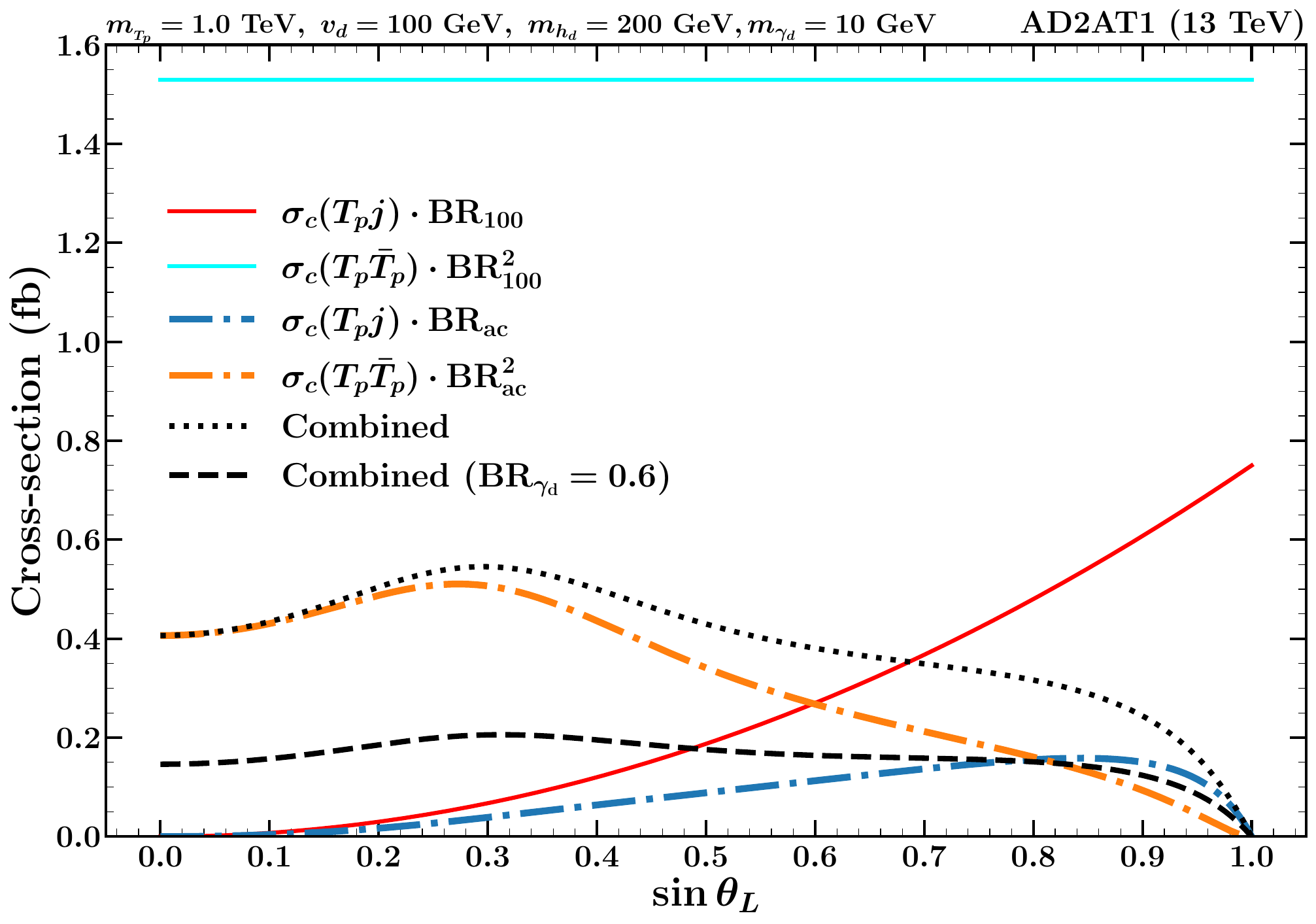}}~
  \subfloat[\label{subfig:bp2_ad2at1_1pt2k}]{\includegraphics[width=0.5\columnwidth]{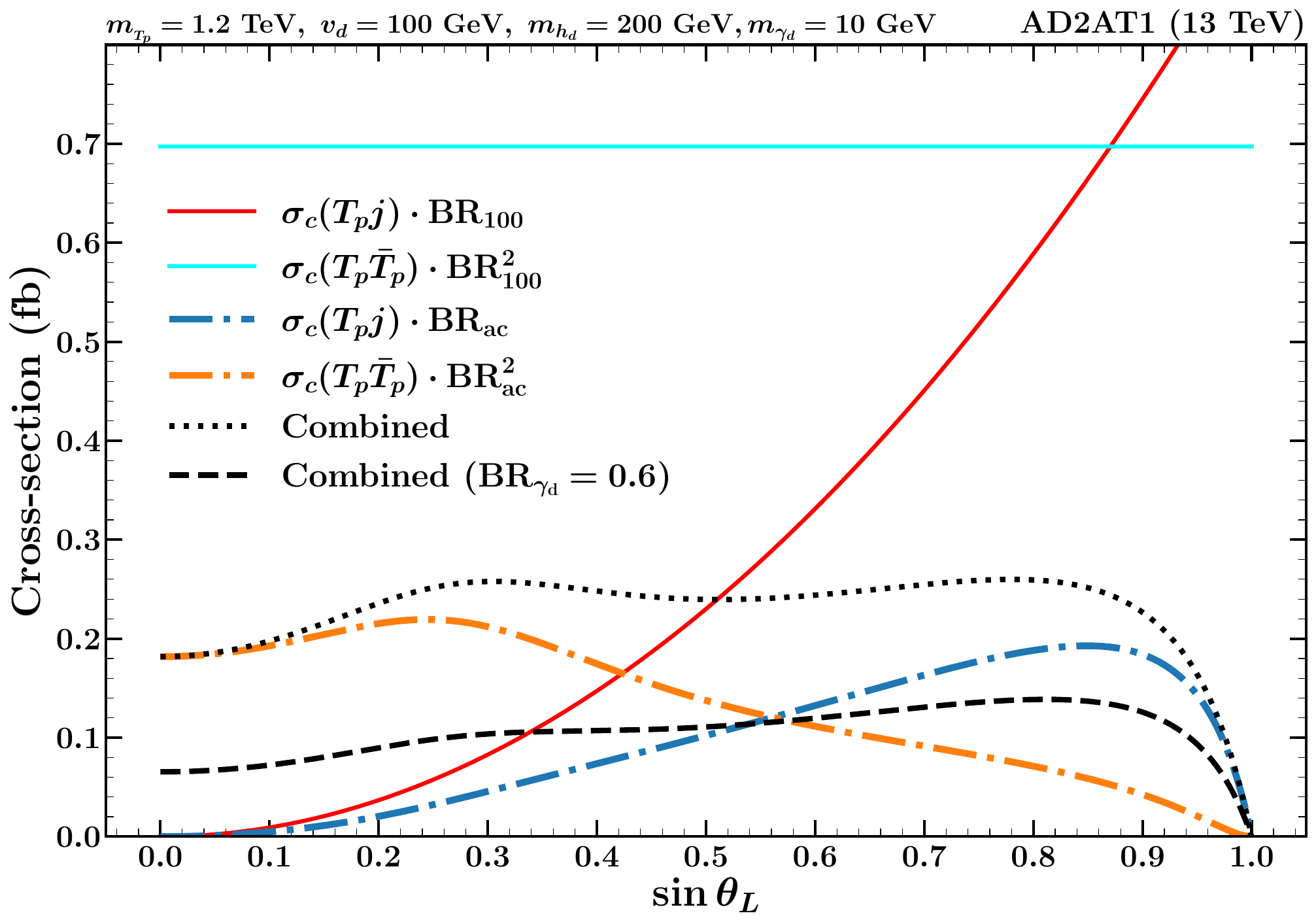}}
  }

  \resizebox{\columnwidth}{!}{
  \subfloat[\label{subfig:bp1_ad2at1_1k}]{\includegraphics[width=0.5\columnwidth]{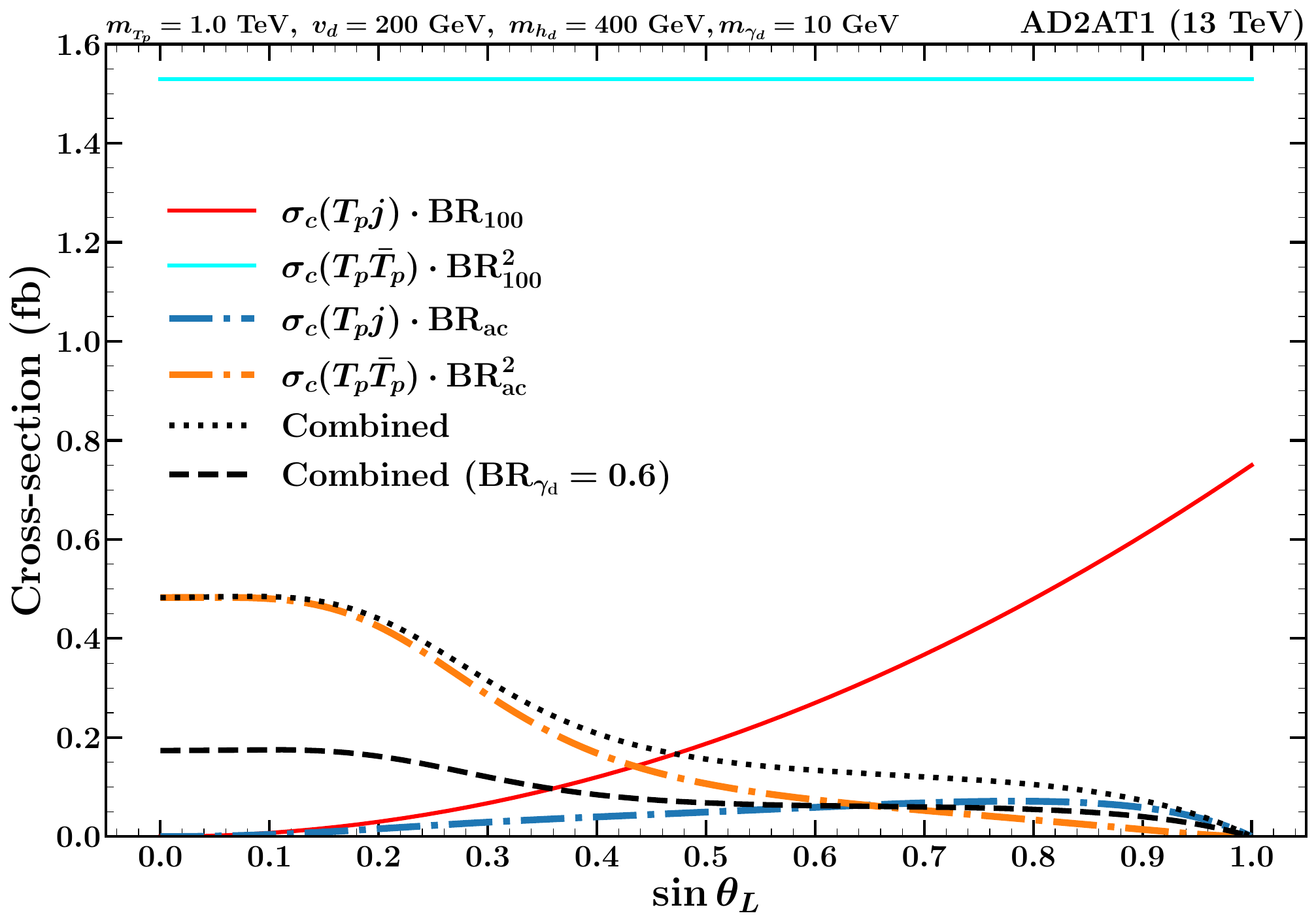}}~
  \subfloat[\label{subfig:bp1_ad2at1_1pt2k}]{\includegraphics[width=0.5\columnwidth]{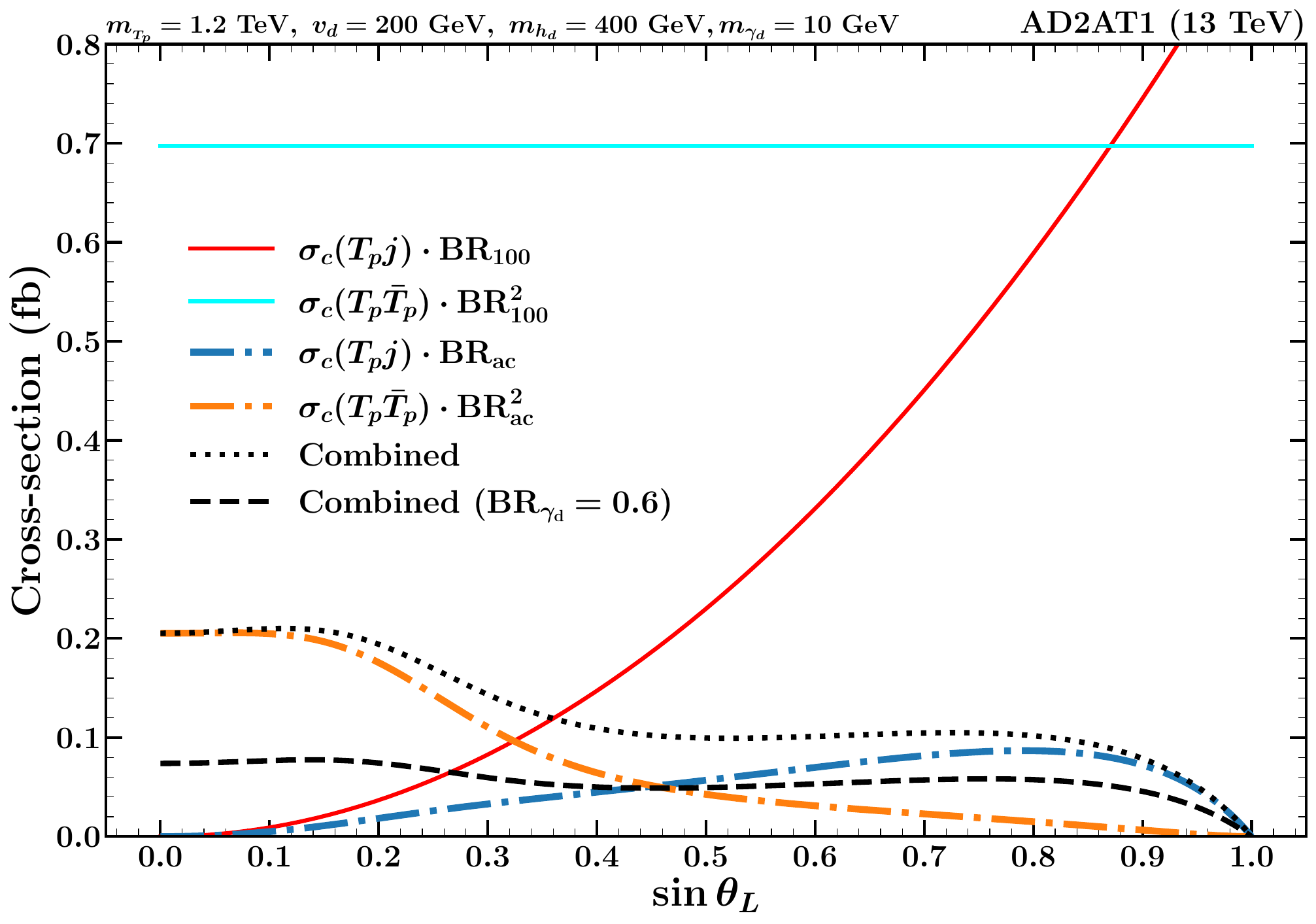}}
  }
  \caption{The solid cyan (red) line represents the contribution to the total {cross section}  in the {\bf AD2AT1} final state coming from $T_{p}\bar{T}_p$ ($T_{p}j$) {subprocess} after applying the $\mtp$ dependent invariant mass cut listed in Table~\ref{table:cuts} assuming BR($T_p \to  t \gamma_d$) = 100\%. The dashed-dot blue (orange) line depicts the same coming from $T_{p}\bar{T}_p$ ($T_{p}j$) after applying the invariant mass cut assuming {\it actual} BR($T_p \to  t \gamma_d$). The dotted black represents the sum of the two contributions with {\it actual} BR($T_p \to  t \gamma_d$). All the curves expect dashed black line correspond to BR($\gamma_d \to hadrons$)= 100\%. The dashed black line corresponds to the combined contributions with BR($\gamma_d \to hadrons$)= 60\%. The {subplots} illustrate four representative benchmark points: (a) and (b) correspond to \( v_d = 100 \) GeV, \( m_{h_d} = 200 \) GeV, and \( m_{\gamma_d} = 10 \) GeV, for $\mtp =$ 1.0 TeV and 1.2 TeV, respectively. Similarly, (c) and (d) represent \( v_d = 200 \) GeV, \( m_{h_d} = 400 \) GeV, and \( m_{\gamma_d} = 10 \) GeV, with $\mtp = $ 1.0 TeV and 1.2 TeV, respectively.}
  \label{fig:excl_explain_ad2at1}
\end{figure}

\begin{figure}[htbp]
  \centering
  \resizebox{\columnwidth}{!}{
  \subfloat[\label{subfig:bp2_ed1et1_1k}]{\includegraphics[width=0.5\columnwidth]{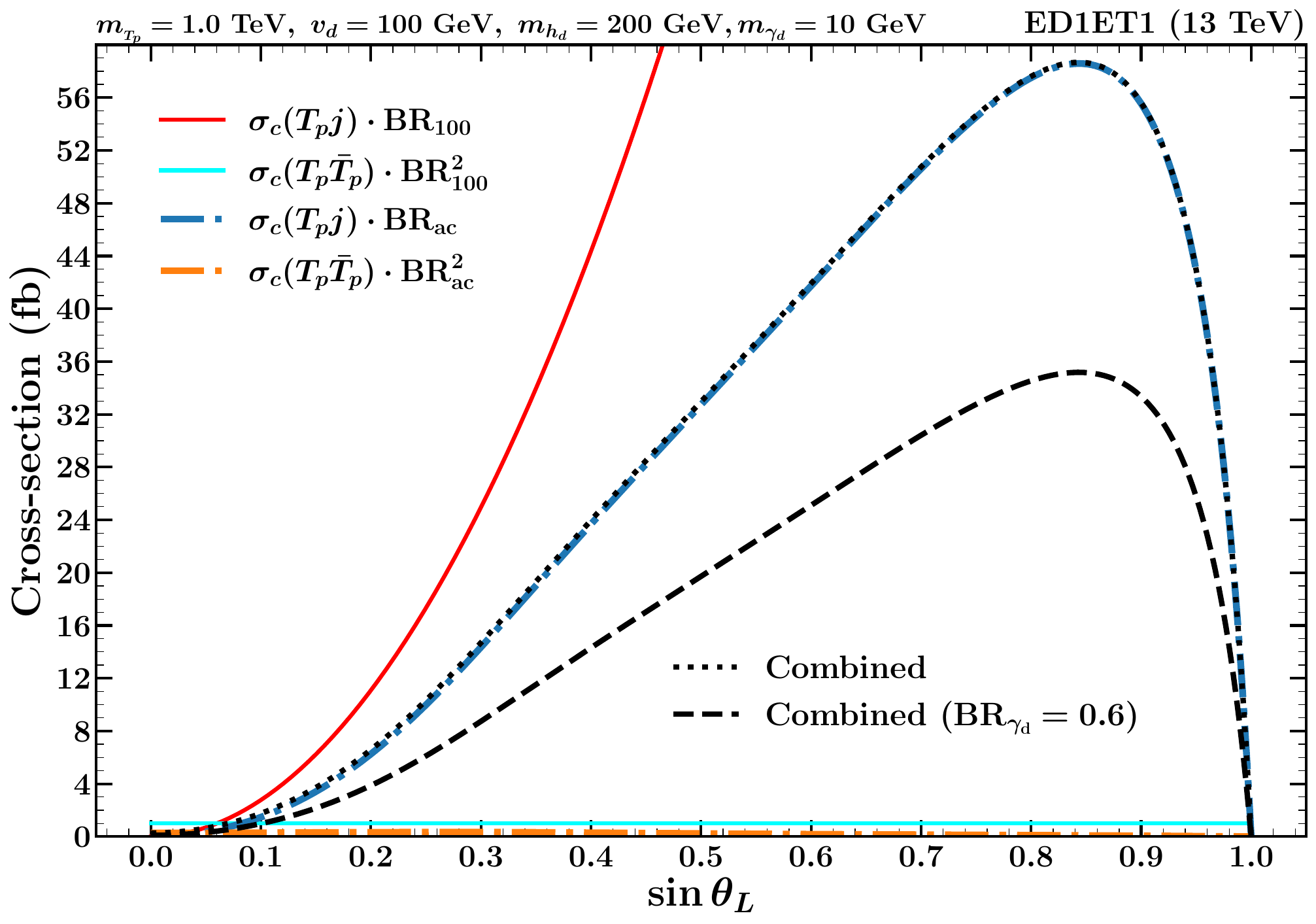}}~
  \subfloat[\label{subfig:bp2_ed1et1_1pt2k}]{\includegraphics[width=0.5\columnwidth]{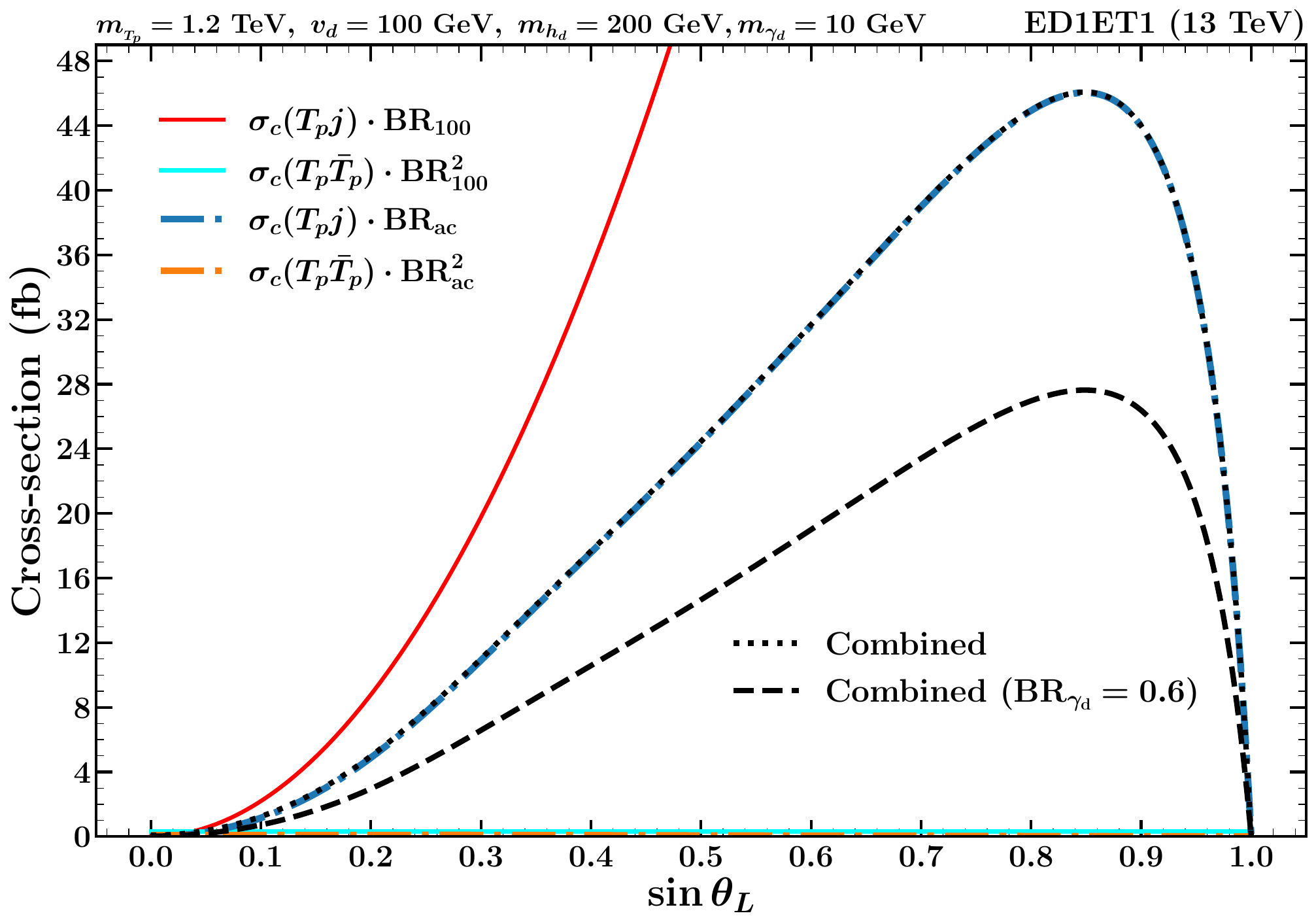}}
  }

  \resizebox{\columnwidth}{!}{
  \subfloat[\label{subfig:bp1_ed1et1_1k}]{\includegraphics[width=0.5\columnwidth]{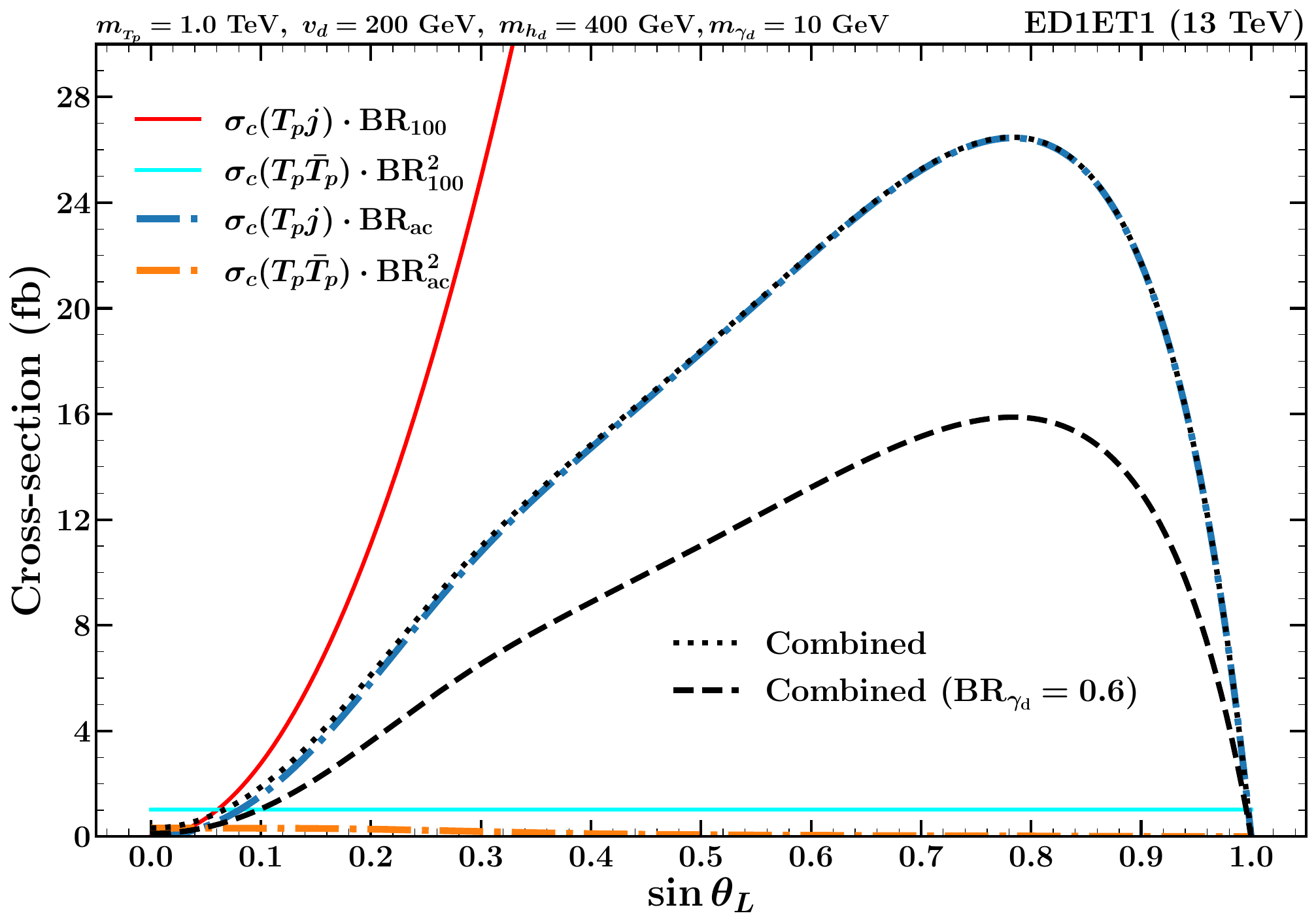}}~
  \subfloat[\label{subfig:bp1_ed1et1_1pt2k}]{\includegraphics[width=0.5\columnwidth]{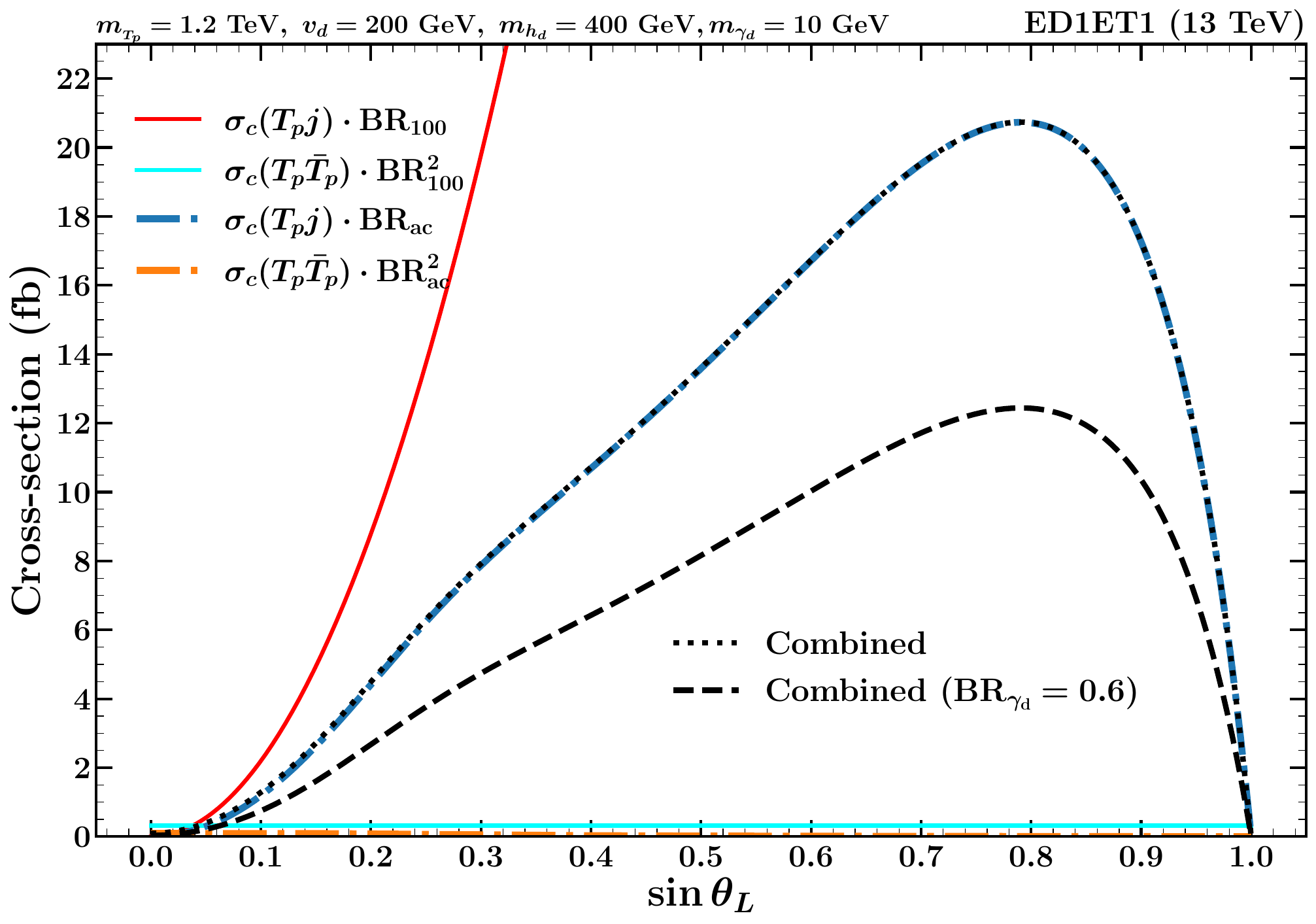}}
  }
  \caption{The solid cyan (red) line represents the contribution to the total {cross section}  in the {\bf ED1ET1} final state coming from $T_{p}\bar{T}_p$ ($T_{p}j$) {subprocess} after applying the $\mtp$ dependent invariant mass cut listed in Table~\ref{table:cuts} assuming BR($T_p \to  t \gamma_d$) = 100\%. The dashed-dot blue (orange) line depicts the same coming from $T_{p}\bar{T}_p$ ($T_{p}j$) after applying the invariant mass cut assuming {\it actual} BR($T_p \to  t \gamma_d$). The dotted black represents the sum of the two contributions with {\it actual} BR($T_p \to  t \gamma_d$). All the curves expect dashed black line correspond to BR($\gamma_d \to hadrons$)= 100\%. The dashed black line corresponds to the combined contributions with BR($\gamma_d \to hadrons$)= 60\%. The {subplots} illustrate four representative benchmark points: (a) and (b) correspond to \( v_d = 100 \) GeV, \( m_{h_d} = 200 \) GeV, and \( m_{\gamma_d} = 10 \) GeV, for $\mtp =$ 1.0 TeV and 1.2 TeV, respectively. Similarly, (c) and (d) represent \( v_d = 200 \) GeV, \( m_{h_d} = 400 \) GeV, and \( m_{\gamma_d} = 10 \) GeV, with $\mtp = $ 1.0 TeV and 1.2 TeV, respectively.
  }
  \label{fig:excl_explain_ed1et1}
\end{figure}
}

\clearpage

\providecommand{\href}[2]{#2}\begingroup\raggedright\endgroup

\end{document}